\documentclass[12pt]{article}
\pdfoutput=1
\usepackage{amsmath,amssymb,amsfonts}
\usepackage{a4wide,epsfig,psfrag,scalefnt}
\usepackage[dvipsnames]{xcolor}
\usepackage{braket}
\usepackage{placeins}
\usepackage{breqn}
\usepackage{slashed}
\usepackage{enumitem}
\usepackage[numbers,sort&compress]{natbib}
\usepackage{caption}
\usepackage{subcaption}

\usepackage{fancyvrb}

\graphicspath{ {figs/} }

\allowdisplaybreaks

\begin{document}

\title{\vskip-3cm{\baselineskip14pt
    \begin{flushleft}
     \normalsize P3H-23-008, TTP23-004, TU-1178, ZU-TH 06/23
    \end{flushleft}} \vskip1.5cm
  Analytic approximations of $2\to 2$ processes with
  massive internal particles
  }

\author{
  Joshua Davies$^{a}$,
  Go Mishima$^{b}$,
  Kay Sch\"onwald$^{c}$,
  Matthias Steinhauser$^{d}$
  \\
  {\small\it (a) Department of Physics and Astronomy, University of Sussex,
    Brighton BN1 9QH, UK}
  \\
  {\small\it (b) Department of Physics, Tohoku University, Sendai, 980-8578, Japan}
  \\
  {\small\it (c) Physik-Institut, Universit\"at Z\"urich, Winterthurerstrasse 190,}\\
  {\small\it 8057 Z\"urich, Switzerland}
  \\
  {\small\it (d) Institut f{\"u}r Theoretische Teilchenphysik,
    Karlsruhe Institute of Technology (KIT),}\\
  {\small\it 76128 Karlsruhe, Germany}
}

\date{}

\maketitle

\thispagestyle{empty}

\begin{abstract}

\noindent
We consider two-loop corrections to $2\to 2$ scattering processes with
massive particles in the final state and massive particles in the
loop.  We discuss the combination of analytic expansions in the
high-energy limit and for small Mandelstam variable~$t$.  For the
example of double Higgs boson production we show that the whole phase
space can be covered and time-consuming numerical integrations can be
avoided.

\end{abstract}


\newpage


\section{\label{sec::intro}Introduction}

In many higher-order calculations of cross sections the virtual corrections
are the bottleneck, particularly if they involve massive
particles propagating in loops.
A prominent example of such a process is Higgs boson pair
production, where the real-radiation contribution with exact dependence on
the top quark mass~\cite{Maltoni:2014eza} was available long before the
corresponding virtual
corrections~\cite{Borowka:2016ehy,Borowka:2016ypz,Baglio:2018lrj}.
One of the reasons is certainly the enormous expressions which are present in
intermediate stages of the calculation, and the complicated integrals
which in general depend on several invariants. Often
a purely numerical approach for the evaluation of the loop integrals is necessary,
which comes with the well-known disadvantages of long run-times and
reduced flexibility in the choice of values for parameters.
In this paper we suggest an alternative approach for the computation
of virtual loop integrals for $2\to 2$ processes. It is based
on the combination expansions in different kinematic regions.

We consider the scattering of two (massless) partons in the initial state with
momenta $q_1$ and $q_2$ into two massive particles in the final state with
momenta $q_3$ and $q_4$.
It is convenient to introduce the Mandelstam variables as
\begin{eqnarray}
  s = (q_1+q_2)^2\,,\qquad t = (q_1+q_3)^2\,,\qquad u = (q_1+q_4)^2\,,
\end{eqnarray}
where all momenta are incoming. Furthermore we have
\begin{eqnarray}
  q_1^2=q_2^2=0\,,\qquad q_3^2=m_X^2\,,\qquad q_4^2=m_Y^2\,,
\end{eqnarray}
where in general $m_X$ and $m_Y$ are allowed to be different
and the transverse momentum of the final-state particles is given by
\begin{eqnarray}
  p_T^2 &=&\frac{u\,t-m_X^2 m_Y^2}{s}\,.
            \label{eq::pT}
\end{eqnarray}
For definiteness we will denote the internal mass by $m_t$, the top quark
mass.

The computation of massive two-loop integrals with the kinematics
described above is a difficult problem.
Purely numerical approaches have been developed and applied to the processes
$gg\to HH$, $gg\to ZZ$, $gg\to ZH$, $gg\to W^+W^-$ (see, e.g.,
Refs.~\cite{Borowka:2016ehy,Borowka:2016ypz,Baglio:2018lrj,Agarwal:2020dye,Chen:2020gae,Bronnum-Hansen:2021olh}).
Usually these computations require a large amount of CPU time for a single phase space
point. Furthermore, it is often necessary to fix numerical values for the top
quark and Higgs boson masses at an early stage of the calculation. Thus a
change of value or renormalization scheme makes it necessary to repeat a large
part of the calculation.

In order to avoid the disadvantages of a purely numerical calculation a number
of analytic approximation methods have been developed. Initially they have
usually been applied to Higgs boson pair production and afterwards also to more
complicated processes. Among the approximations for $gg\to HH$ are large
top quark mass expansions~\cite{Dawson:1998py,Grigo:2013rya,Degrassi:2016vss},
high-energy expansions~\cite{Davies:2018ood,Davies:2018qvx}, small
transverse-momentum expansions~\cite{Bonciani:2018omm} and expansions around
the top quark threshold~\cite{Grober:2017uho}.  In Refs.~\cite{Xu:2018eos,Wang:2020nnr} a
method has been developed where the two-loop amplitude is expanded for small
Higgs boson mass with a subsequent numerical evaluation.

Since such approximations are only valid in a restricted phase space
it is tempting to combine different approaches. 
A first example of such a combination has been presented 
in Ref.~\cite{Davies:2019dfy} where the exact numerical results from
Refs.~\cite{Borowka:2016ehy,Borowka:2016ypz} were combined with the
high-energy expansion of Refs.~\cite{Davies:2018ood,Davies:2018qvx}. The
CPU-time expensive calculations were only necessary for relatively small
values of the Higgs transverse momentum, say below
$p_T\approx 200$~GeV, and the fast evaluation of the analytic
high-energy expansions could be used for the remaining phase space.

A similar approach to the one proposed in this paper has been discussed in
Refs.~\cite{Bellafronte:2022jmo,Degrassi:2022mro} where the analytic small
$p_T$ and high-energy expansions are ``merged''.  For both expansions Pad\'e
approximations are constructed, however, only to low order ($[1/1]$ and
$[6/6]$, respectively).  The Pad\'e approximants are constructed from the
analytic expression and kept fixed, thus there is no estimation of the
uncertainty due to this approach.
In our approach high-order Pad\'e approximants are constructed
numerically in the high-energy region and the approach of
Ref.~\cite{Davies:2020lpf} is used to
determine an uncertainty estimate.  Furthermore, instead of an expansion in
$p_T$ we perform an expansion in the Mandelstam variable $t$. We believe that
our approach leads to simpler expressions in intermediate steps.
Note that in~\cite{Bellafronte:2022jmo,Degrassi:2022mro} only terms up to $m_H^2$
have been used in the high-energy approximation. This introduces a systematic
uncertainty of up to a few percent, as we will discuss below. In this work we will
include quartic corrections which reduces this uncertainty below the percent level.

In this paper we review the high-energy expansion method developed in
Refs.~\cite{Davies:2018qvx,Davies:2019dfy,Davies:2020lpf}. An improvement in
the method allows us to obtain significantly deeper expansions in
$m_t^2/s$, $m_t^2/t$ and $m_t^2/u$ which includes terms up to about
$m_t^{120}$ (see also Ref.~\cite{Davies:2022ram}) (instead of $m_t^{32}$ as
in~\cite{Davies:2018qvx,Davies:2020lpf}).  The deeper expansions combined with
the construction of Pad\'e approximants extends the range of validity to even
smaller values of $\sqrt{s}$ and $p_T$. We will provide details regarding this approach in
Section~\ref{sub::high_energy}.

In Section~\ref{sub::t0} we will describe our approach for the
expansion around $t\to 0$. It is based on the observation that for
this limit a simple Taylor expansion can be performed, rather than a complicated
asymptotic expansion.
We can thus reduce the calculation to integrals which
only depend on $m_t^2/s$. These integrals are obtained with the help
of differential equations using the ``expand and match'' approach
developed in Refs.~\cite{Fael:2021xdp,Fael:2022miw}.  The boundary conditions are
obtained from the large-$m_t$ limit, in which the integrals are simple and
can be computed analytically.

In Section~\ref{sec::gghh} we will use the process $gg\to HH$ to illustrate
the methods of Sections~\ref{sub::high_energy} and~\ref{sub::t0}.
However, the approach is more general and with straightforward modifications
it can also be applied to other processes as, e.g., $gg\to ZH$.  We
will show that we can cover the whole kinematic phase space which we
parametrize in terms of $\sqrt{s}$ and $p_T$. A summary of our findings
together with a discussion of
possible bottlenecks are discussed in Section~\ref{sec:conclusions}.



\section{\label{sec::ana_exp}Analytic expansions}

We begin by performing a Taylor expansion in the masses of the final-state
particles. This is always possible for diagrams where the final-state particles
couple to massive internal lines.
This produces an amplitude in terms of four-point functions which depend
on $s$, $t$ and $m_t$, but not on $m_X$ or $m_Y$.
We then proceed by considering analytic expansions of the amplitude in the
following limits:
\begin{itemize}
\item[A.] high energy
\item[B.] $t\to 0$
\end{itemize}

In both cases we perform an exact reduction of the amplitude to master
integrals, which we then expand in the relevant limit. The reduction is
the same for both cases, leading to the same master integrals. For the
process $gg\to HH$ this step was first done in
Refs.~\cite{Davies:2018ood,Davies:2018qvx} and leads to 161 two-loop
master integrals.
In the following subsections we briefly discuss the
features of methods~A and~B in more detail.

It is also possible to perform an asymptotic expansion in the limit of a large
top quark mass. In this case it is not necessary to expand in the masses of
the final state particles. Such an expansion is automated in the program
{\tt exp}~\cite{Seidensticker:1999bb,Harlander:1997zb} and the approach is
well established; results for the $gg\to HH$ form factors at three loops
can be found in Refs.~\cite{Grigo:2013rya,Davies:2019djw}.
In this work we use the results of this approach to provide boundary
conditions for the differential equations considered in method B described above.
We also show some numerical values for the form factors in this approximation
in Section~\ref{sub::1lFF}, however our proposed procedure to approximate the
two-loop form factors requires only the high-energy and small-$t$ expansions.


\subsection{\label{sub::high_energy}High-energy expansion}

The method of high-energy expansion, including a subsequent Pad\'e
approximant--based improvement,
has been developed in
Ref.~\cite{Davies:2018ood,Davies:2018qvx,Davies:2019dfy,Davies:2020lpf,Davies:2020drs}.
We improve this approach by constructing a deeper expansion of the master
integrals, which includes 120 terms in the small-$m_t$ expansion.
Such an expansion is obtained in the following way: 
\begin{enumerate}
  \item We insert an ansatz for the expansion of each master integral $M_i$, $i=1,...,161$
  \begin{eqnarray}
    M_i(\epsilon,s,t,m_t) &=& 
    \sum\limits_{a=-3}^{a_{i,\text{max}}}
    \sum\limits_{b=-3}^{b_{\text{max}}}
    \sum\limits_{c=0}^{4+a} c_{abc}^{(i)}(s,t) \: \epsilon^a \:
                              \left(\frac{m_t}{\sqrt{s}}\right)^b \:
                              \ln\left(\frac{m_t^2}{s}\right)^c ~,
  \end{eqnarray}
  into the system of differential equations for the master integrals, with
  respect to $m_t$. $a_{i,\text{max}}$ is a master integral--specific value
  determined by the $\epsilon$ order required to produce the amplitude to
  $\epsilon^0$ and we choose $b_{\text{max}}=120$ for each master integral.
  The planar master integrals depend only on even powers of $m_t$, while the
  non-planar integrals also have contributions from odd powers as was shown
  in Ref.~\cite{Davies:2018qvx}.
\item By comparing the coefficients of powers of $\epsilon$, $m_t$ and $\ln(m_t)$
  we establish a system of linear equations for the expansion coefficients
  $c_{abc}^{(i)}(s,t)$, which depend on the Mandelstam variables $s$ and $t$.
  We solve this
  system in terms of a small number of boundary constants by making use of
  the \texttt{reduce_user_defined_system} feature of
  \texttt{Kira}~\cite{Klappert:2020nbg}.
  Solving over finite fields with subsequent rational reconstruction
  using \texttt{FireFly}~\cite{Klappert:2020aqs,Klappert:2019emp}
  is much faster than solving symbolically using \texttt{Fermat}~\cite{fermat}.
  It is this method of solving the system of equations which allows us to
  expand much more deeply than Ref.~\cite{Davies:2018qvx}, which expands only
  up to $b_{\text{max}}=32$.
\item The boundary constants can be fixed using the solutions from
  Refs.~\cite{Davies:2018ood,Davies:2018qvx}, where these constants were
  computed using the method of regions and Mellin-Barnes techniques, see also
  Ref.~\cite{Mishima:2018olh} for more details.
\end{enumerate}

The expansion coefficients of the master integrals are then exported to a
{\tt FORM} {\tt Tablebase} which is used to efficiently insert the expansions
into the amplitude, which is also expanded in $\epsilon$ and $m_t$ to the
required depth.

The subsequent Pad\'e approximation is performed numerically following
Refs.~\cite{Davies:2019dfy,Davies:2020lpf}. For convenience we repeat the
important steps in the following.
The starting point is a form factor as an
expansion in $m_t$, i.e., numerical values for all other kinematic variables
and masses are inserted. We then apply the replacements
$m_t^{2k} \to m_t^{2k} x^k$ and $m_t^{2k-1} \to m_t^{2k-1} x^k$ to pair
together the even and odd powers of $m_t$, yielding a degree-$N$ polynomial
in the variable $x$, with half the maximum degree of the $m_t$ expansion.

Next we construct Pad\'e approximants in the variable $x$ and write the
form factor as a rational function of the form
\begin{eqnarray}
  [n/m](x)
  &=& 
      \frac{a_0 + a_1 x + \ldots + a_n x^n}{1 + b_1 x + \ldots + b_m x^m}
      \,,
      \label{eq::Pade}
\end{eqnarray}
where the coefficients $a_i$ and $b_i$ are determined 
by comparing the coefficients of $x^k$
after expanding the right-hand side of Eq.~(\ref{eq::Pade}) around the point $x=0$.
Evaluation of this rational function at $x=1$ yields the Pad\'e approximated
value for the form factor.

The numerator and denominator degrees $(n,m)$ in Eq.~(\ref{eq::Pade}) are free
parameters; one only must ensure that $n+m\le N$ such that a sufficient
number of expansion terms are available to determine the coefficients $a_i$
and $b_i$. We define $N_{\rm low}$ and $N_{\rm high}$ and 
include Pad\'e approximations in our analysis which fulfil
\begin{align}
	N_{\rm low}\le n+m \le N_{\rm high}\quad
	\textnormal{and}\quad N_{\rm low} \le n + m - | n - m | \,.
	\label{eq::N_low_high}
\end{align}
Our default choice is $N_{\rm low}=49$ and $N_{\rm high}=56$ which leads to 28
different Pad\'e approximants\footnote{While the master integrals are determined
up to $N=60$ ($m_t^{120}$), negative powers of $m_t$ in the amplitude coefficients
mean that the expansion of the form factors can be produced up to
$N=56$ ($m_t^{112}$).}.
They are combined using three different criteria:
\begin{itemize}
\item The rational function in Eq.~(\ref{eq::Pade}) develops poles at the roots
  of the
  denominator.  We give more weight to those Pad\'e approximants which have
  poles further away from the evaluation point $x=1$ (``pole-distance re-weighted'' Pad\'e
  approximation).
\item We give more weight to Pad\'e approximants which are derived from a
  larger number of expansion terms.
\item We give more weight to ``near-diagonal'' Pad\'e approximants.
\end{itemize}

We combine the weights from each criterion for each of the Pad\'e approximants, and use
the combined weight to produce a central value and corresponding uncertainty for
the phase-space point under consideration.
Explicit formulae for the individual steps of the construction
are given in Section~4 of Ref.~\cite{Davies:2020lpf}.
In the supplementary material~\cite{progdata}
to this paper we provide {\tt Mathematica} code which can be used to construct,
for a given polynomial in $x$, an approximation based on the
procedure described above, including an uncertainty estimate.

\begin{figure}[t]
	\centering
  \includegraphics[height=0.25\textwidth]{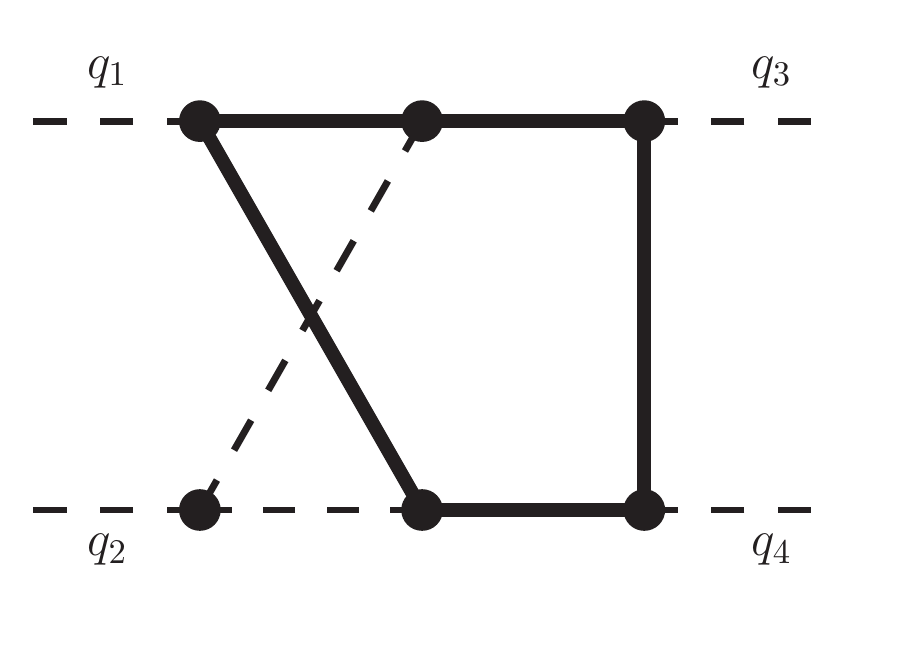}
  \caption{The two-loop Feynman diagram $G_{59}(1,1,1,1,1,1,1,-1,0)$ (see
    Appendix~A of Ref.~\cite{Davies:2018qvx} for more details). Solid and dashed lines correspond to
    massive and massless propagators. All external momenta are massless.}
  \label{fig::diag_non-pl_MI}
\end{figure}

\begin{figure}[t]
    \includegraphics[width=0.485\textwidth]{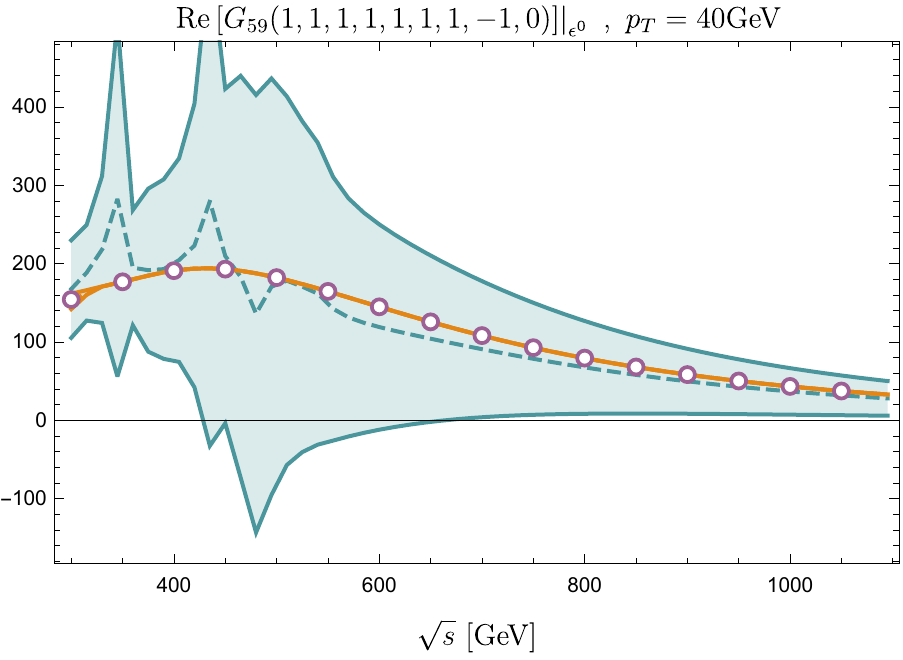}
    \includegraphics[width=0.485\textwidth]{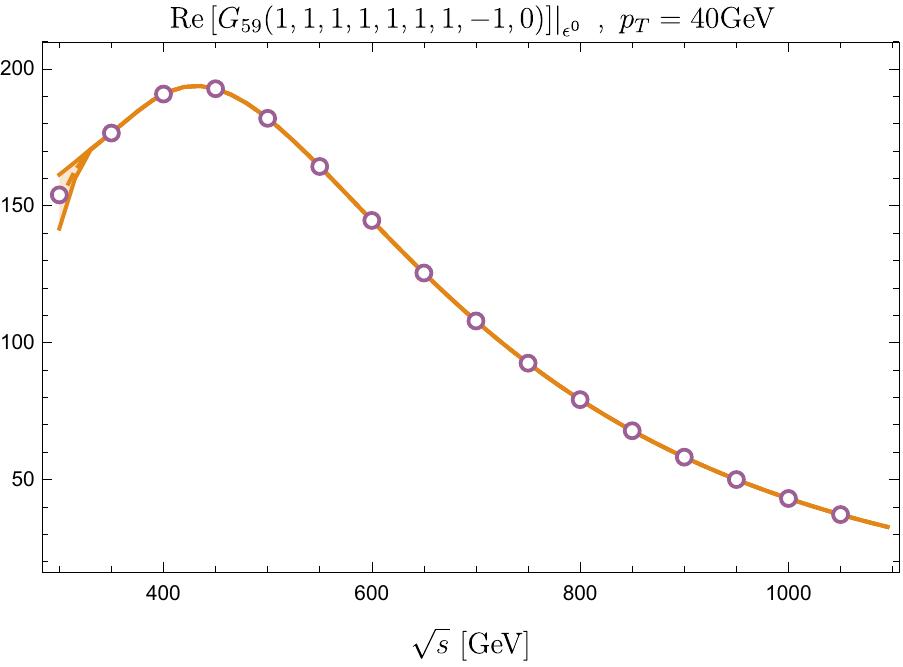}
    \\
    \centering
    \includegraphics[height=0.04\textwidth]{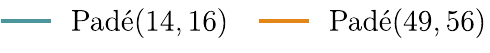}
    \includegraphics[height=0.04\textwidth]{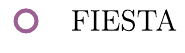}
    \\
    \flushleft
    \vspace{-5mm}\hspace{0.24\textwidth}(a)\hspace{0.47\textwidth}(b)
    \caption{Comparison of Pad\'{e}-based approximations constructed from different
       expansion depths ($N_{\rm low},N_{\rm high}$) with numerical results obtained using {\tt FIESTA}, for the
       non-planar master integral shown in Fig.~\ref{fig::diag_non-pl_MI}, with a numerator.
       \label{fig::non-pl_MI}
     }
\end{figure}

We have demonstrated this approach applied to a single planar master integral in
Ref.~\cite{Davies:2022ram} and the comparison to (exact) numerical results can
be found in Fig.~7(a) of that reference.
In Fig.~\ref{fig::non-pl_MI} we discuss results for the
non-planar integral shown in Fig.~\ref{fig::diag_non-pl_MI}. We choose
$p_T=40$~GeV and vary $\sqrt{s}$ between 300~GeV and 1100~GeV. In
Fig.~\ref{fig::non-pl_MI}(a) we compare Pad\'e results constructed from
expansions up to $m_t^{32}$ and $m_t^{112}$, which are shown by the green and
orange bands, respectively.  One observes a dramatic reduction of the
uncertainty. At the same time it is reassuring to see that the uncertainty
estimate of the Pad\'e procedure is reliable, when comparing to the numerical
values obtained using FIESTA~\cite{Smirnov:2021rhf}.
In Fig.~\ref{fig::non-pl_MI}(b) we focus on the comparison of the orange band
with the results from FIESTA; we observe good agreement within uncertainties
in the whole plotted range of $\sqrt{s}$, even very close to the threshold
for the production of two top quarks.

\subsection{\label{sub::t0}\boldmath Expansion for $t\to 0$}

In this subsection we aim for an expansion of the original 161 master integrals around
$t=0$ such that the amplitude can be expanded in this limit. This complements the
the high-energy expansion, i.e.~we aim for a good description
in the region around the threshold where $s\approx 4 m_t^2$ and the high-energy
expansion breaks down.
However, as we will
see below, good results are also obtained for larger values of $\sqrt{s}$, in
particular for smaller values of $p_T$. The expansion is performed as follows.
\begin{itemize}\setlength\itemsep{-.2em}

\item As for the high-energy expansion, we first expand in the masses of the
  final-state particles. For $gg\to HH$ it
  is sufficient to expand up to $m_H^4$ to obtain a precision below the
  percent level.
  We are left with integral families which depend on $s,t$ and $m_t$.
  Here we note that the expansion in $m_H$ generates
  spurious $1/t$ terms which cancel after inserting the $t$-expansion of the
  master integrals.

  As discussed previously, this expansion is a simple Taylor expansion in
  cases where the final-state particles couple to massive internal lines;
  otherwise, a more involved asymptotic expansion must be performed.

\item Establish differential equations, with respect to $t$, for the master
  integrals of the $2\to 2$ problem where all external lines are massless.
  The master integrals, and thus the resulting $t$-differential equations,
  are the same as in the high-energy case discussed in Section~\ref{sub::high_energy}.

\item We use the differential equations to obtain, for each master integral, a
  generic Taylor expansion around $t=0$. This is achieved by expanding the coefficients of the differential
  equations around $t\to 0$ and for each master integral, inserting an ansatz
  of the form
  \begin{eqnarray*}
    M_i(\epsilon,s,t,m_t) &=& \sum_{a=-3}^{a_{i,\text{max}}} \sum_{b\ge 0}
       c_{ab}^{(i)}(s, m_t^2) \:
       \epsilon^a \:
       \left(\frac{t}{m_t^2}\right)^b 
            \,,
  \end{eqnarray*}
  where the (unknown) coefficients $c_{ab}^{(i)}(s, m_t^2)$ are functions
  of $s$ and $m_t^2$.

\end{itemize}

Note that for $t\to 0$ some of the propagators of the original integral
families (see Appendix~A of Refs.~\cite{Davies:2018ood}
and~\cite{Davies:2018qvx}) become linearly dependent.  After a partial fraction
decomposition we can define new integral families which contain fewer
propagators. In terms of these new families, the number of master integrals
in the $t\to 0$ limit reduces from 161 to 48.
One of the resulting topologies has been studied in
Ref.~\cite{vonManteuffel:2017hms}, where it was shown that two master
integrals are elliptic and cannot be expressed in terms of iterated integrals.
These master integrals depend on two different square roots.

We have calculated all 46 non-elliptic master integrals analytically by
solving the associated differential equations in the variable $s/m_t^2$
following the algorithms outlined in Ref.~\cite{Ablinger:2018zwz} implemented
with the help of the packages \texttt{Sigma}~\cite{Schneider:2007},
\texttt{OreSys}~\cite{ORESYS} and \texttt{HarmonicSums}~\cite{HarmonicSums}.
The boundary conditions have been fixed in the large-$m_t$ limit, where the
integrals can be calculated by performing a large mass expansion,
implemented in
\texttt{q2e}/\texttt{exp}~\cite{Seidensticker:1999bb,Harlander:1997zb}.  Our
final result can be expressed in terms of iterated integrals over letters
which contain the three square roots
$\sqrt{x}\sqrt{4-x},\sqrt{x}\sqrt{4+x},\sqrt{4-x}\sqrt{4+x}$.  However, we
find that this representation is not well suited for numerical evaluation
for several reasons:
\begin{enumerate}
\item Some of the iterated integrals depend on two square-root valued 
  letters at the same time, which cannot easily be rationalized simultaneously.
\item The iterated integrals have spurious poles at $s/m_t^2=1$ and 
  $s/m_t^2=4$, which require analytic continuation.
\item The analytic results for the two elliptic integrals are rather involved.
\end{enumerate}   
Therefore, we calculate all 48 master integrals using the semi-analytic
approach developed in Refs.~\cite{Fael:2021kyg,Fael:2022rgm}.
For each master integral, we provide a deep expansion of 50 terms around different
values of $s/m_t^2$, with high-precision numerical coefficients.
In particular we construct expansions around 18 values of $s/m_t^2$ to cover
values of $s$ between 0 and $\infty$.
Our starting point for the construction of the approximations is the expansion
around $s=0$ where all master integrals can be computed analytically.
As a by-product we extend the large-$m_t$ expansion of these master integrals
(but only at $t=0$).

This method has a number of advantages compared to purely numerical approaches.
Since the value of $m_t$ is only inserted into the final expression, it is
possible to easily change the value or renormalization scheme used for $m_t$.
It is straightforward to take derivatives w.r.t.~$m_t$ of the one-loop
expressions in order to generate the corresponding counterterm contributions.


\FloatBarrier

\section{\label{sec::gghh}Application to Higgs boson pair production}

In this section we apply the expansion methods discussed above to the particular
case of the $gg\to HH$ amplitude.
We start by examining the $m_H$ and $t$ expansions of one-loop master
integrals by comparing to numerical results obtained with {\tt
  FIESTA}~\cite{Smirnov:2021rhf} and {\tt Package-X}~\cite{Patel:2016fam}. We
show that the Taylor expansion in $m_H$ produces good agreement with the
exact result, even for smaller values of $\sqrt{s}$ close to the Higgs pair
production threshold at $\sqrt{s}=2 m_H$.
Afterwards we
discuss results for the one- and two-loop form factors.  Finally we compare
the virtual corrections to the Higgs pair production cross section with the
numerical results obtained in Ref.~\cite{Davies:2019dfy}.

For the numerical evaluations we use input values for the top quark and
Higgs boson masses of $m_t = 173.21$~GeV and $m_H=125.1$~GeV, respectively.

For completeness we provide in the following the definition of the form
factors for Higgs boson pair production.
The amplitude for the process $g(q_1)g(q_2)\to H(q_3)H(q_4)$
can be decomposed into two Lorentz structures
($a$ and $b$ are adjoint colour indices)
\begin{eqnarray}
  {\cal M}^{ab} &=& 
  \varepsilon_{1,\mu}\varepsilon_{2,\nu}
  {\cal M}^{\mu\nu,ab}
  \,\,=\,\,
  \varepsilon_{1,\mu}\varepsilon_{2,\nu}
  \delta^{ab} X_0 s 
  \left( F_1 A_1^{\mu\nu} + F_2 A_2^{\mu\nu} \right)
  \,,
                    \label{eq::M}
\end{eqnarray}
where
\begin{eqnarray}
  A_1^{\mu\nu} &=& g^{\mu\nu} - {\frac{1}{q_{12}}q_1^\nu q_2^\mu
  }\,,\nonumber\\
  A_2^{\mu\nu} &=& g^{\mu\nu}
                   + \frac{1}{{p_T^2} q_{12}}\left(
                   q_{33}    q_1^\nu q_2^\mu
                   - 2q_{23} q_1^\nu q_3^\mu
                   - 2q_{13} q_3^\nu q_2^\mu
                   + 2q_{12} q_3^\mu q_3^\nu \right)\,.
\end{eqnarray}
Here we have introduced the abbreviation $q_{ij} = q_i\cdot q_j$
and $p_T$ is given in Eq.~(\ref{eq::pT}). The prefactor $X_0$ is given by
\begin{eqnarray}
  X_0 &=& \frac{G_F}{\sqrt{2}} \frac{\alpha_s(\mu)}{2\pi} T_F \,,
\end{eqnarray}
where $T_F=1/2$, $G_F$ is Fermi's constant and $\alpha_s(\mu)$ is the strong
coupling constant evaluated at the renormalization scale $\mu$.

We define the expansion in $\alpha_s$ of the form factors as
\begin{eqnarray}
  F &=& F^{(0)} + \frac{\alpha_s(\mu)}{\pi} F^{(1)} + \cdots
  \,,
  \label{eq::F}
\end{eqnarray}
and decompose the functions $F_1$ and $F_2$ introduced in Eq.~(\ref{eq::M})
into ``triangle'' and ``box'' form factors. We thus cast the one- and two-loop
corrections in the form ($k=0,1$)
\begin{eqnarray}
  F_1^{(k)} &=& \frac{3 m_H^2}{s-m_H^2} F^{(k)}_{\rm tri}+F^{(k)}_{\rm box1}
                + \delta_{k1} F^{(1)}_{\rm dt1}\,, \nonumber\\
  F_2^{(k)} &=& F^{(k)}_{\rm box2} + \delta_{k1}F^{(1)}_{\rm dt2}\,.
                \label{eq::F_12}
\end{eqnarray}
$F^{(1)}_{\rm dt1}$ and $F^{(1)}_{\rm dt2}$ denote the contribution from
one-particle reducible double-triangle diagrams, see, e.g.~Fig.~1(f) of
Ref.~\cite{Davies:2019dfy}. The main focus in this paper is on
$F^{(1)}_{\rm box1}$ and $F^{(1)}_{\rm box2}$. Analytic results for the
leading-order form factors are available
from~\cite{Glover:1987nx,Plehn:1996wb}
and the two-loop triangle form factors have been computed in
Refs.~\cite{Harlander:2005rq,Anastasiou:2006hc,Aglietti:2006tp}. The
results for the double-triangle contribution
can be found in~\cite{Degrassi:2016vss}.


\subsection{\boldmath Expansion of a one-loop master integral in $m_H$}
\label{sub::mh_exp}

\begin{figure}[tb]
	\centering
  \includegraphics[height=0.25\textwidth]{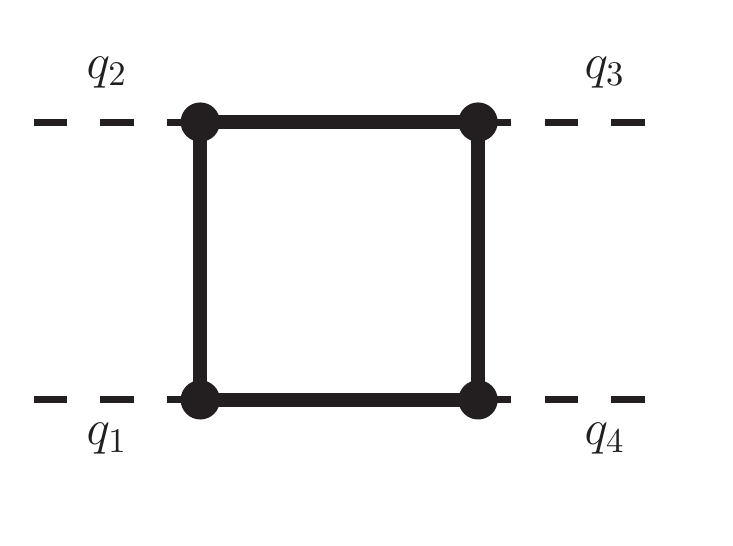}
  \caption{The one-loop master integral $G_2(1,1,1,1)$, where all internal lines are massive and
    for the external lines we have $q_1^2=q_2^2=0$ and $q_3^2=q_4^2=m_H^2$.}
  \label{fig::G2}
\end{figure}

\begin{figure}[t]
  \includegraphics[width=0.49\textwidth]{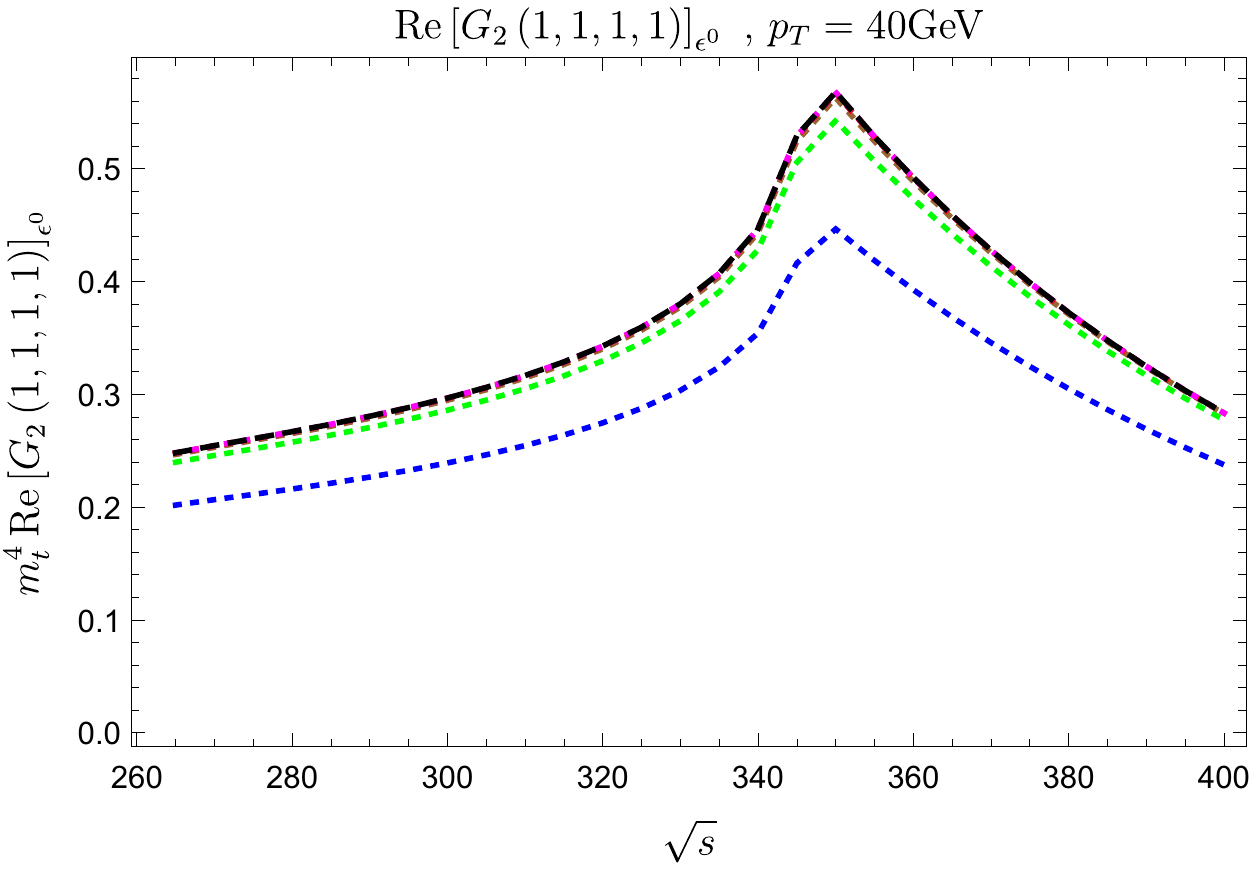}
  \includegraphics[width=0.49\textwidth]{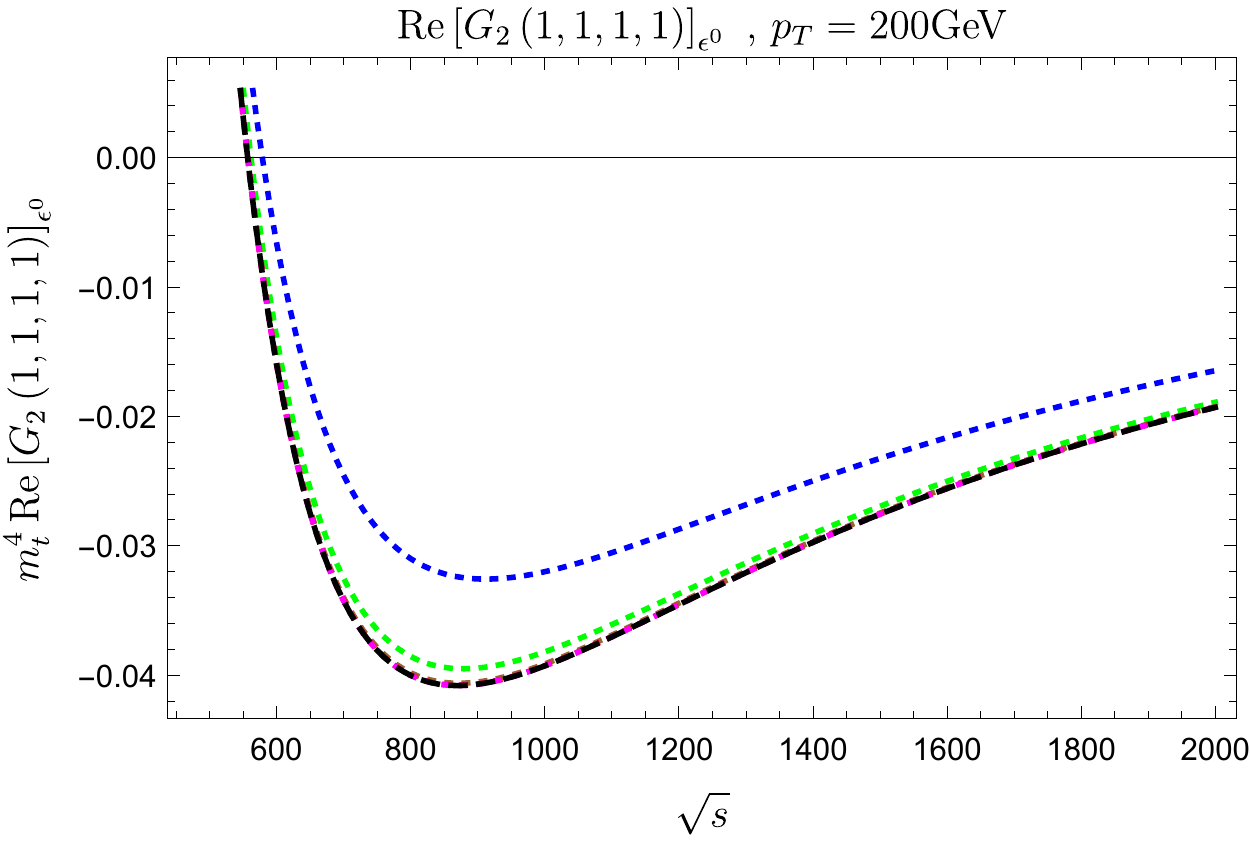}
  \includegraphics[width=0.49\textwidth]{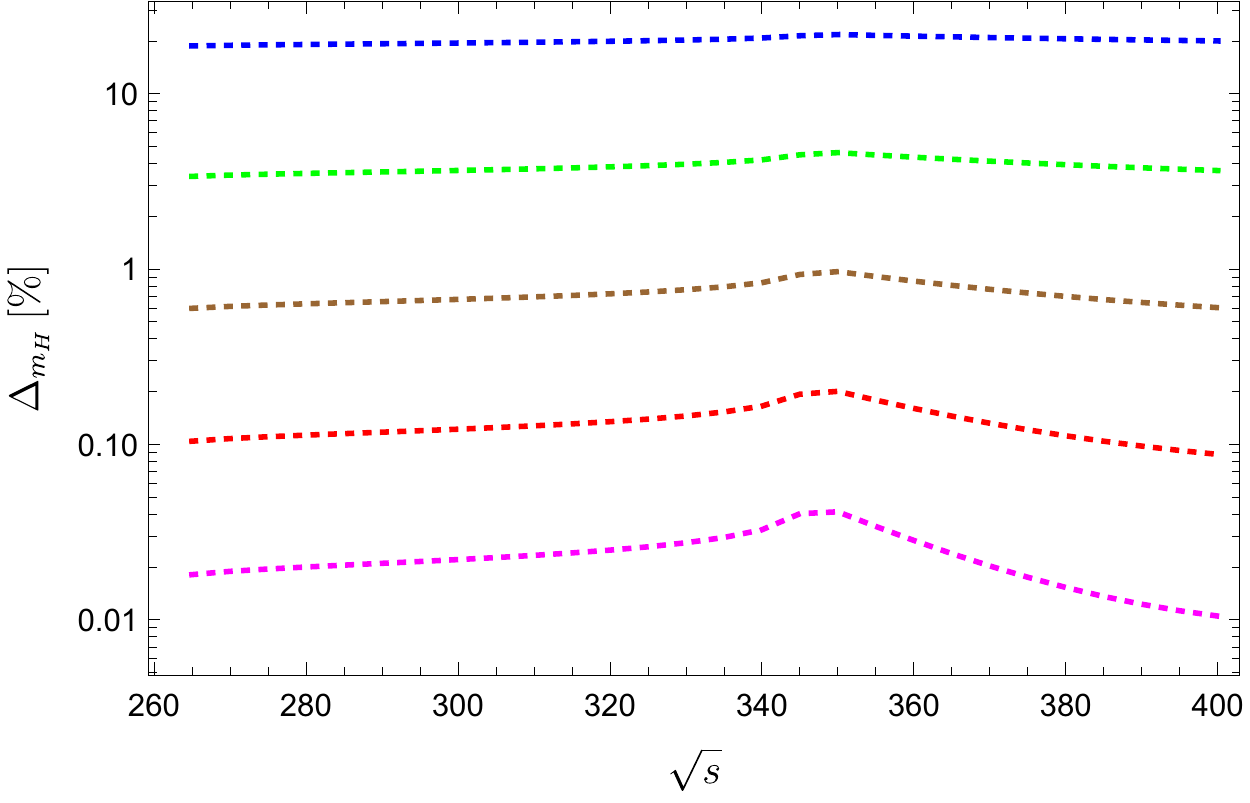}
  \includegraphics[width=0.49\textwidth]{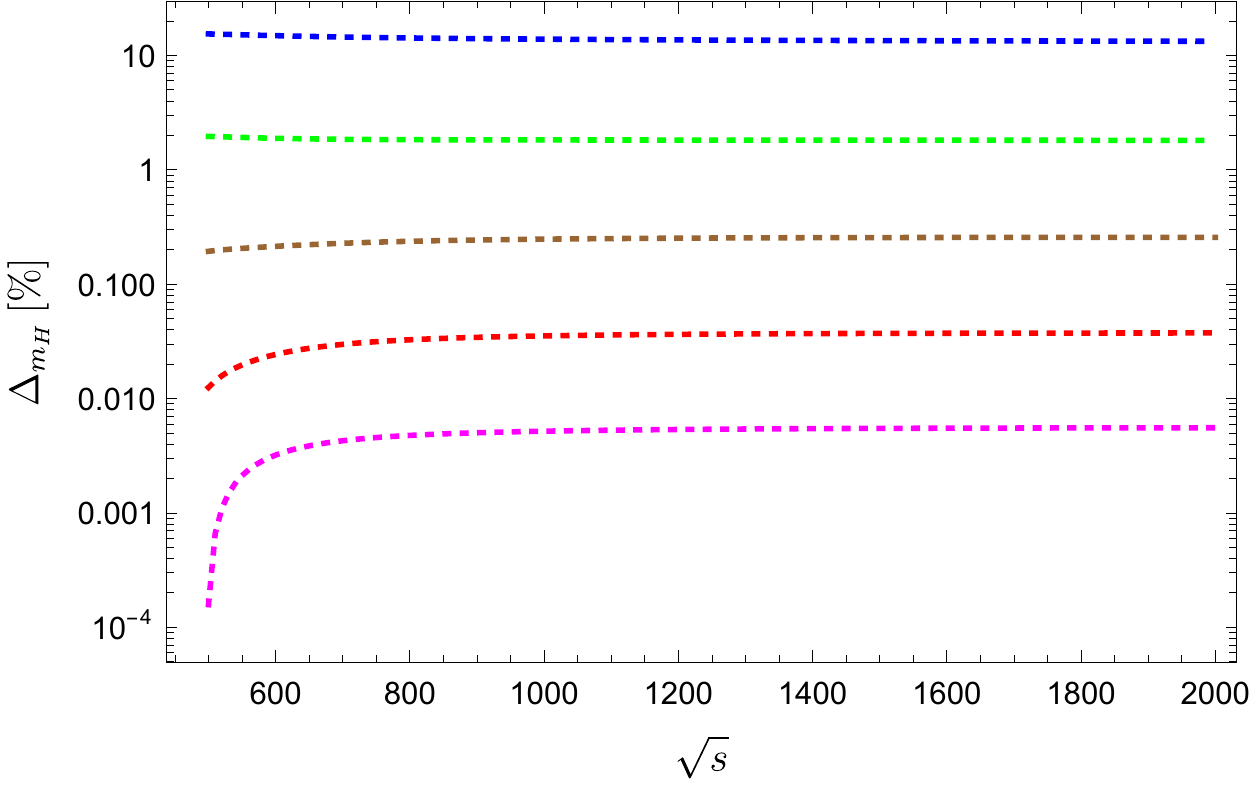}
  \\
  \centering \includegraphics[width=0.8\textwidth]{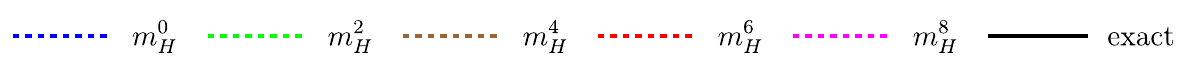}
  \caption{Real part of the master integral $G_2(1,1,1,1)$ as a function of
    $\sqrt{s}$ for $p_T=40$~GeV (left) and $p_T=200$~GeV (right).
    The coloured lines include expansions in $m_H$ up to the indicated orders.
    The exact result is shown in black. The lower panels show the relative error
    between the expansions and the exact curve.}
  \label{fig::mhexp}
\end{figure}

In Fig.~\ref{fig::mhexp} we show, as a function of $\sqrt{s}$, the real part
of the one-loop box master integral $G_2(1,1,1,1)$ (see Appendix~A of
Ref.~\cite{Davies:2018ood} for details on the notation), which is depicted
in Fig.~\ref{fig::G2}.
The left and right panels correspond to $p_T=40$~GeV and
$p_T=200$~GeV, respectively. The coloured lines show expansions in $m_H^2$ up
to fourth order, and the black line represents the exact result.
After the
Taylor expansion in $m_H$ a reduction to master integrals is necessary. It has
been performed with {\tt LiteRed}~\cite{Lee:2013mka} and for the numerical evaluation
of the resulting master integrals we have used {\tt Package-X}~\cite{Patel:2016fam}.

The upper row shows the results for the master integral and the lower row shows
the relative error between the expansions and the exact curve. One observes
that the $m_H^0$ curves do not describe the exact result particularly well, with
differences at the 15-20\% level, however including the quadratic and quartic terms
provide a description below the 5\% level and 1\% level, respectively; these
observations are largely independent of the values of $\sqrt{s}$ and $p_T$.

\FloatBarrier

\subsection{\boldmath Expansion of a one-loop master integral in $t$}

\begin{figure}[t]
  \includegraphics[width=0.49\textwidth]{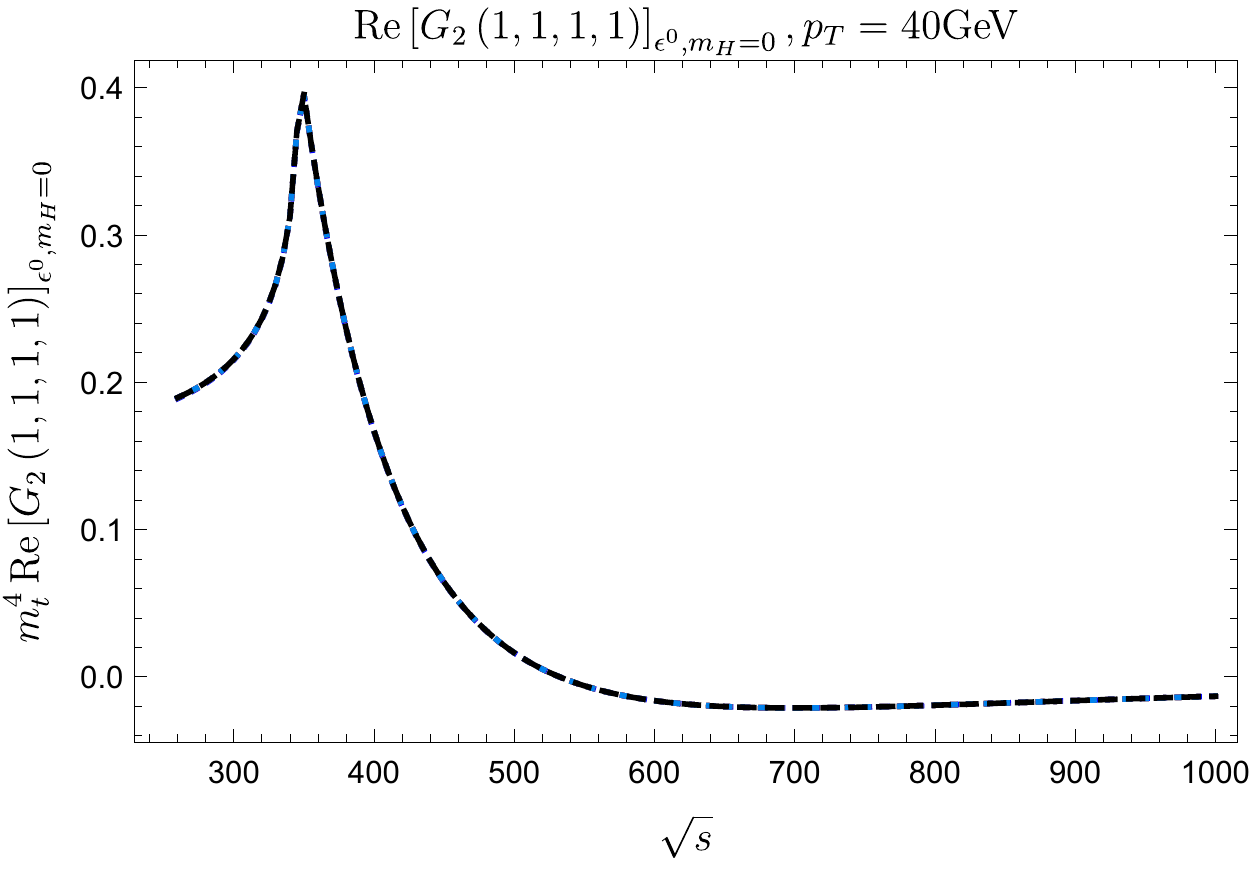}
  \includegraphics[width=0.49\textwidth]{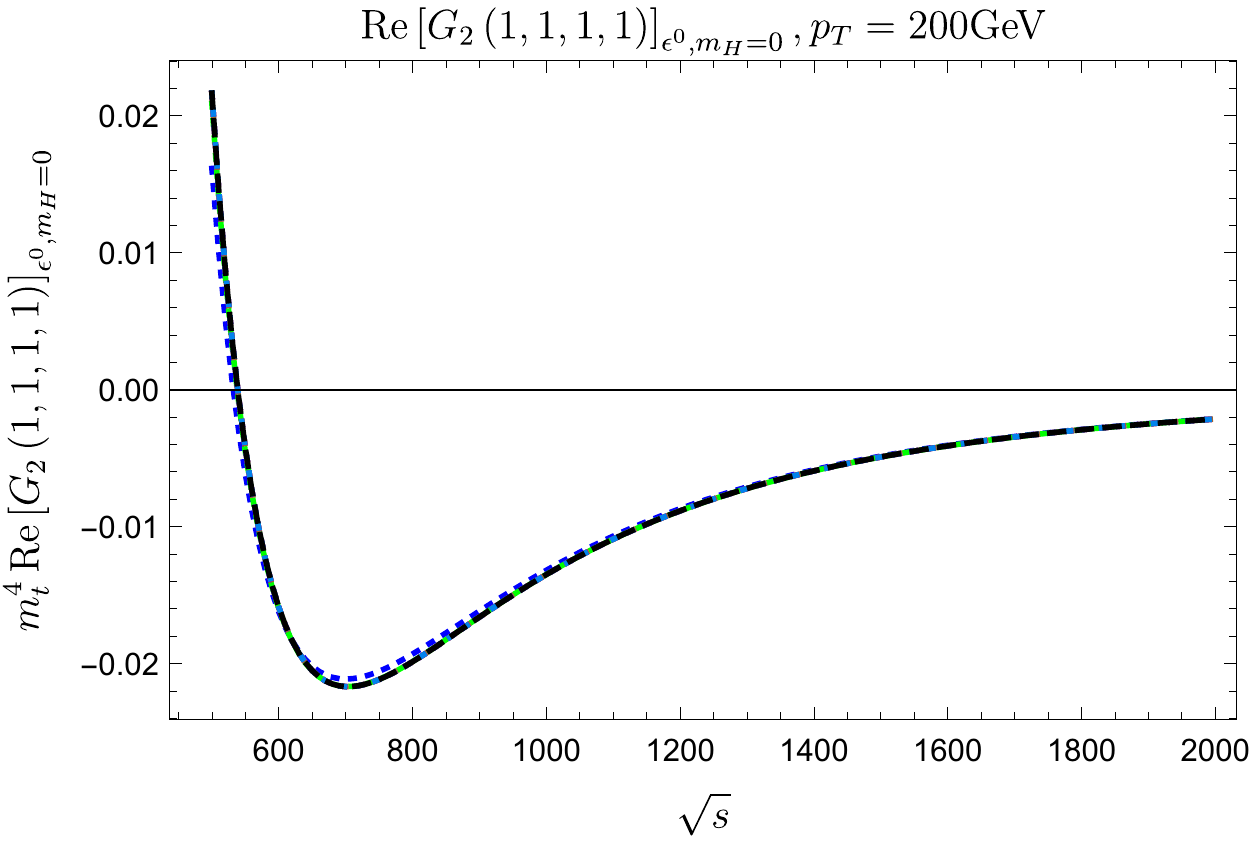}
  \includegraphics[width=0.49\textwidth]{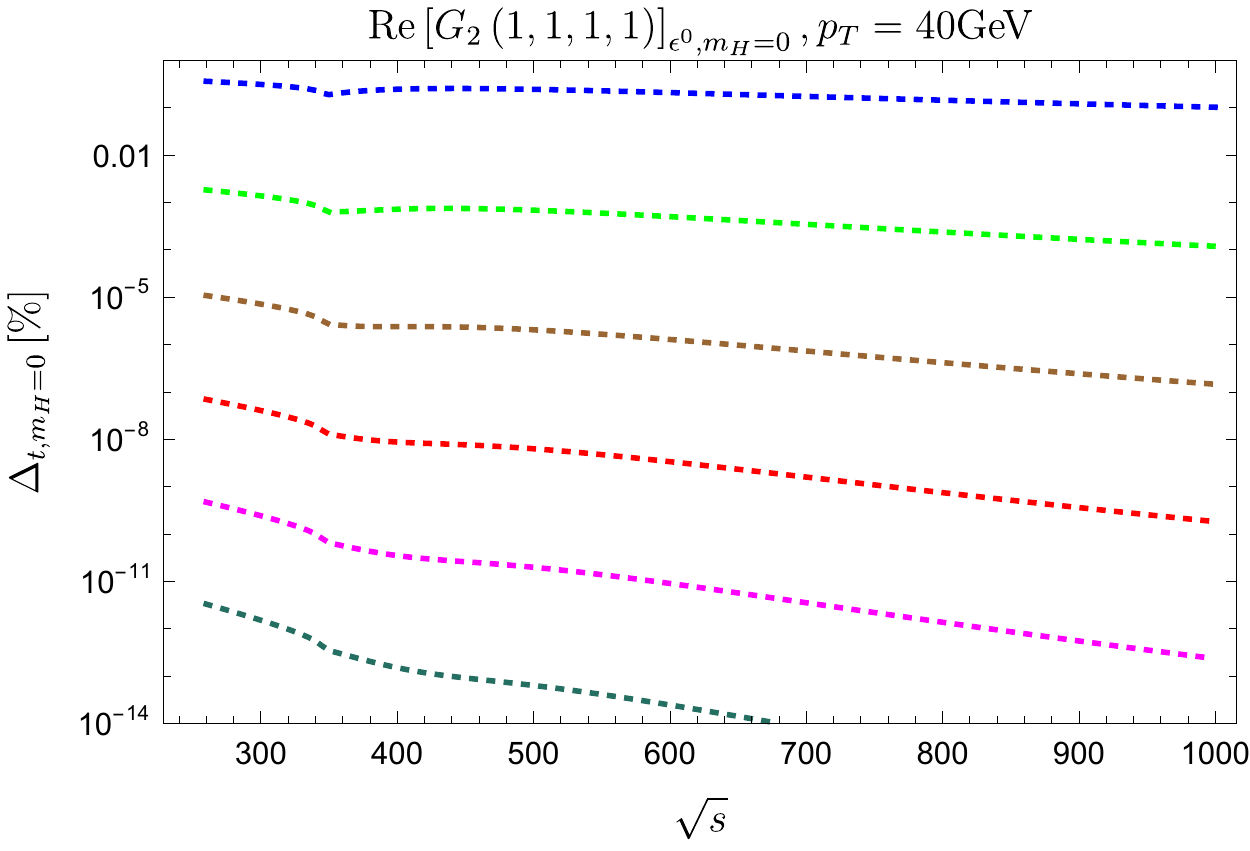}
  \includegraphics[width=0.49\textwidth]{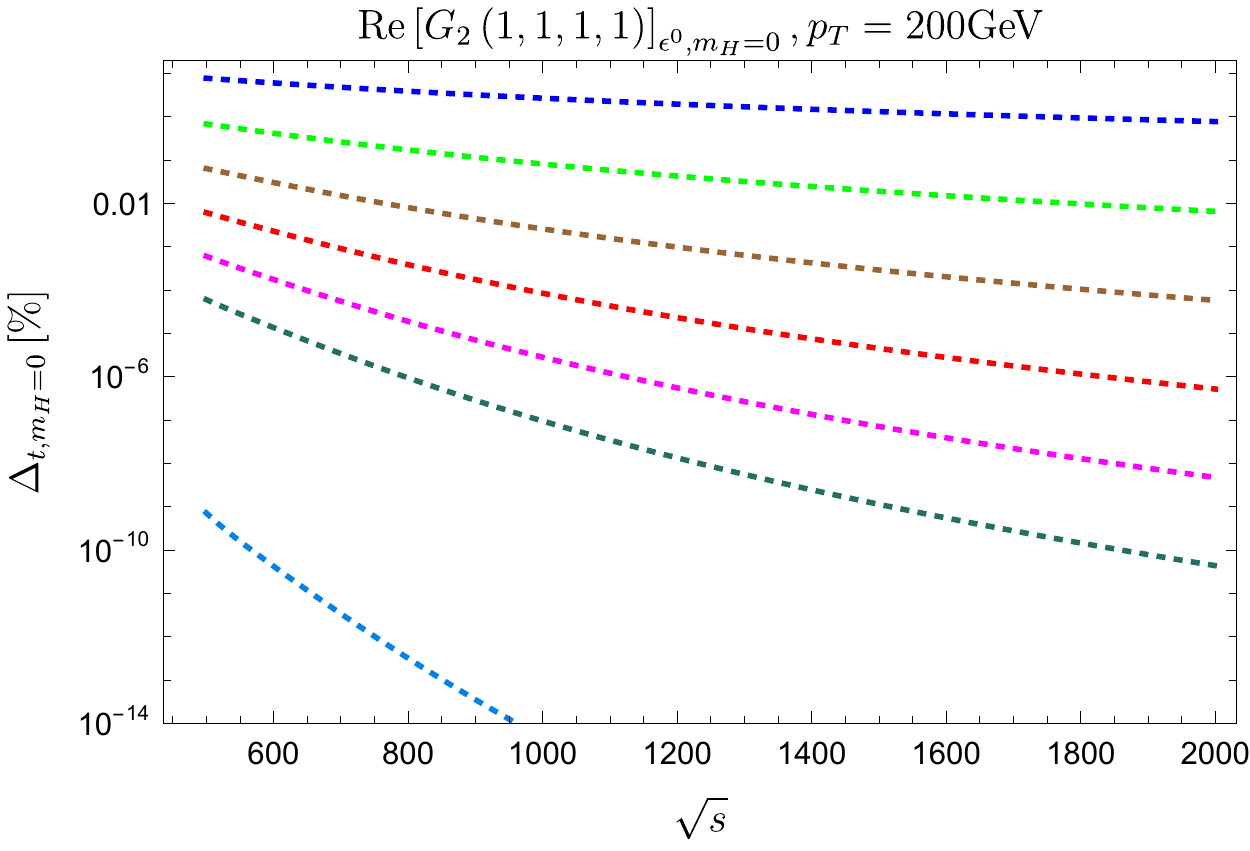}
  \\
  \centering \includegraphics[width=0.8\textwidth]{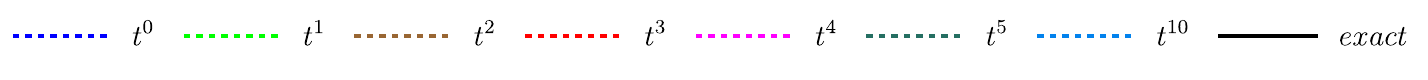}
  \caption{\label{fig::texp}Real part of the master integral $G_2(1,1,1,1)$
    as a function of $\sqrt{s}$ for $p_T=40$~GeV (left) and $p_T=200$~GeV
    (right).  The coloured lines include expansions in $t$ up to $t^{10}$.
    The exact result is shown in black. The lower panels show the relative error
    between the expansions and the exact curve.}
\end{figure}

Next we study the $t\to 0$ expansion of the same one-loop box master integral,
$G_2(1,1,1,1)$. For this purpose we choose $m_H=0$, i.e., the leading term of
the expansion discussed in Section~\ref{sub::mh_exp}.
We perform the expansion
in $t$ using {\tt LiteRed}~\cite{Lee:2013mka} and then map the resulting
integrals to new integral families which have only three propagators and depend only
on $s/m_t^2$.
For these integrals we establish a system of differential equations which can be
solved analytically, incorporating boundary conditions from the $s\to 0$ limit.
The resulting coefficients of the polynomial in $t$ can be written in terms of
Harmonic Polylogarithms~\cite{Remiddi:1999ew}, which we evaluate using the 
{\tt Mathematica} package {\tt HPL.m}~\cite{Maitre:2005uu,Maitre:2007kp}.

In Fig.~\ref{fig::texp} we show the convergence of the $t$ expansion for
the values $p_T=40$~GeV and $p_T=200$~GeV in the left and right columns, respectively.
The lower row shows the relative error between the expansion and the exact curve.
For the smaller value of $p_T=40$~GeV, we observe that the leading expansion term ($t^0$)
already reproduces the exact result at the percent level. For $p_T=200$~GeV the
leading term does not perform so well, however by including higher-order terms the
expansion converges on the exact result very quickly.



\subsection{\label{sub::1lFF}Expansion of the one-loop form factors}

We now discuss the high-energy and small-$t$ expansions at the level of the
one-loop form factors $F_{\rm box1}^{(0)}$ and $F_{\rm box2}^{(0)}$, and compare them to
the exact results.

\begin{figure}[t]
  \begin{tabular}{ccc}
    \includegraphics[width=0.31\textwidth]{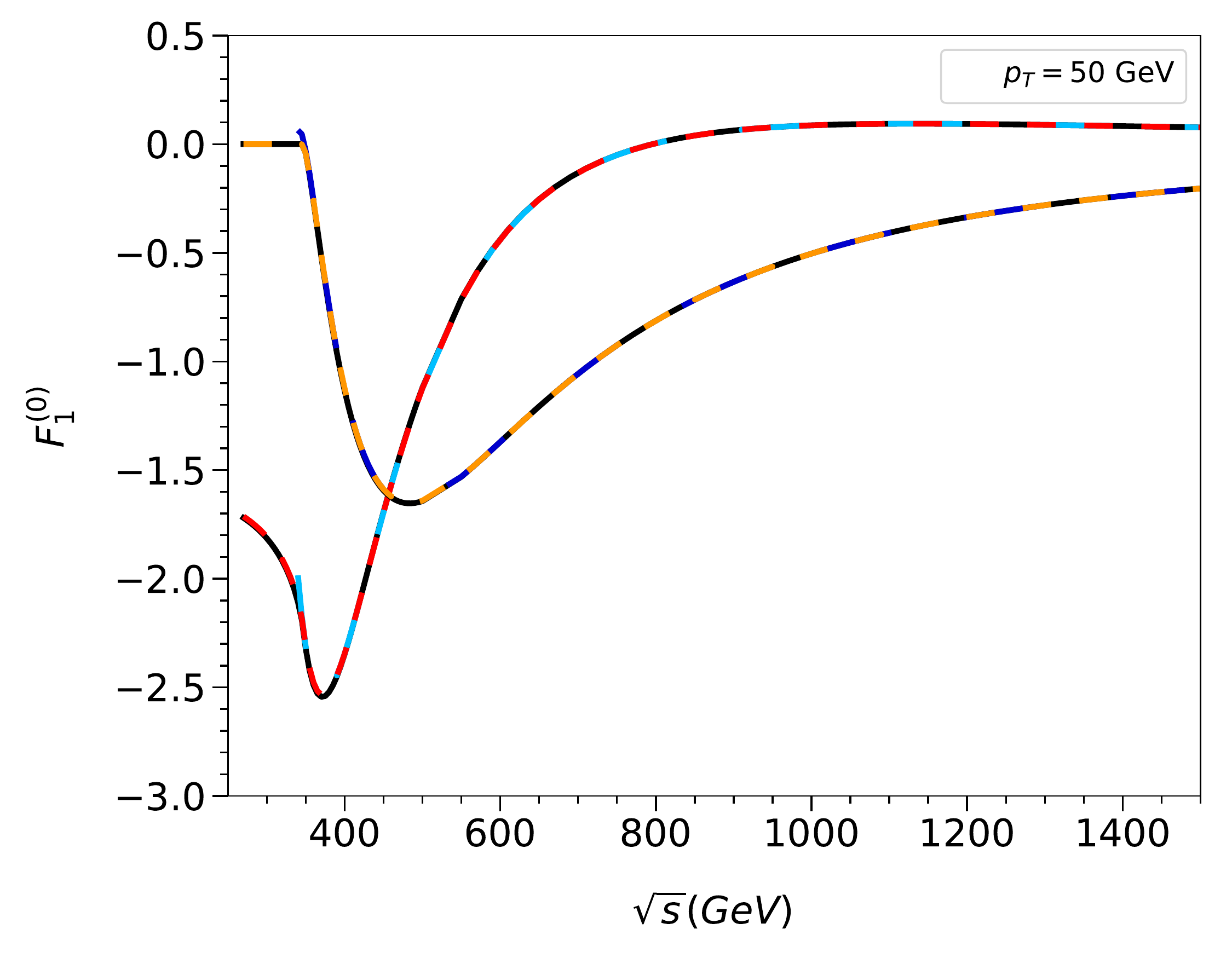} &
    \includegraphics[width=0.31\textwidth]{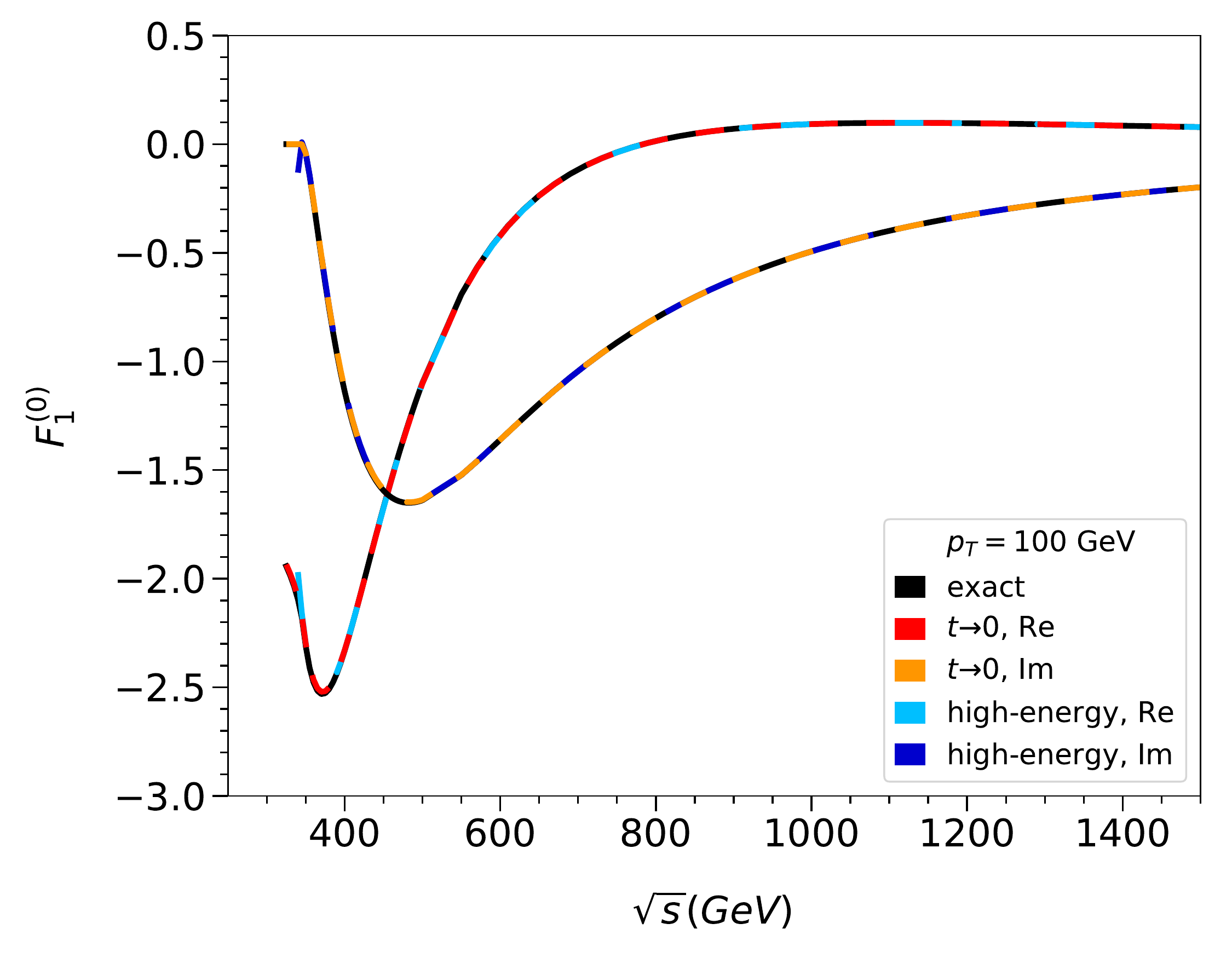} &
    \includegraphics[width=0.31\textwidth]{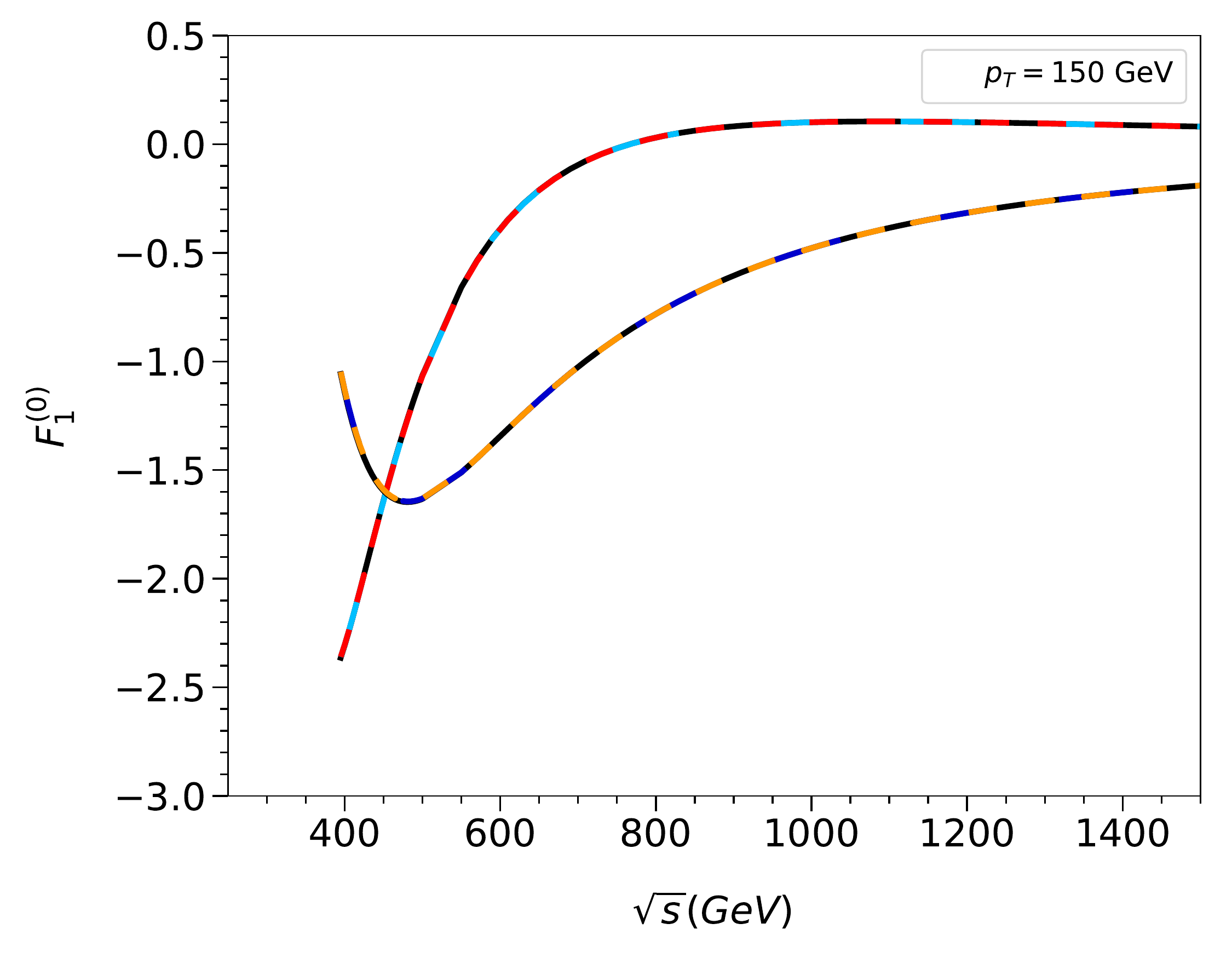}
    \\
    \includegraphics[width=0.31\textwidth]{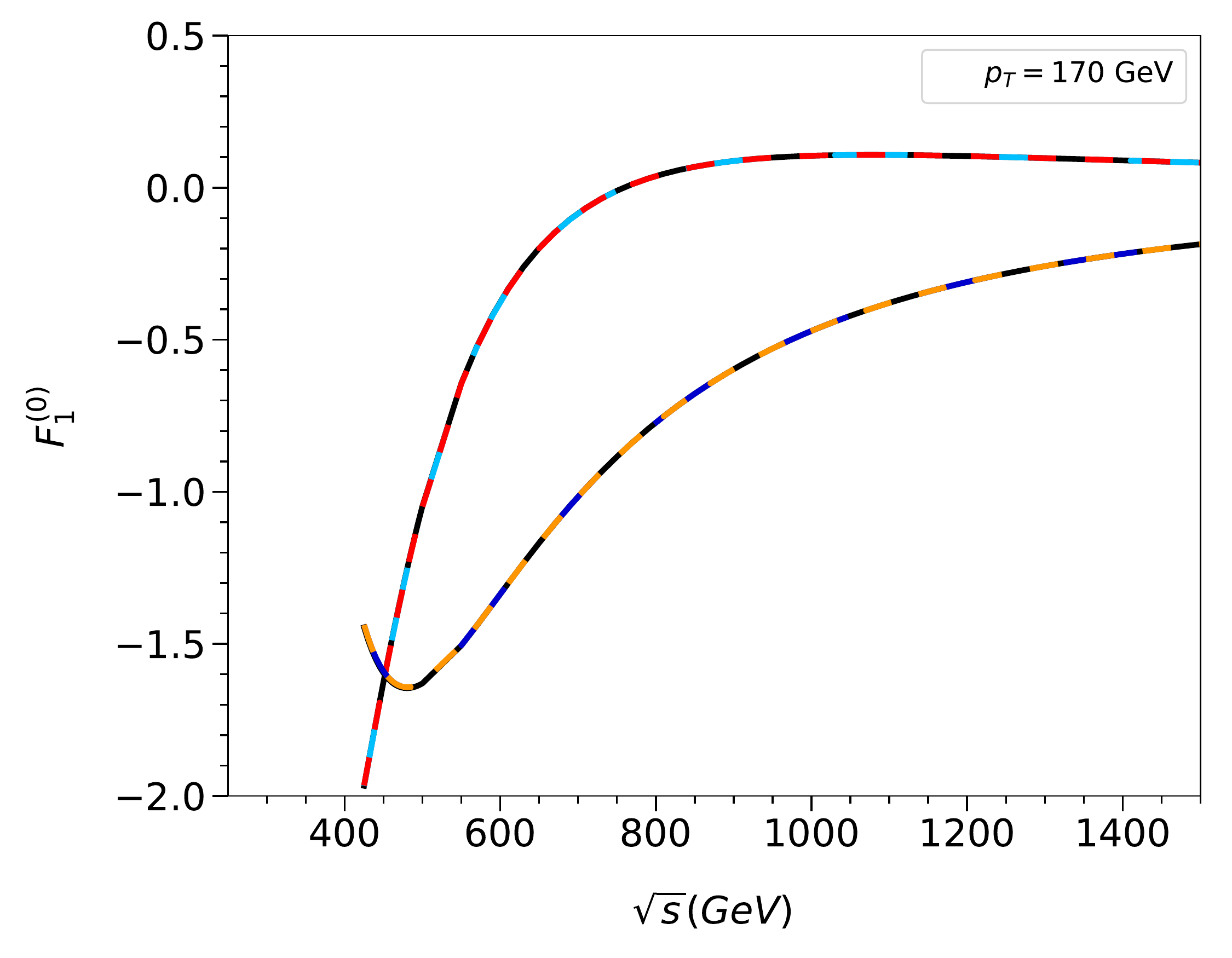} &
    \includegraphics[width=0.31\textwidth]{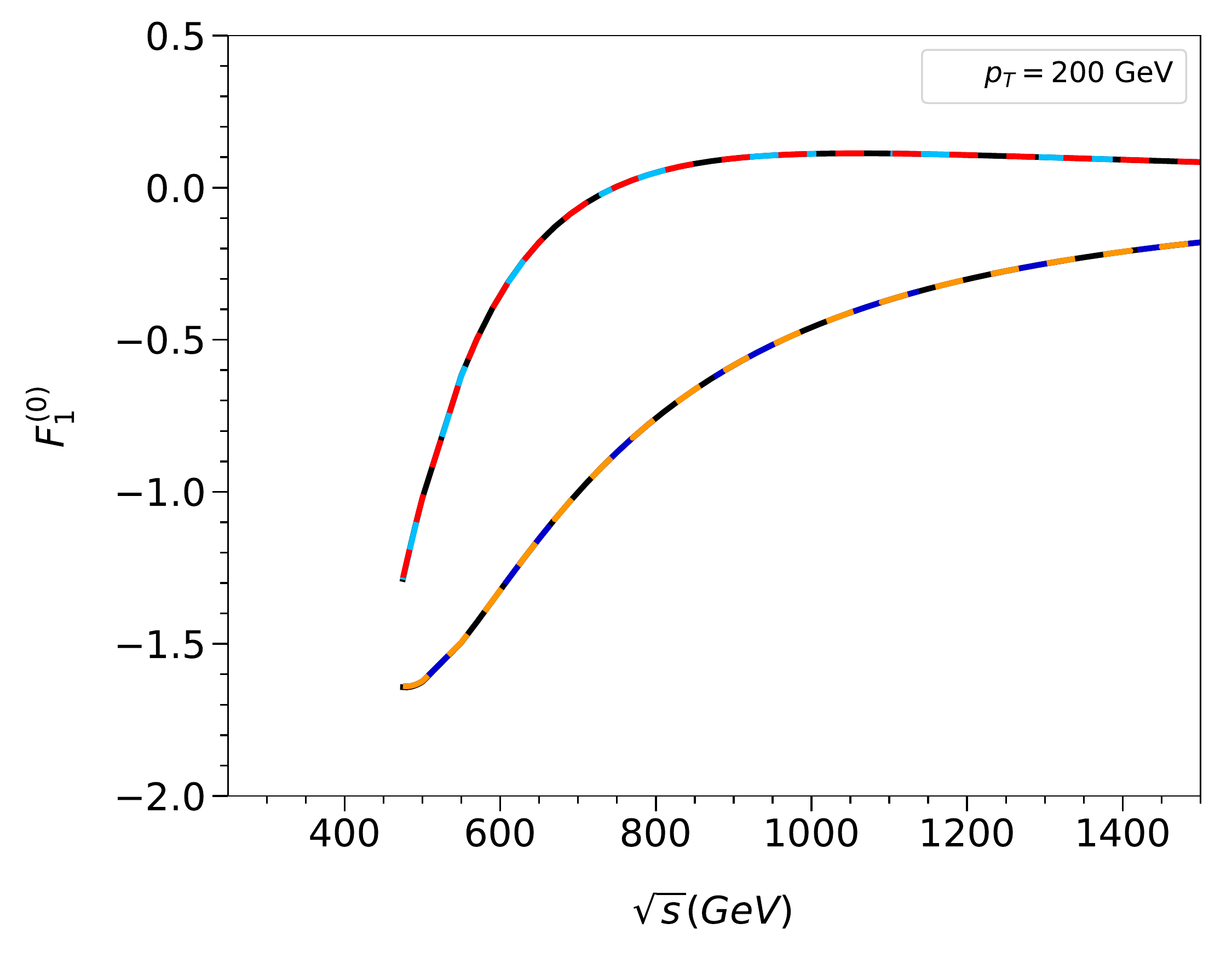} &
    \includegraphics[width=0.31\textwidth]{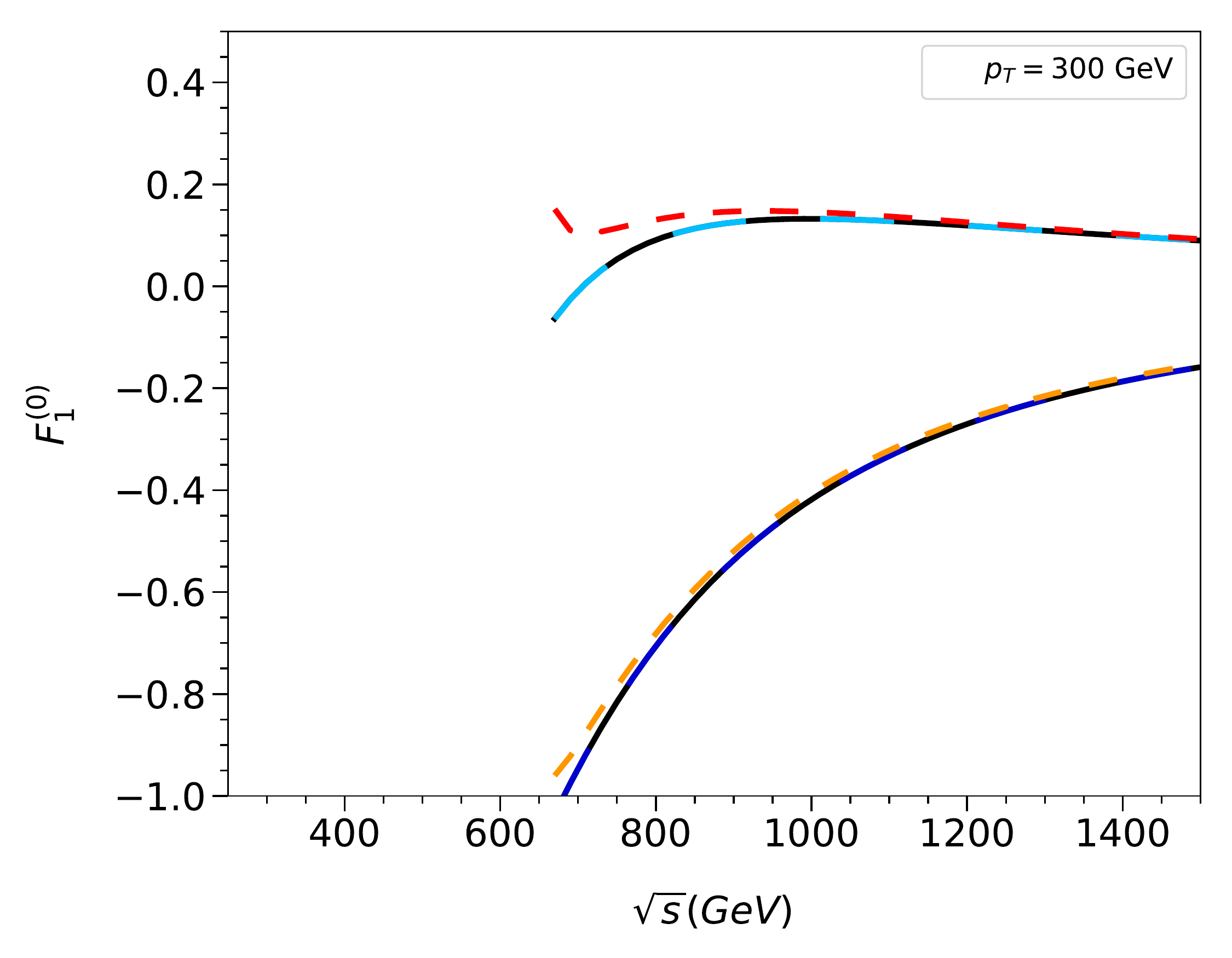} 
  \end{tabular}
  \caption{\label{fig::F11l}One-loop form factor
    $F_{\rm box1}^{(0)}$ as a function of $\sqrt{s}$
    for various values of $p_T$.}
\end{figure}

\begin{figure}[t]
  \begin{tabular}{ccc}
    \includegraphics[width=0.31\textwidth]{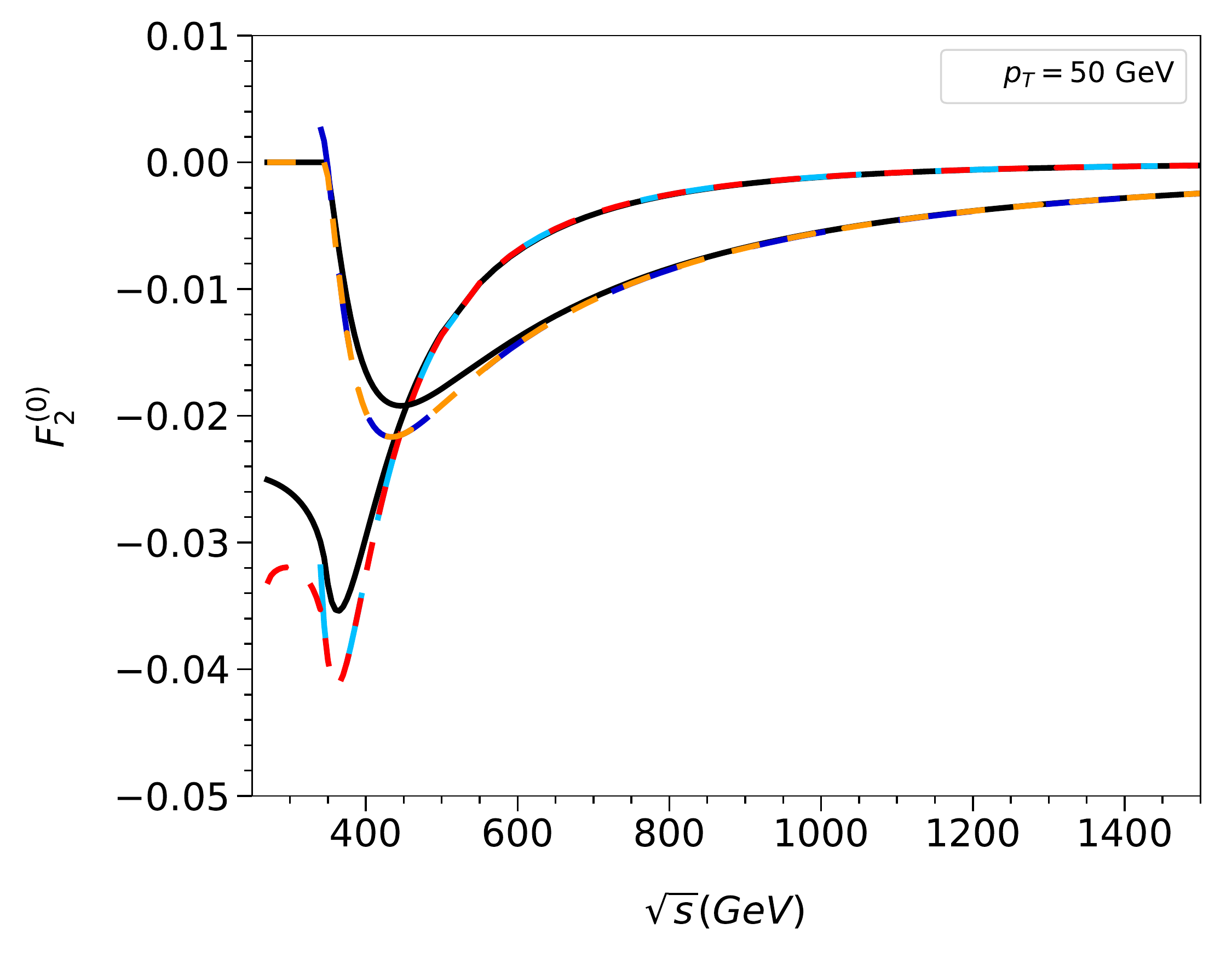} &
    \includegraphics[width=0.31\textwidth]{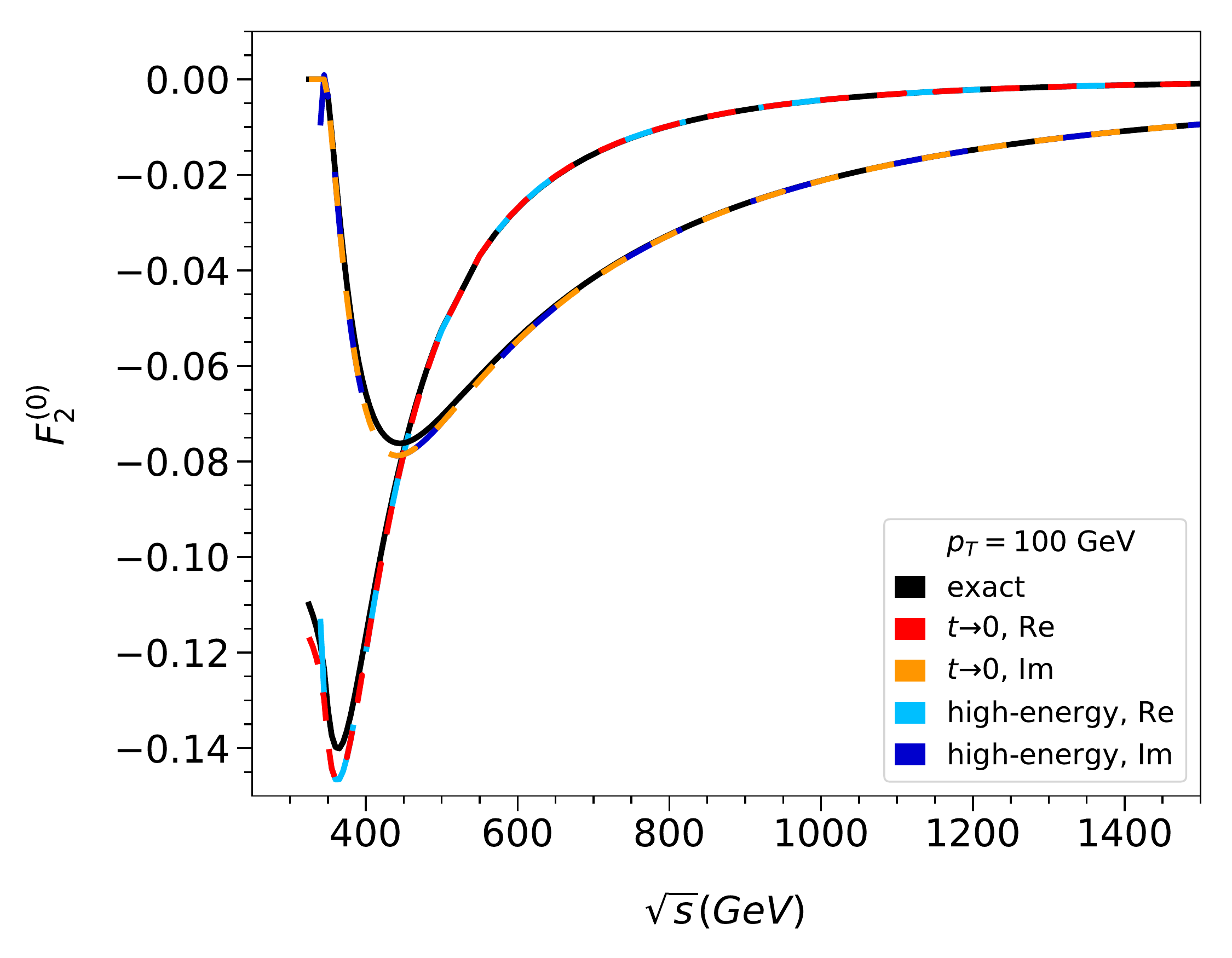} &
    \includegraphics[width=0.31\textwidth]{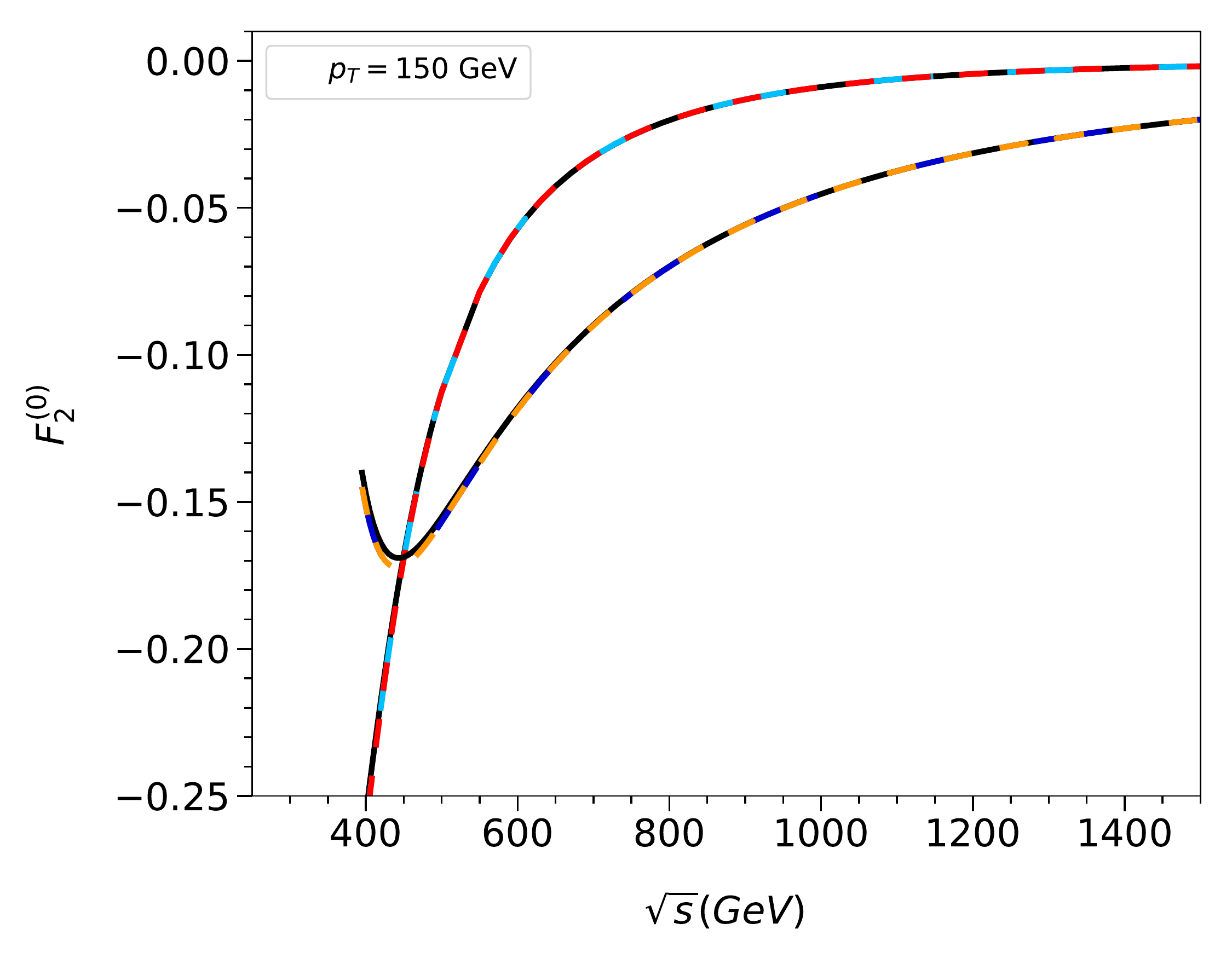} 
    \\
    \includegraphics[width=0.31\textwidth]{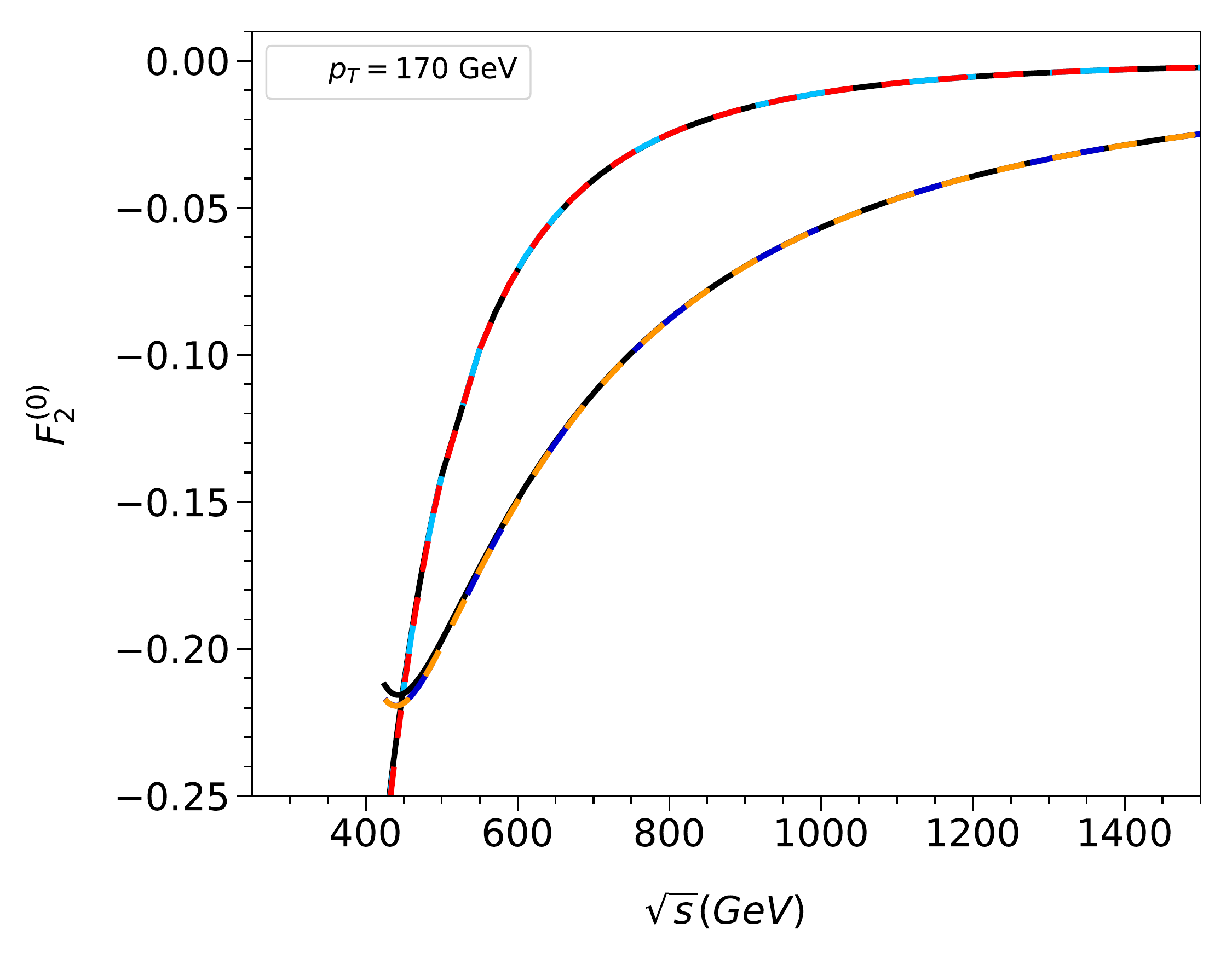} &
    \includegraphics[width=0.31\textwidth]{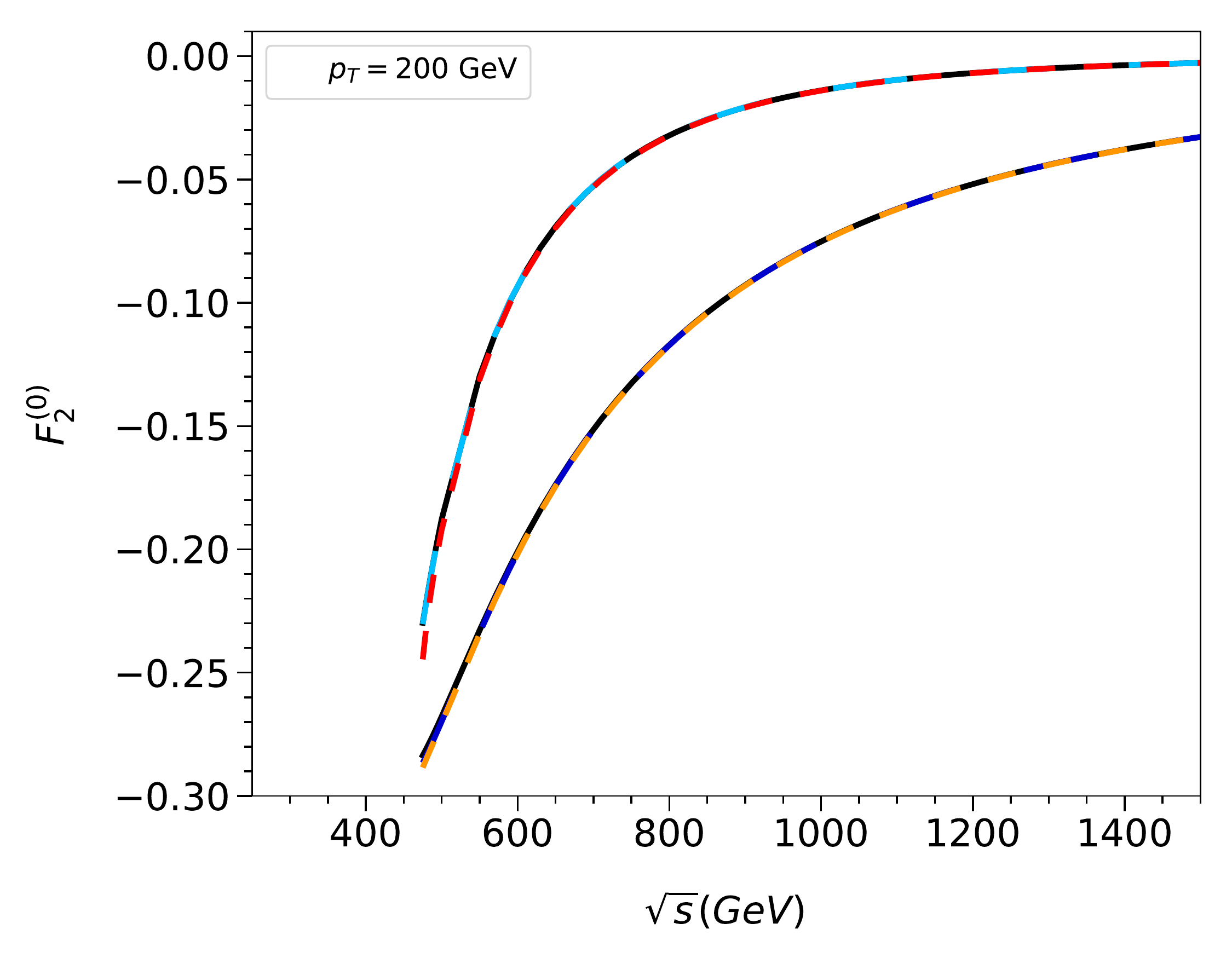} &
    \includegraphics[width=0.31\textwidth]{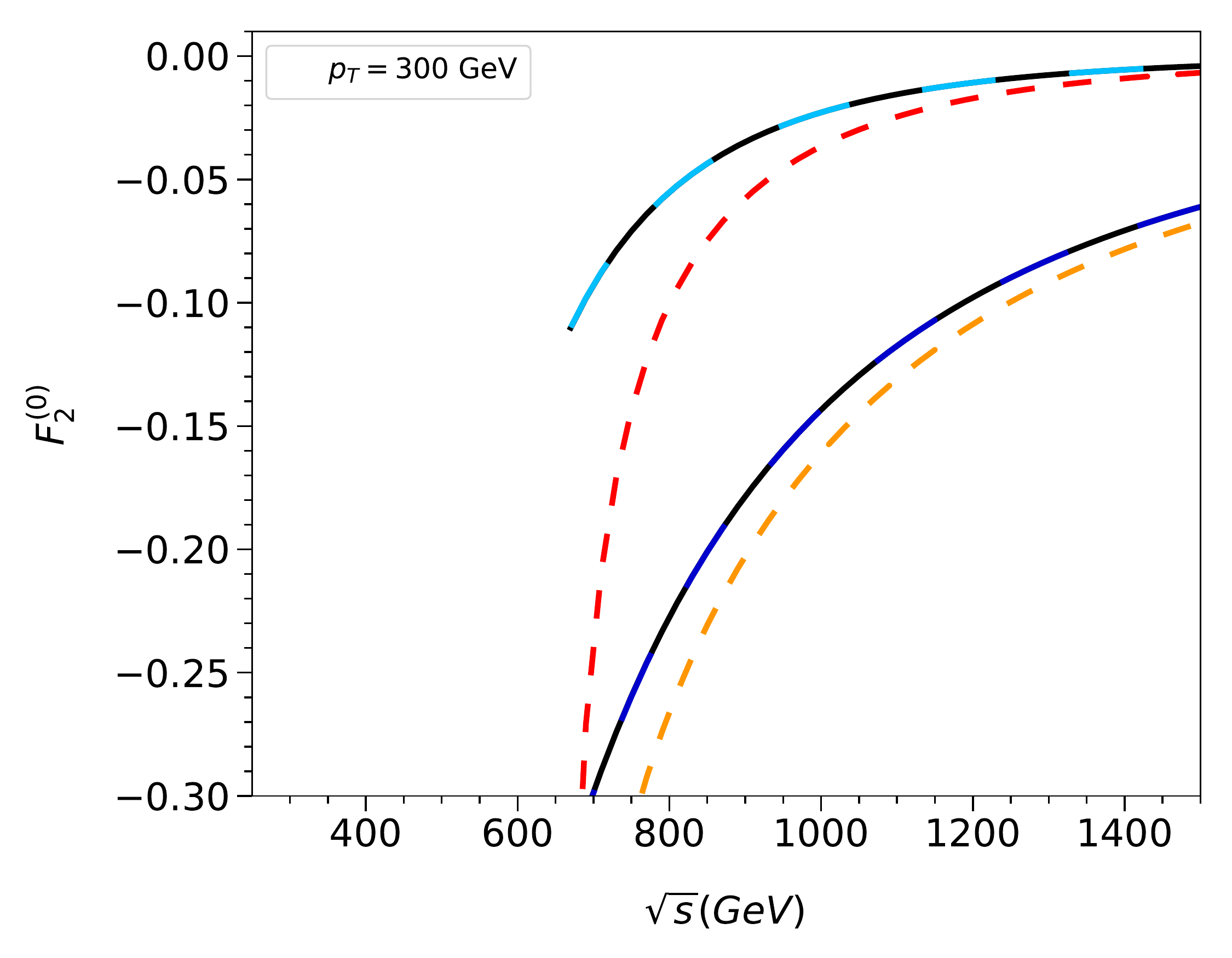} 
  \end{tabular}
  \caption{\label{fig::F21l}One-loop form factor
    $F_{\rm box2}^{(0)}$ as a function of $\sqrt{s}$
    for various values of $p_T$.}
\end{figure}

In Figs.~\ref{fig::F11l} and~\ref{fig::F21l} we
show, for various values of $p_T$, the results for the form factors
$F_{\rm box1}^{(0)}$ and $F_{\rm box2}^{(0)}$ as a function of $\sqrt{s}$.
The
high-energy and small-$t$ expansions are shown as coloured dashed lines; the
solid black line (in the background) corresponds to the exact result.  For
these plots we have incorporated quartic expansion terms in $m_H$, the order
which is also available at the two-loop level.
Furthermore, for the small-$t$ expansion
terms up to $t^5$ are taken into account and the high-energy expansion
includes Pad\'e approximations which include terms up to at least $(m_t^2)^{49}$
and at most $(m_t^2)^{56}$.

Above the top quark pair threshold we observe that both expansions agree with
the exact result even for values of $p_T$ as small as $50$~GeV and as large as
$200$~GeV.  For larger values of $p_T$ the small-$t$ expansion starts to deviate from
the black curve, as can be seen in the panel for $p_T=300$~GeV, whereas the
high-energy approximation agrees very well, as expected.  On the other hand,
for values of $p_T$ below $50$~GeV the small-$t$ expansion provides
an excellent approximation. From the panels in
Figs.~\ref{fig::F11l} and~\ref{fig::F21l} one observes that for
$100~\mbox{GeV}\lesssim p_T \lesssim 200$~GeV both approximations work well
for $\sqrt{s}\gtrsim 350$~GeV.

Below the top quark pair threshold we observe that the small-$t$ expansion
provides an excellent description of the exact result, whereas the high-energy
expansion deviates; this is expected since
it does not contain any information about the threshold.
Values $\sqrt{s}\lesssim 2m_t$ are kinematically only allowed for $p_T\lesssim
120$~GeV. 

To quantify the quality of the approximations we show in
Tabs.~\ref{tab::F_1loop_50},~\ref{tab::F_1loop_200} and~\ref{tab::F_1loop_10},
for three different values of $p_T$, results for the real part of
$F_{\rm box1}^{(0)}$ for various values of $\sqrt{s}$. We show the exact results,
the results for the small-$t$ expansion for different expansion depths in
$m_H$, the high-energy expansion including terms up to $m_H^4$, and results
for the large-$m_t$ expansion (LME) up to $1/m_t^{12}$~\cite{Davies:2019djw}.

Let us start the discussion with Tab.~\ref{tab::F_1loop_50} ($p_T=50$~GeV)
where we observe the following:
\begin{itemize}
\item If we restrict ourselves to the approximation which includes quartic
  $m_H$ terms, in the region above the top quark threshold we observe an
  agreement of at least 3 significant digits between the small-$t$ and
  high-energy expansions.
\item
  The agreement between the exact result and the approximations based on an
  expansion in $m_H$ up to quartic order is well below the percent level.
\item 
  Including expansion terms in $m_H$, beyond the quartic terms, for the small-$t$ expansion
  improves the agreement with the exact result.
\end{itemize}

Similar conclusions also hold for
$p_T=200$~GeV, as can be seen in Tab.~\ref{tab::F_1loop_200}.
In practical applications the high-energy expansion
can be used for such values of $p_T$.

The purpose of Tab.~\ref{tab::F_1loop_10} is to show that
the small-$t$ expansion also works for small values of $p_T$
and large values of $\sqrt{s}$. It is impressive that for
such small values of $p_T$ the high-energy expansion still
provides good approximations for $\sqrt{s}$ values around
$400$~GeV. This demonstrates the power of a deep expansion
in $m_t$ combined with a Pad\'e improvement. For larger values of
$\sqrt{s}$ the high-energy expansion breaks down, because it is
no longer the case that $|t|\gg m_t^2$.

For $F_{\rm box2}^{(0)}$ the comparison is not so straightforward, as can be seen
in the first two panels of Fig.~\ref{fig::F21l} and in Tab.~\ref{tab::F2_1loop_50}.
We observe that the expansion in $m_H$ does not converge sufficiently quickly for
the quartic terms to provide a good description of the exact curve for
$p_T\lesssim 100$~GeV. While including terms to $m_H^8$ in the small-$t$ expansion
again provides good agreement, such expansion terms are not available at two loops.

We show in Tab.~\ref{tab::F2_1loop_50} that below the top quark pair production threshold,
the large top quark expansion of Ref.~\cite{Davies:2019djw}
(including expansion terms to $1/m_t^{12}$)
provides a good approximation of
the exact result and can be used instead in this region. However, $F_{\rm box2}^{(0)}$
is numerically much smaller than $F_{\rm box1}^{(0)}$; we have verified that the use of
the large top quark expansion in this region does not affect the results and conclusions
of Section~\ref{sub::Vfin}.

\begin{table}[t]
\begin{tabular}{cc|llllll}
$\sqrt{s}$ (GeV)& &$ 270$ &$ 300$ &$ 350$ &$ 400$ &$ 610$ &$ 990$\\
\hline
exact& &$ -1.72013$ &$ -1.81435$ &$ -2.32246$ &$ -2.34773$ &$ -0.393996$ &$ 0.0855054$\\
\hline
small-$t$&$m_H^0$ &$ -1.44108$ &$ -1.52523$ &$ -1.92423$ &$ -2.01154$ &$ -0.420989$ &$ 0.0626770$\\
&$m_H^2$ &$ -1.67642$ &$ -1.77026$ &$ -2.25482$ &$ -2.30931$ &$ -0.404100$ &$ 0.0837986$\\
&$m_H^4$ &$ -1.71321$ &$ -1.80759$ &$ -2.31050$ &$ -2.34518$ &$ -0.395265$ &$ 0.0854682$\\
&$m_H^6$ &$ -1.71902$ &$ -1.81331$ &$ -2.32026$ &$ -2.34808$ &$ -0.394063$ &$ 0.0855094$\\
&$m_H^8$ &$ -1.71995$ &$ -1.81419$ &$ -2.32204$ &$ -2.34793$ &$ -0.393990$ &$ 0.0855057$\\
\hline
high-en.&$m_H^4$ &--- &--- &$ -2.31129$ &$ -2.34521$ &$ -0.395262$ &$ 0.0854694$\\
\hline
LME & &$-1.71813$&$-1.80468$&$-2.08865$&$-2.76874$&---&---\\
\end{tabular}
\caption{\label{tab::F_1loop_50} Real part of $F_{\rm box1}^{(0)}$ for $p_T=50$~GeV.}
\end{table}

\begin{table}[t]
\begin{tabular}{cc|llllll}
$\sqrt{s}$ (GeV)& &$ 610$ &$ 990$\\
\hline
exact& &$ -0.311182$ &$ 0.110469$\\
\hline
small-$t$&$m_H^0$ &$ -0.340443$ &$ 0.089788$\\
&$m_H^2$ &$ -0.319571$ &$ 0.109173$\\
&$m_H^4$ &$ -0.311692$ &$ 0.110538$\\
&$m_H^6$ &$ -0.310705$ &$ 0.110570$\\
&$m_H^8$ &$ -0.310651$ &$ 0.110567$\\
\hline
high-energy&$m_H^4$ &$ -0.312218$ &$ 0.110440$\\
\end{tabular}
\caption{\label{tab::F_1loop_200} Real part of $F_{\rm box1}^{(0)}$ for $p_T=200$~GeV.}
\end{table}

\begin{table}[t]
\begin{tabular}{cc|llllll}
$\sqrt{s}$ (GeV)& &$ 270$ &$ 300$ &$ 350$ &$ 400$ &$ 610$ &$ 990$\\
\hline
exact& &$ -1.72358$ &$ -1.81816$ &$ -2.32666$ &$ -2.35282$ &$ -0.400246$ &$ 0.0835134$\\
\hline
small-$t$&$m_H^0$ &$ -1.44780$ &$ -1.52956$ &$ -1.92815$ &$ -2.01570$ &$ -0.426920$ &$ 0.0605334$\\
&$m_H^2$ &$ -1.68133$ &$ -1.77444$ &$ -2.25910$ &$ -2.31430$ &$ -0.410425$ &$ 0.0817808$\\
&$m_H^4$ &$ -1.71707$ &$ -1.81151$ &$ -2.31474$ &$ -2.35027$ &$ -0.401533$ &$ 0.0834753$\\
&$m_H^6$ &$ -1.72257$ &$ -1.81714$ &$ -2.32446$ &$ -2.35317$ &$ -0.400314$ &$ 0.0835175$\\
&$m_H^8$ &$ -1.72342$ &$ -1.81800$ &$ -2.32624$ &$ -2.35302$ &$ -0.400239$ &$ 0.0835137$\\
\hline
high-en.&$m_H^4$ &--- &--- &$ -2.32046$ &$ -2.35382$ &$ -0.464921$ &$ -0.539285$\\
\hline
LME & &$-1.72158$&$-1.80854$&$-2.09373$&$-2.77895$&---&---\\
\end{tabular}
\caption{\label{tab::F_1loop_10} Real part of $F_{\rm box1}^{(0)}$ for $p_T=10$~GeV.}
\end{table}

From the considerations above, we propose the following selection criteria
for the choice of expansion in the different regions of
the $\{\sqrt{s},p_T\}$ plane:
\begin{itemize}
\item Below $p_T=150$~GeV: use small-$t$ expansion for all values of $\sqrt{s}$.
\item For $150$~GeV$\lesssim p_T \lesssim 200$~GeV either approximation can be used.
\item Above $p_T=200$~GeV use the high-energy expansion for all values of $\sqrt{s}$.
\end{itemize}
As a consequence, below $\sqrt{s}=2 m_t$ the small-$t$ expansion is always selected.
The fact that the high-energy and small-$t$ expansions agree with each other (and with the exact
result) in the region $150$~GeV$\lesssim p_T \lesssim 200$~GeV increases our confidence
in the accuracy of the expansions; we will check for this agreement at two loops, where
no exact analytic result for the form factors is available.

\begin{table}[t]
\begin{tabular}{cc|llllll}
$\sqrt{s}$ (GeV)& &$ 270$ &$ 300$ &$ 350$ &$ 400$ &$ 610$ &$ 990$\\
\hline
exact& &$ -0.025050$ &$ -0.026046$ &$ -0.033323$ &$ -0.029569$ &$ -0.006633$ &$ -0.001207$\\
\hline
small-$t$&$m_H^0$ &$ -0.111991$ &$ -0.072393$ &$ -0.064400$ &$ -0.050849$ &$ -0.009550$ &$ -0.001571$\\
&$m_H^2$ &$ -0.069277$ &$ -0.058082$ &$ -0.061193$ &$ -0.048812$ &$ -0.008496$ &$ -0.001339$\\
&$m_H^4$ &$ -0.033254$ &$ -0.031982$ &$ -0.039319$ &$ -0.032503$ &$ -0.006558$ &$ -0.001190$\\
&$m_H^6$ &$ -0.026450$ &$ -0.027041$ &$ -0.034525$ &$ -0.029807$ &$ -0.006603$ &$ -0.001206$\\
&$m_H^8$ &$ -0.025286$ &$ -0.026208$ &$ -0.033565$ &$ -0.029543$ &$ -0.006631$ &$ -0.001207$\\
\hline
high-en.&$m_H^4$ &--- &--- &$ -0.039369$ &$ -0.032504$ &$ -0.006558$ &$ -0.001189$\\
\hline
LME & &$-0.024977$&$-0.025767$&$-0.028531$&$-0.034309$&---&---\\
\end{tabular}
\caption{\label{tab::F2_1loop_50} Real part of $F_{\rm box2}^{(0)}$ for $p_T=50$~GeV.}
\end{table}

\FloatBarrier

\subsection{Two-loop form factors}

In the following we present results for the two-loop box form factors
where for the ultra-violet renormalization and infra-red subtraction we
follow Ref.~\cite{Davies:2018qvx}. In particular, we renormalize the top
quark mass in the on-shell scheme.

\begin{figure}[t]
  \begin{tabular}{ccc}
    \includegraphics[width=0.31\textwidth]{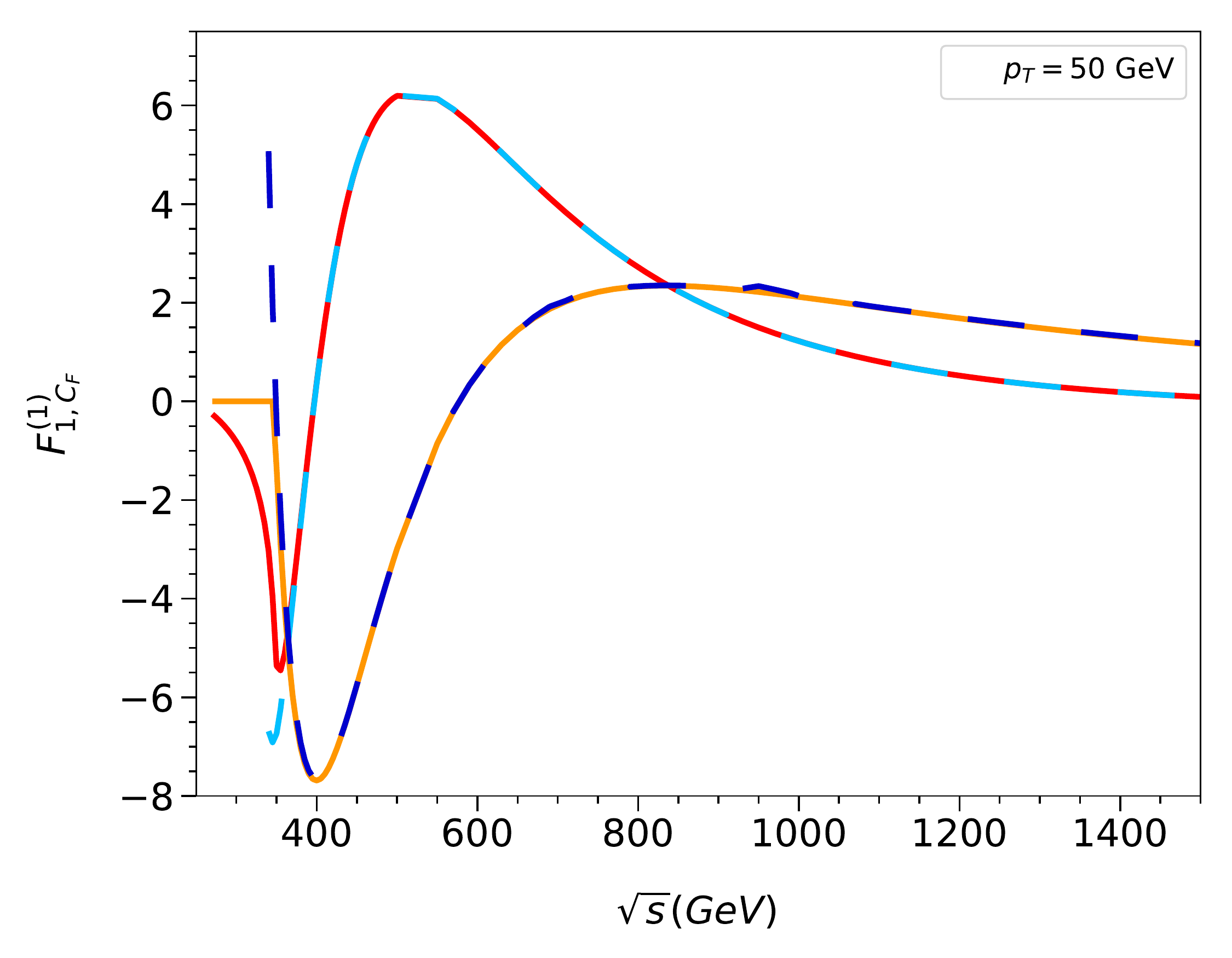} &
    \includegraphics[width=0.31\textwidth]{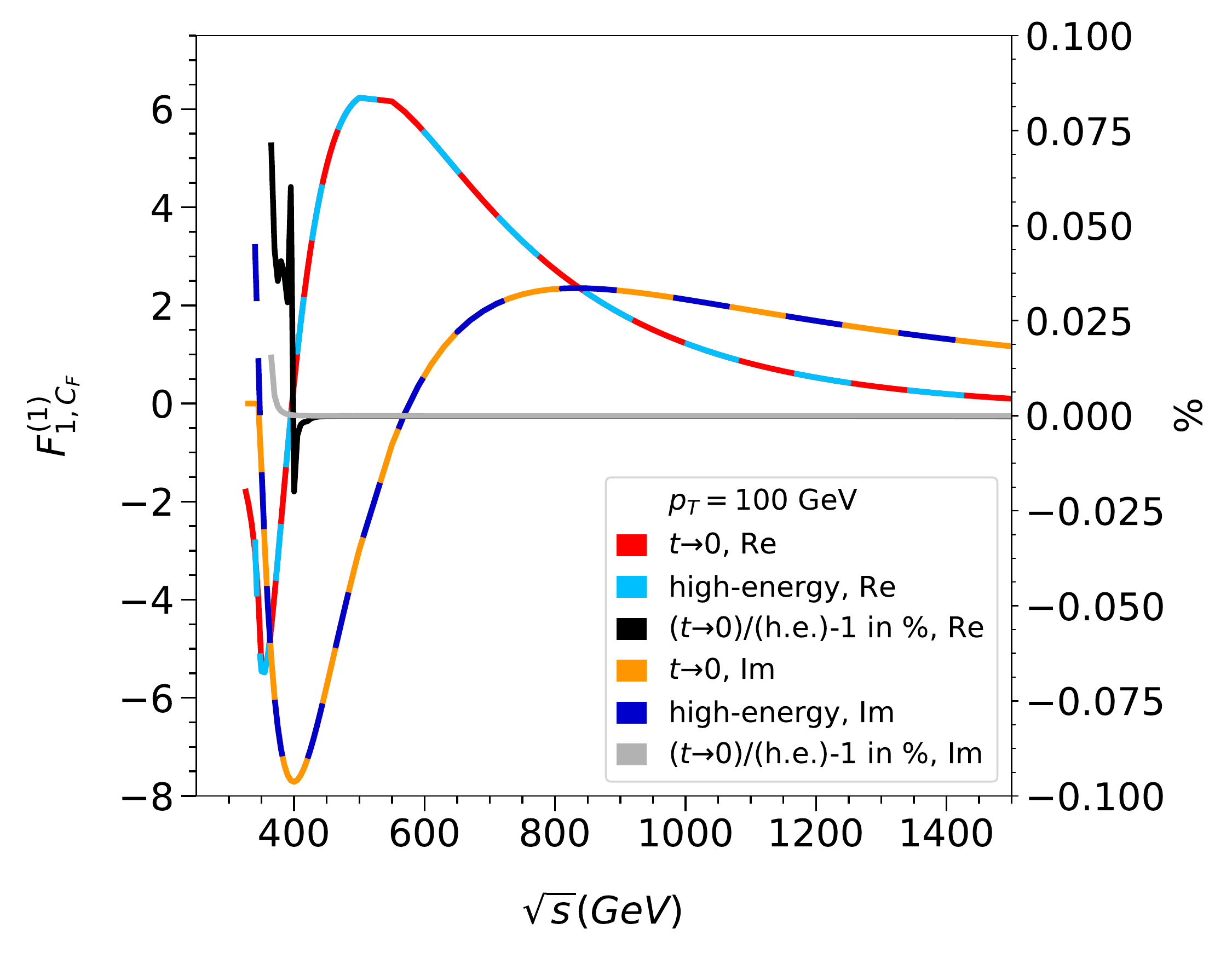} &
    \includegraphics[width=0.31\textwidth]{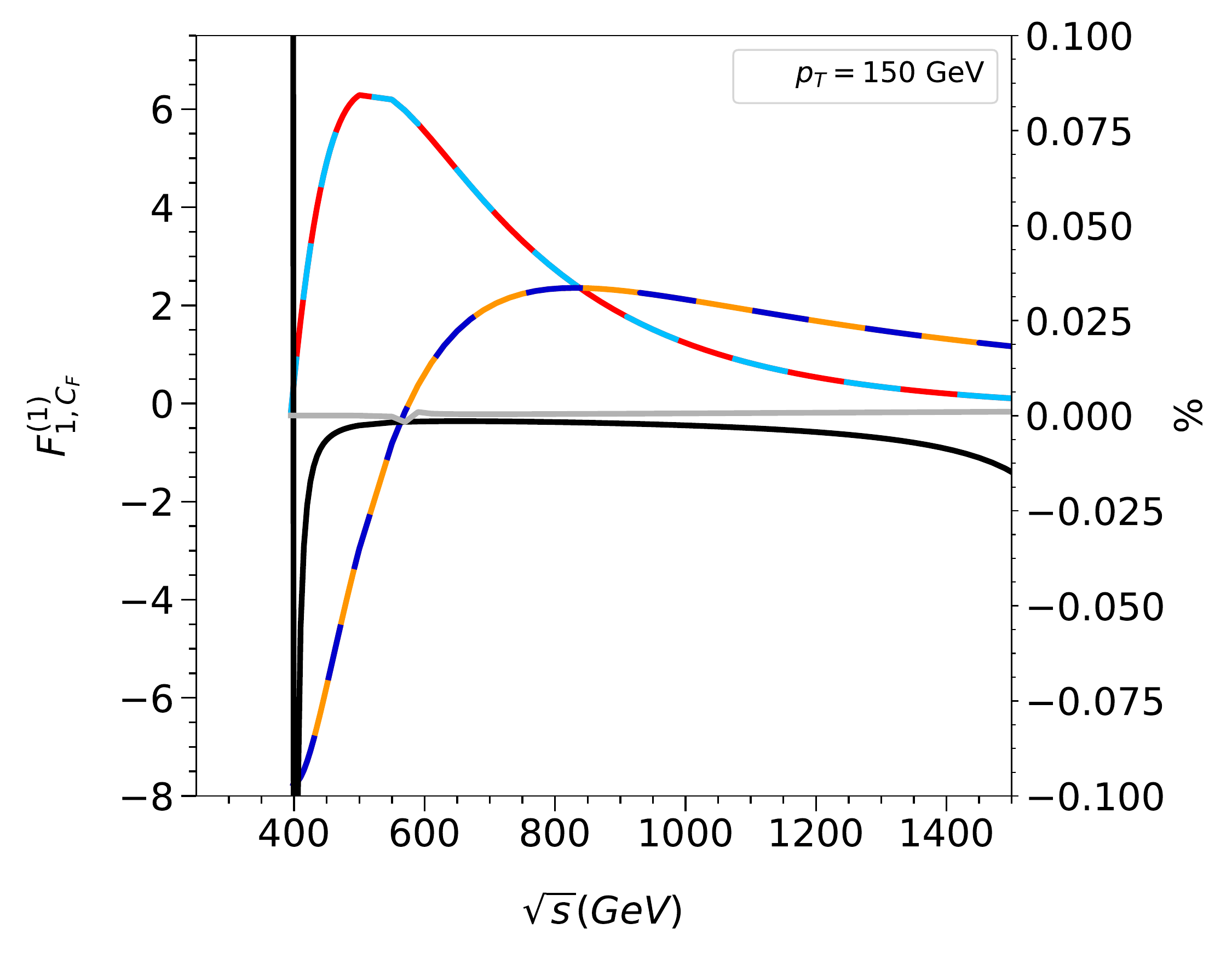} 
    \\
    \includegraphics[width=0.31\textwidth]{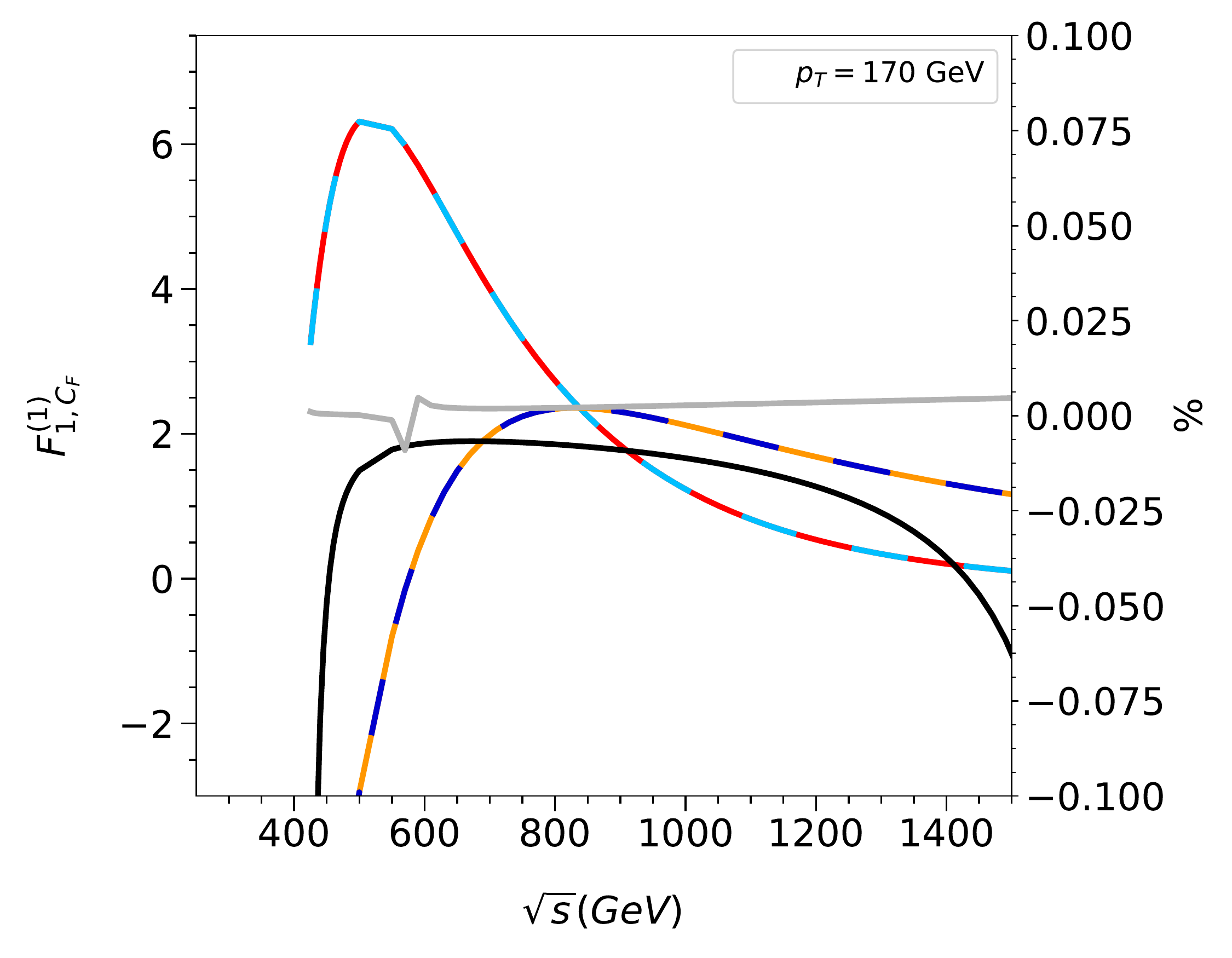} &
    \includegraphics[width=0.31\textwidth]{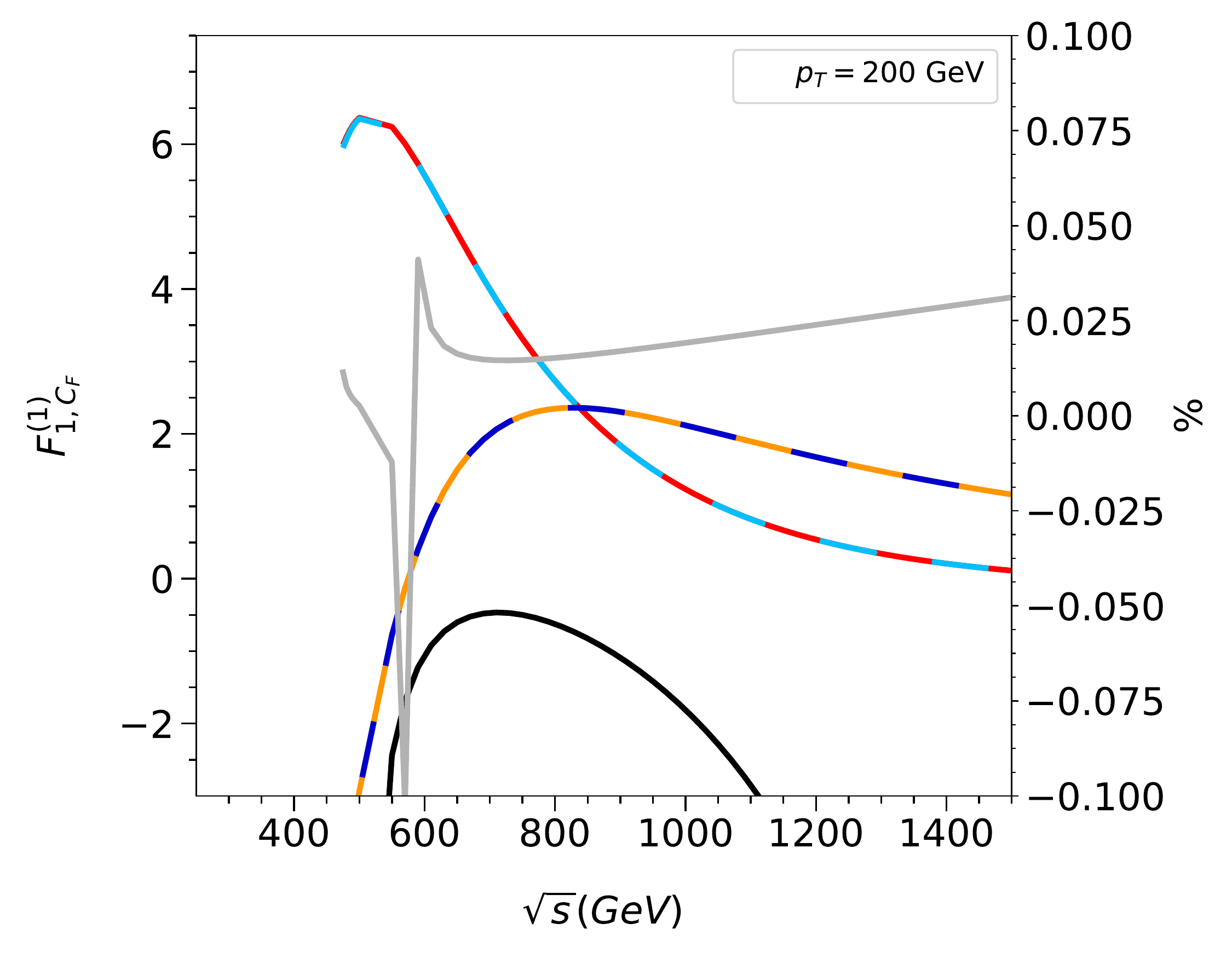} &
    \includegraphics[width=0.31\textwidth]{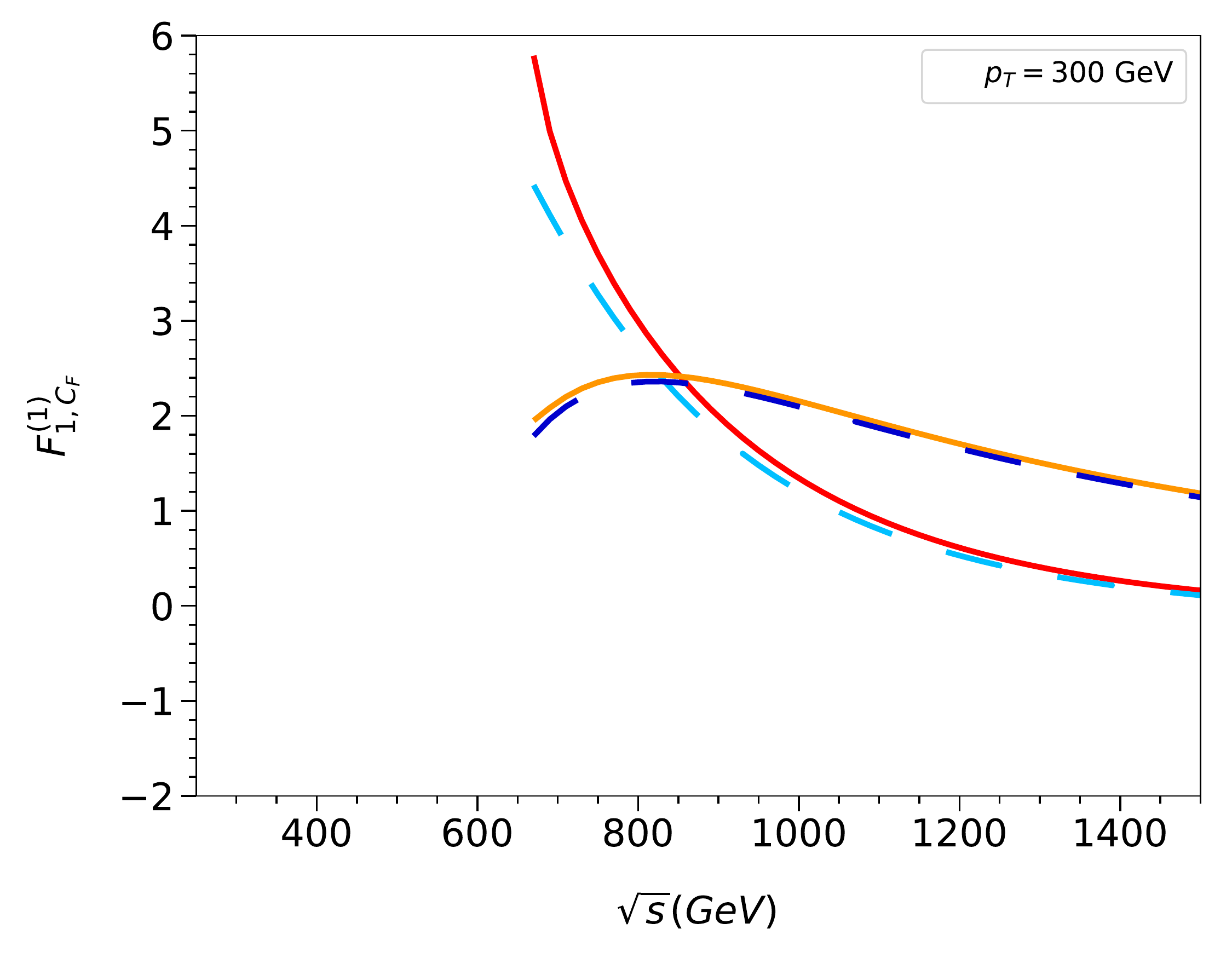} 
  \end{tabular}
  \caption{\label{fig::F12l_cR}$C_F$ contribution to the two-loop form factor
    $F_{\rm box1}^{(1)}$ as a function of $\sqrt{s}$
    for various values of $p_T$.}
\end{figure}

\begin{figure}[t]
  \begin{tabular}{ccc}
    \includegraphics[width=0.31\textwidth]{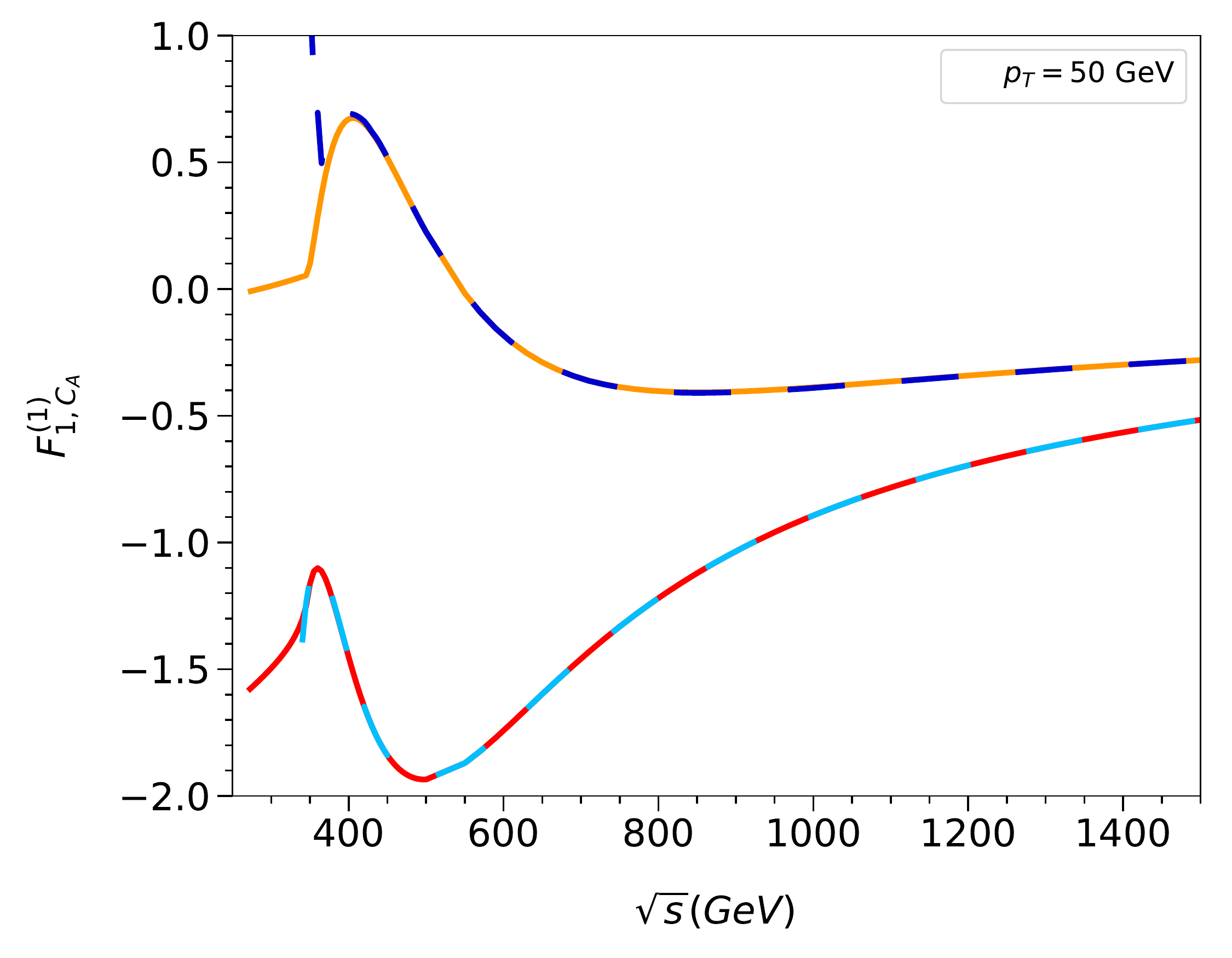} &
    \includegraphics[width=0.31\textwidth]{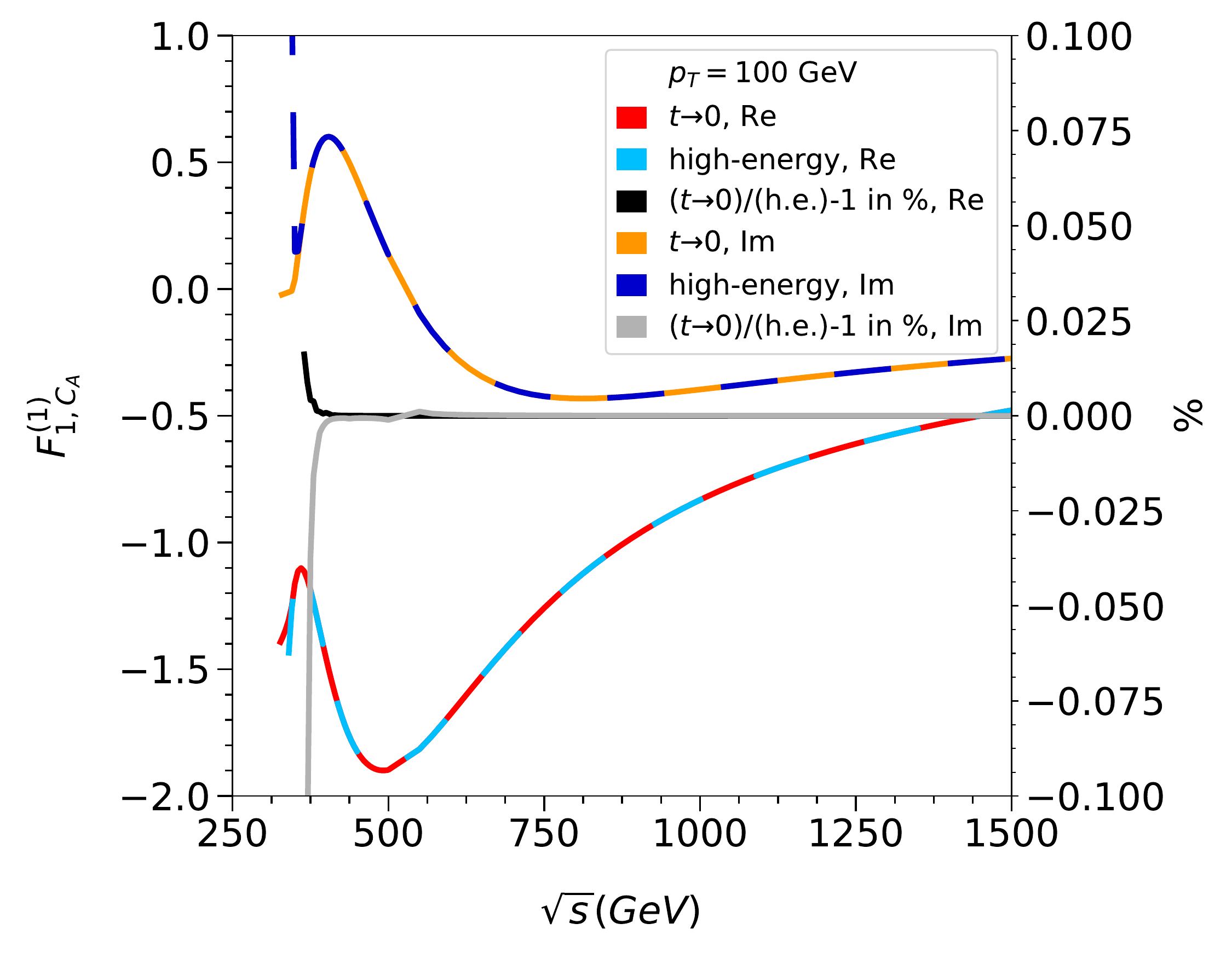} &
    \includegraphics[width=0.31\textwidth]{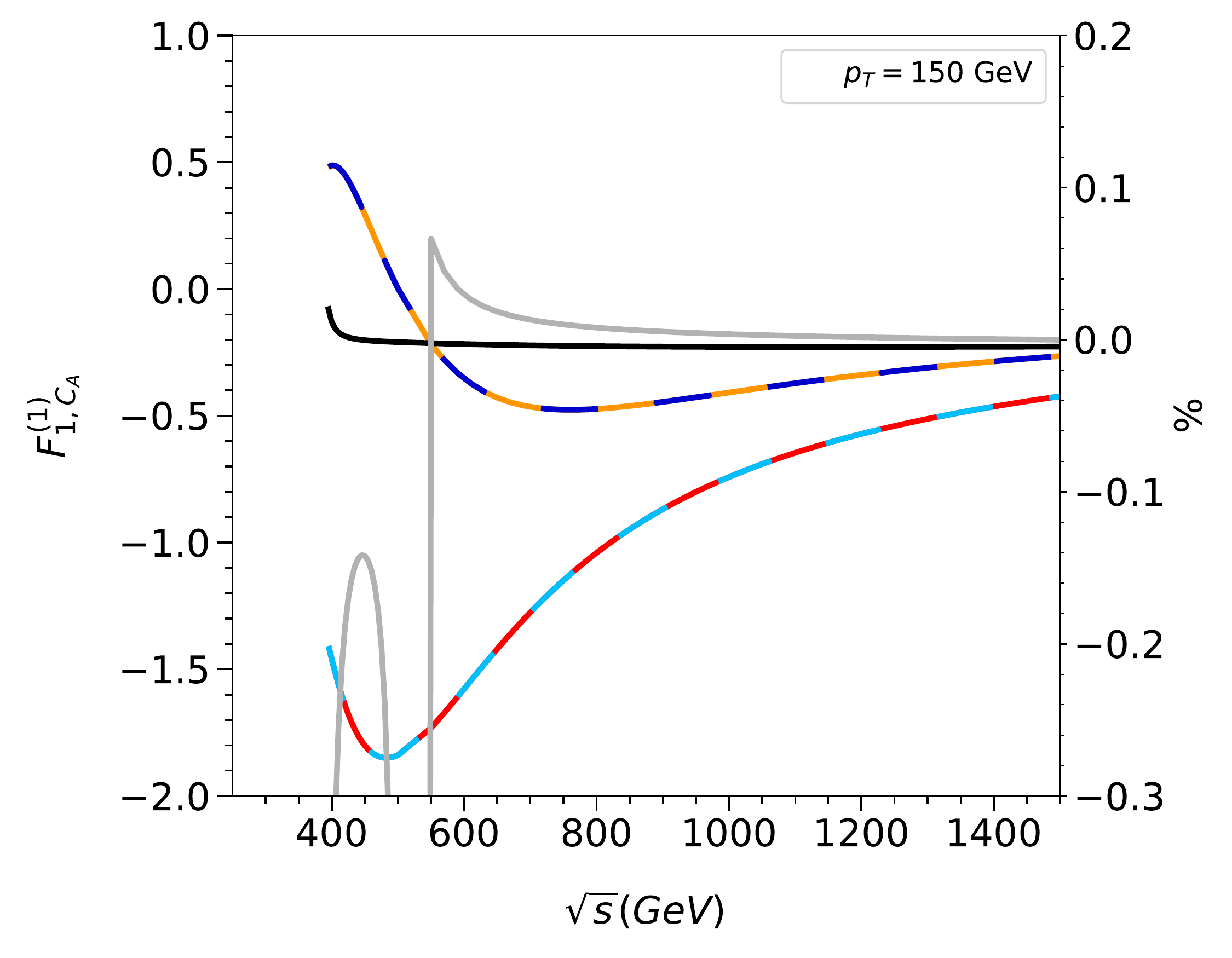} 
    \\
    \includegraphics[width=0.31\textwidth]{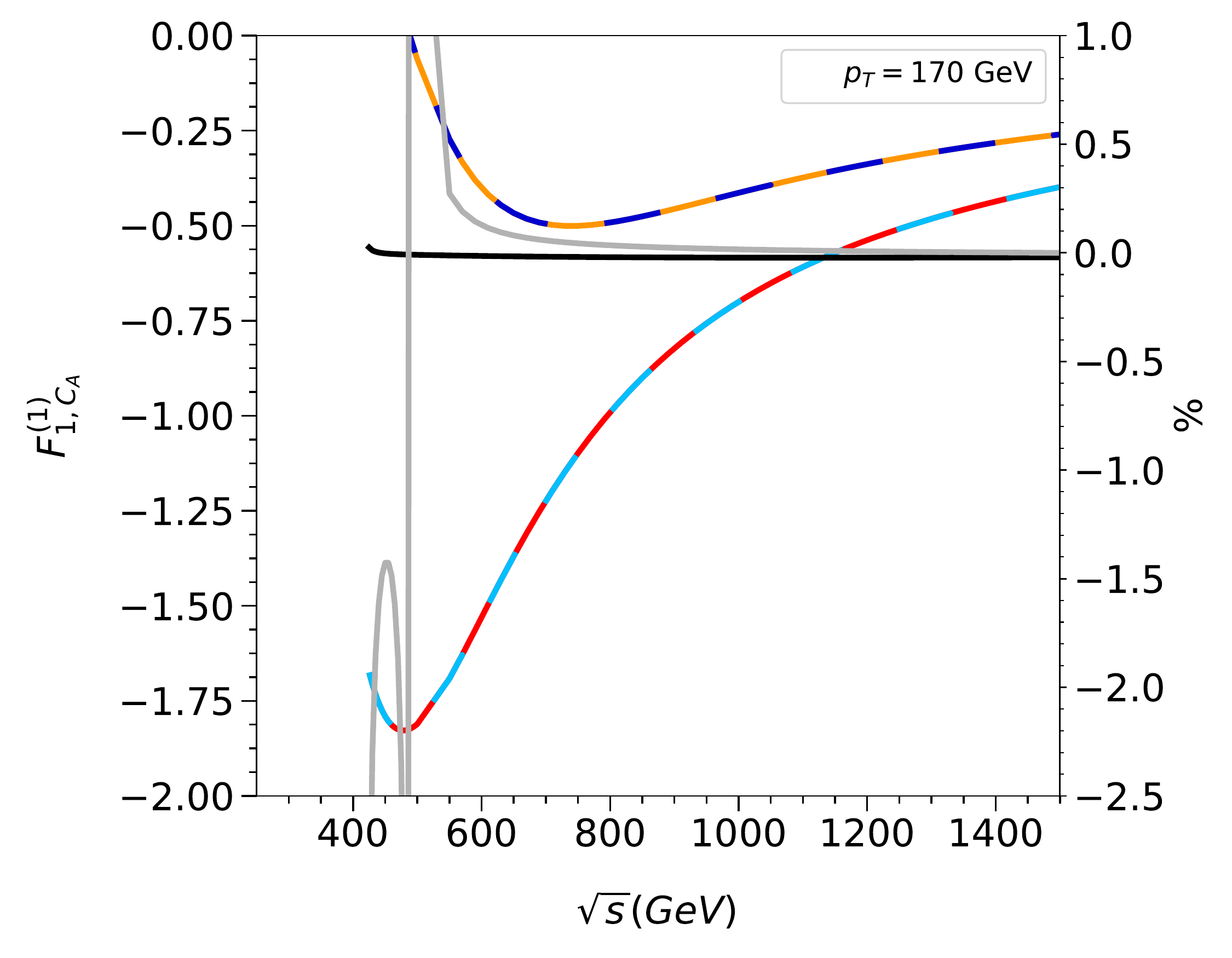} &
    \includegraphics[width=0.31\textwidth]{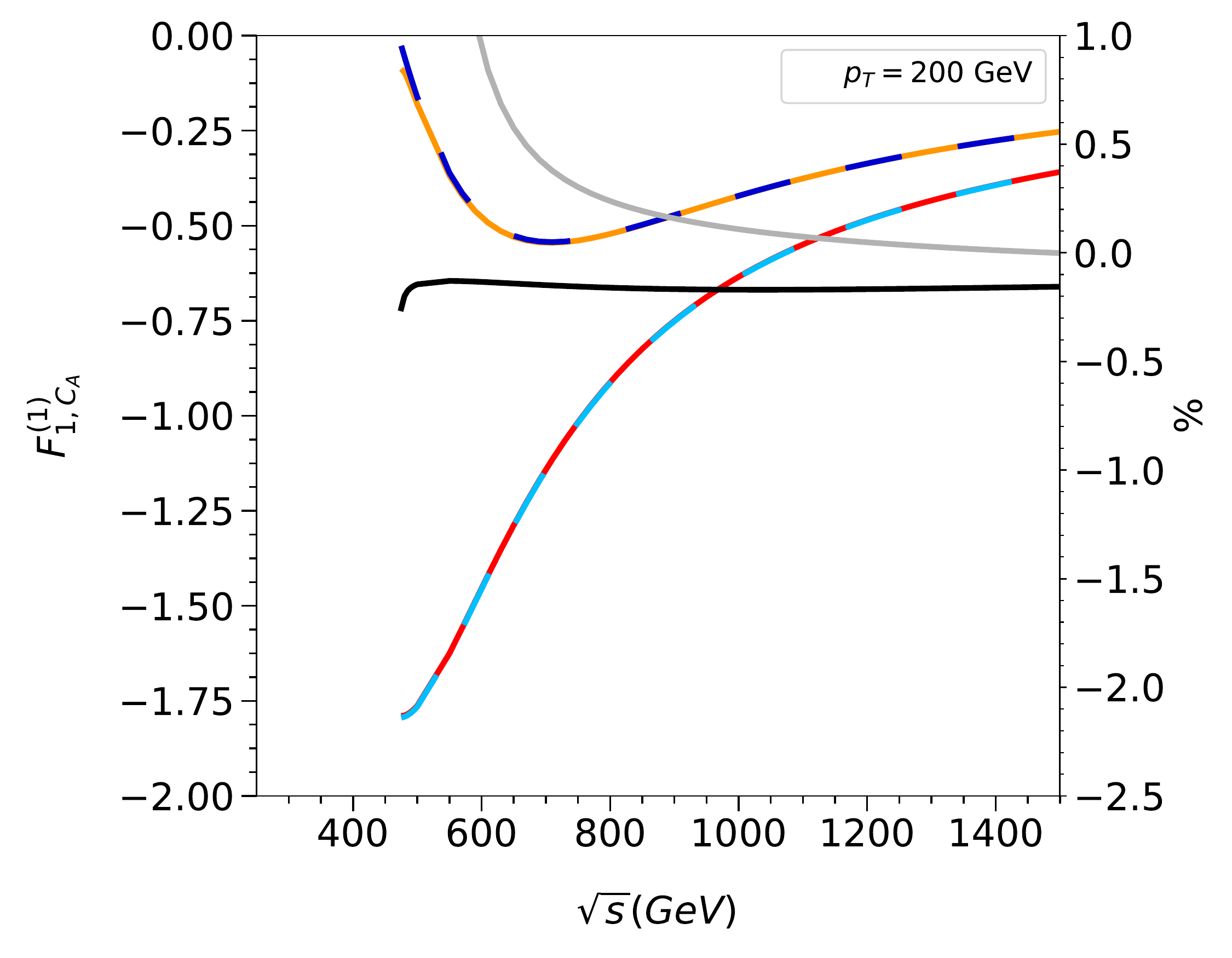} &
    \includegraphics[width=0.31\textwidth]{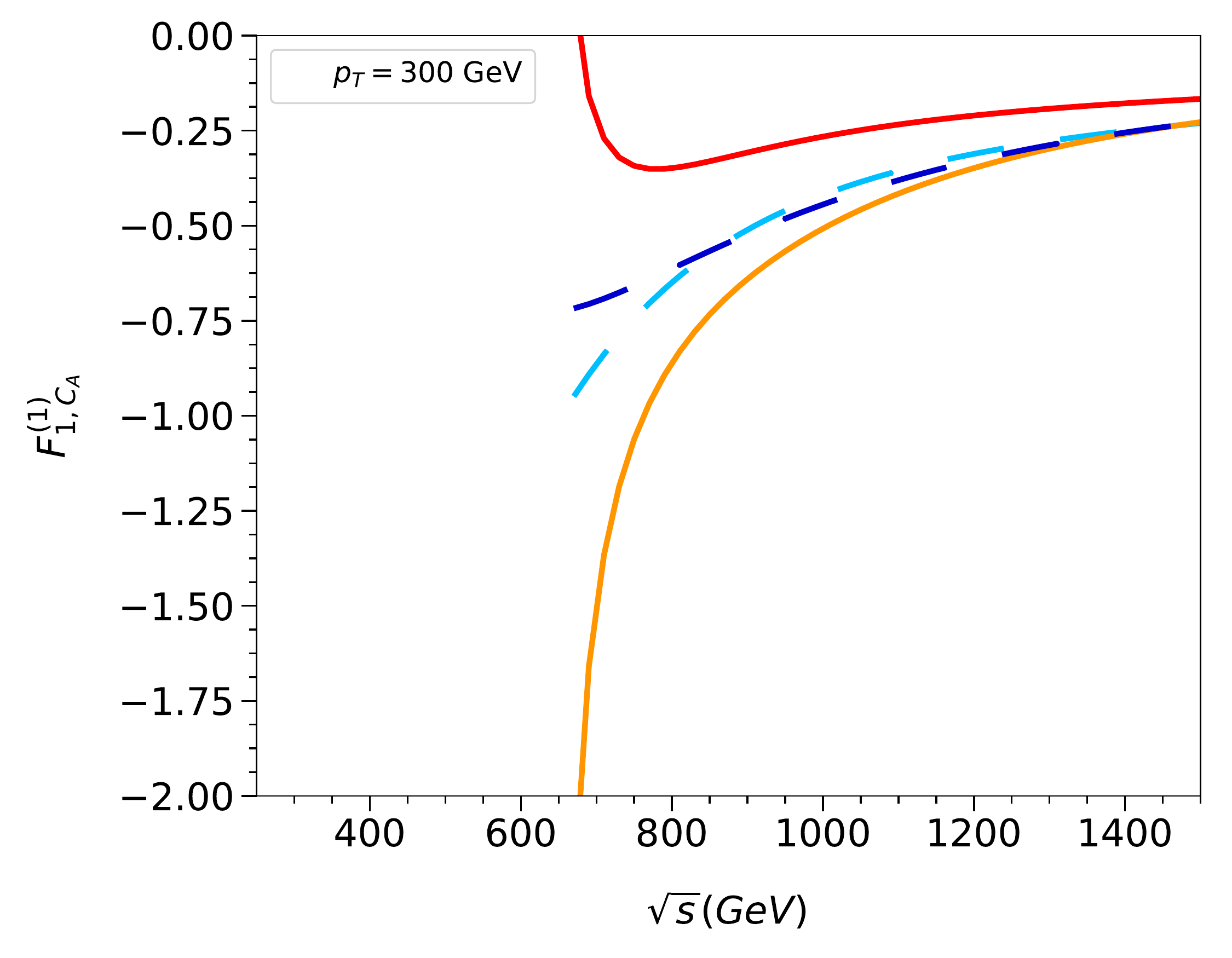} 
  \end{tabular}
  \caption{\label{fig::F12l_cA}$C_A$ contribution to the two-loop form factor
    $F_{\rm box1}^{(1)}$ as a function of $\sqrt{s}$
    for various values of $p_T$.}
\end{figure}

\begin{figure}[t]
  \begin{tabular}{ccc}
    \includegraphics[width=0.31\textwidth]{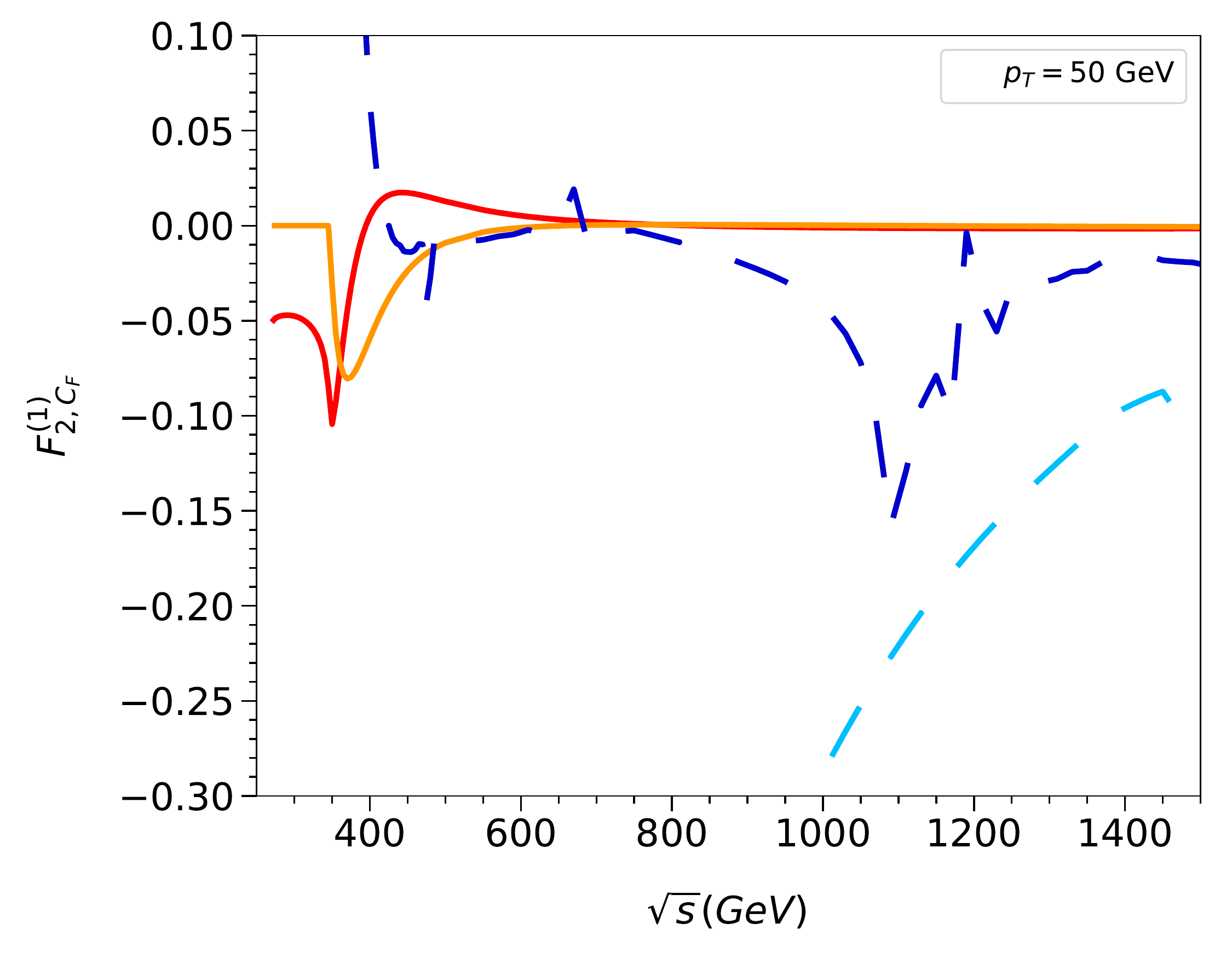} &
    \includegraphics[width=0.31\textwidth]{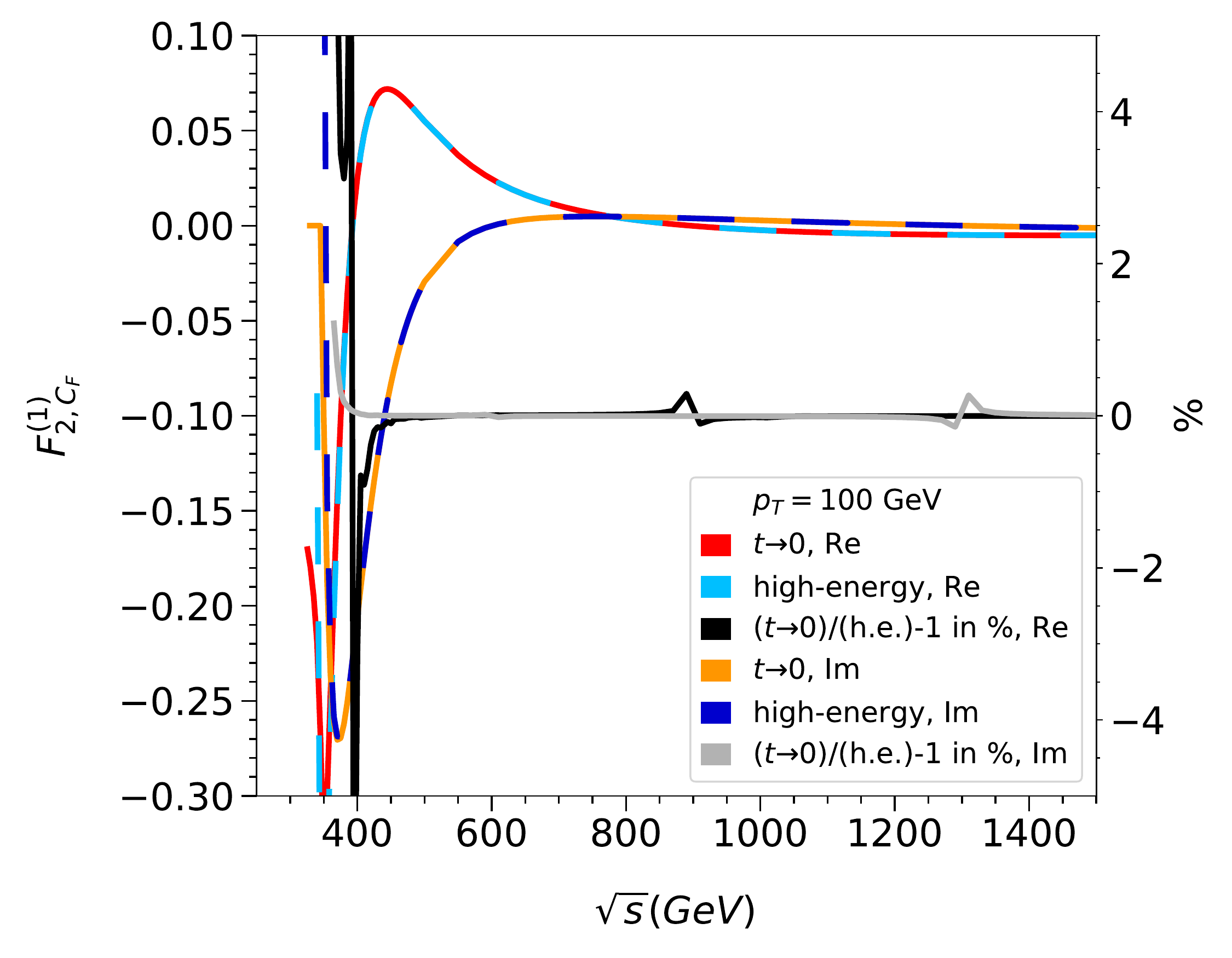} &
    \includegraphics[width=0.31\textwidth]{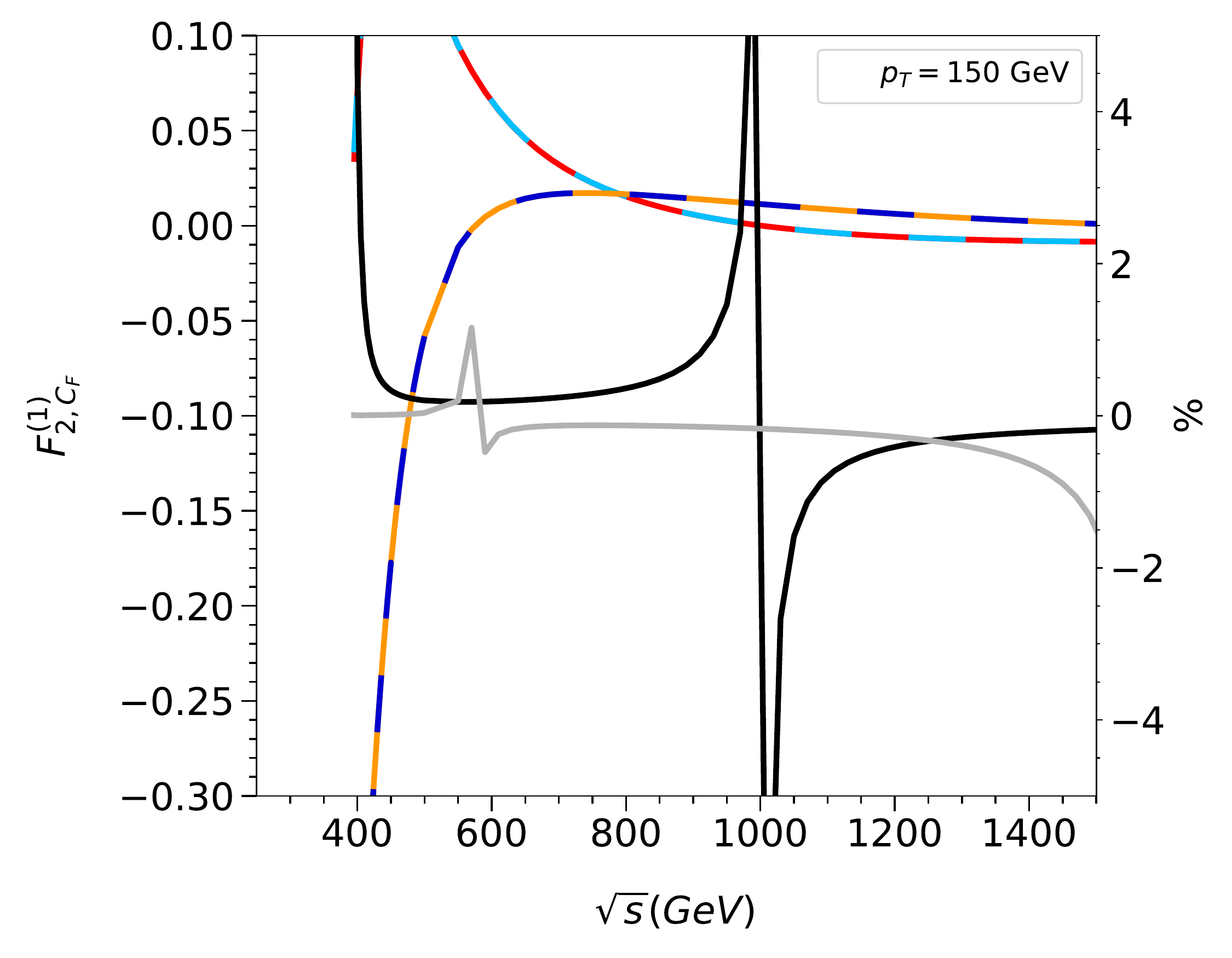} 
    \\
    \includegraphics[width=0.31\textwidth]{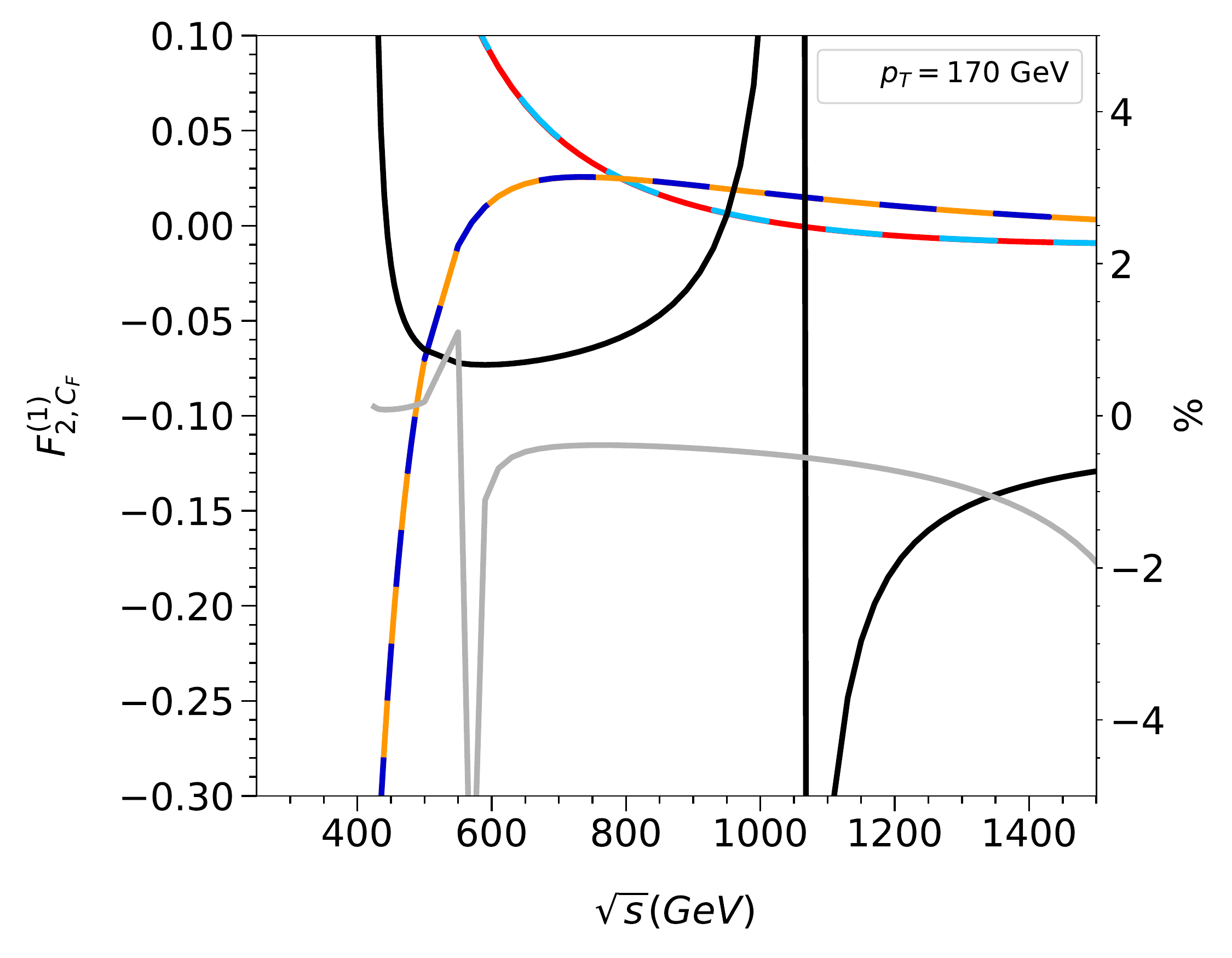} &
    \includegraphics[width=0.31\textwidth]{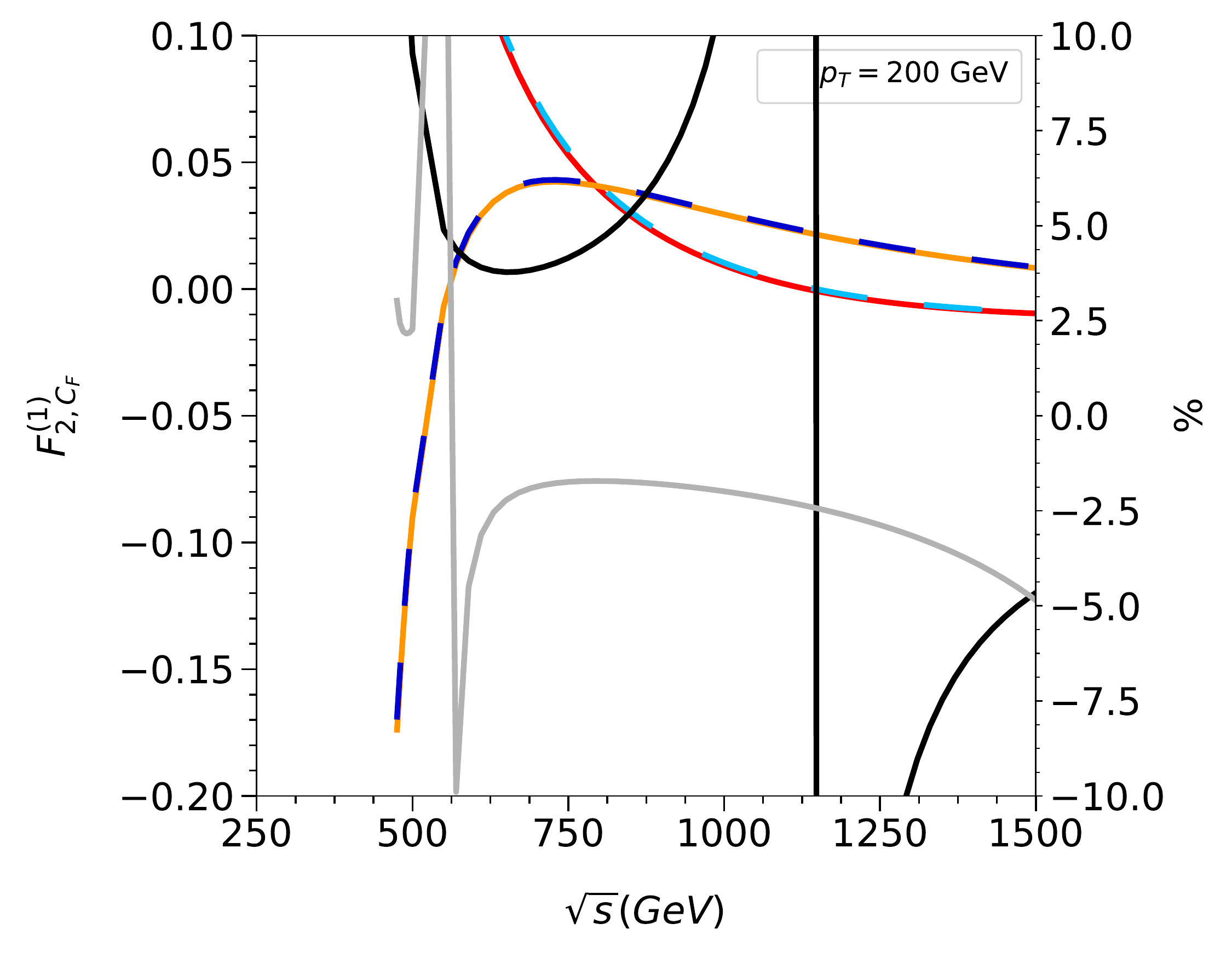} &
    \includegraphics[width=0.31\textwidth]{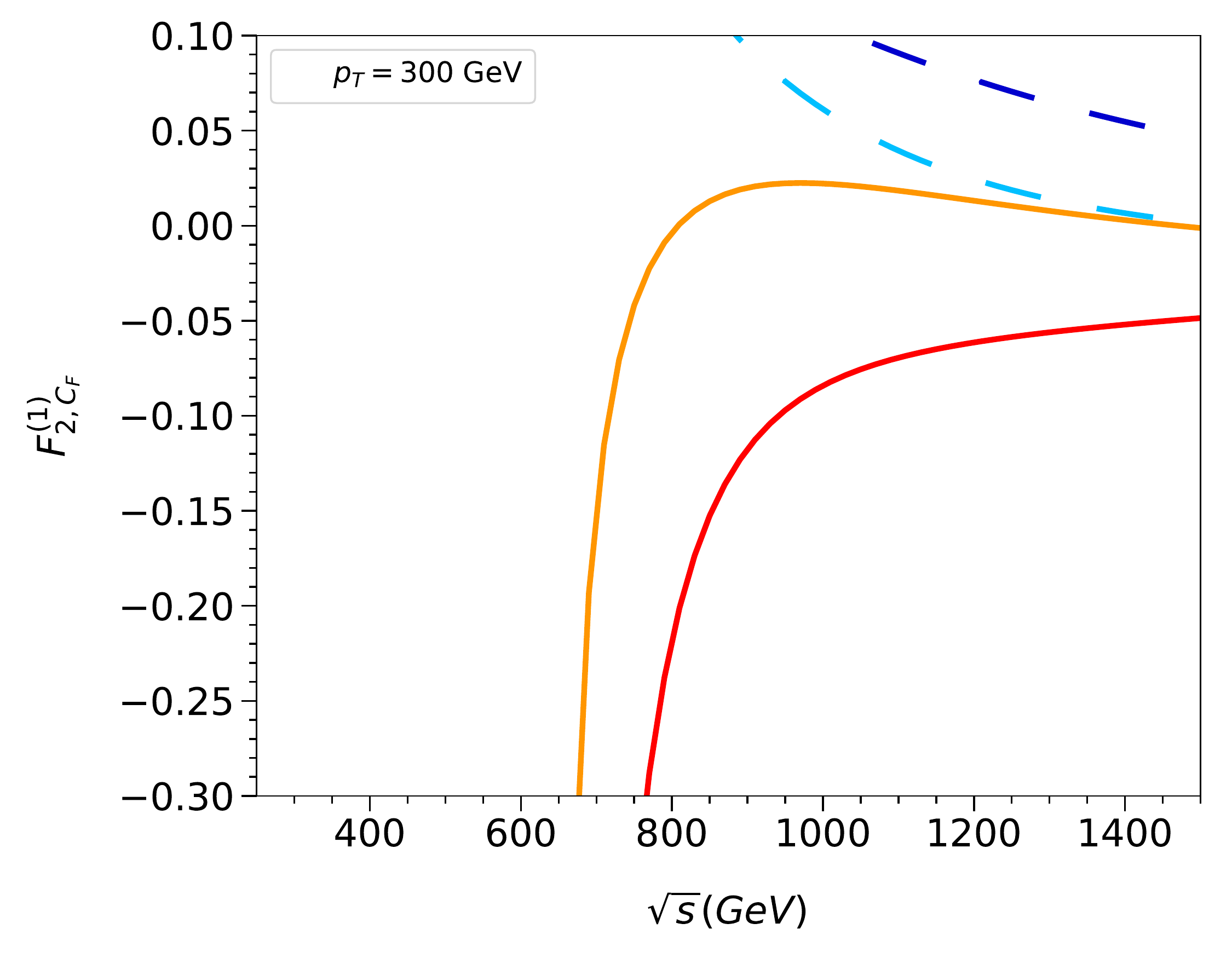} 
  \end{tabular}
  \caption{\label{fig::F22l_cR}$C_F$ contribution to the two-loop form factor
    $F_{\rm box2}^{(1)}$ as a function of $\sqrt{s}$
    for various values of $p_T$.}
\end{figure}

\begin{figure}[t]
  \begin{tabular}{ccc}
    \includegraphics[width=0.31\textwidth]{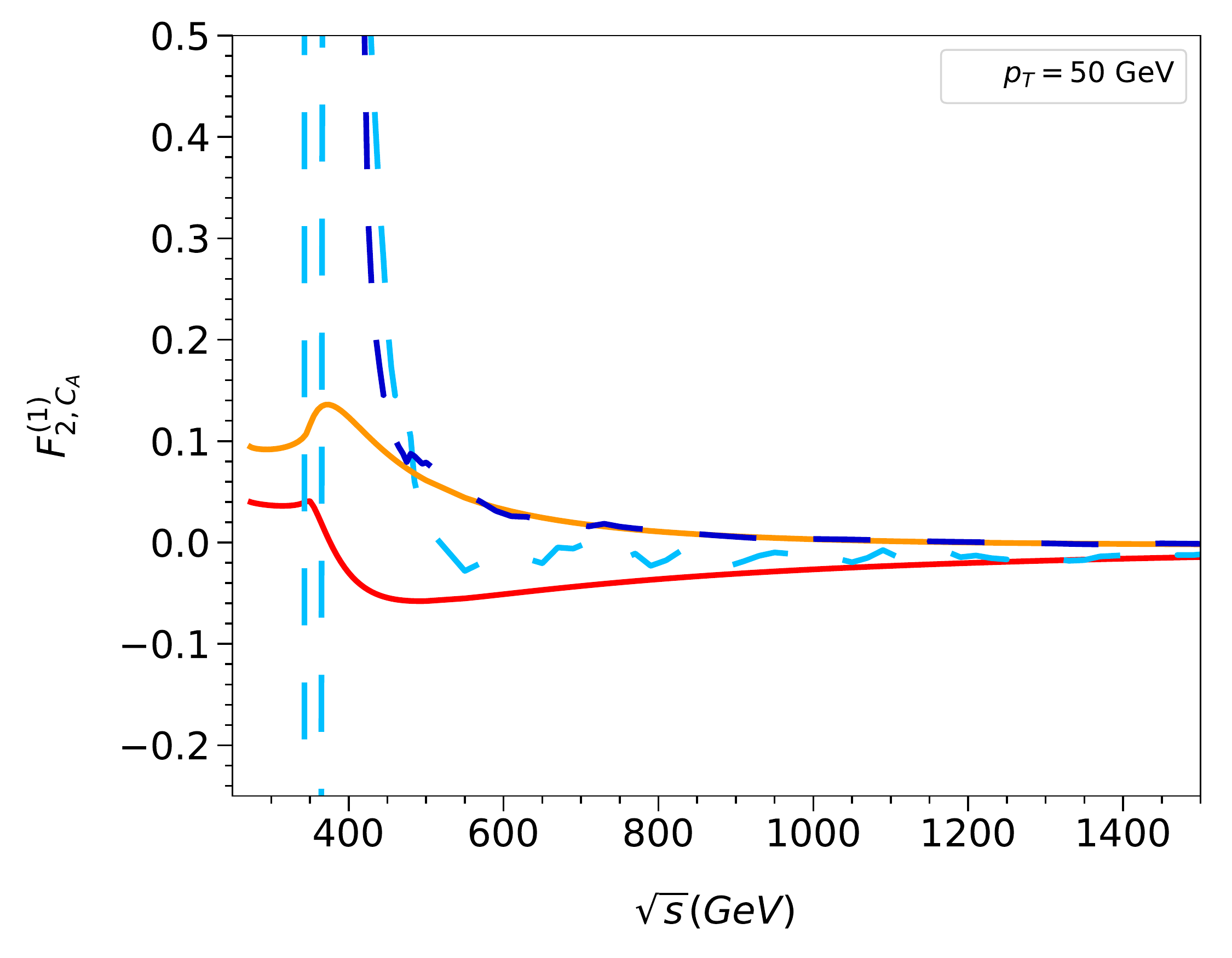} &
    \includegraphics[width=0.31\textwidth]{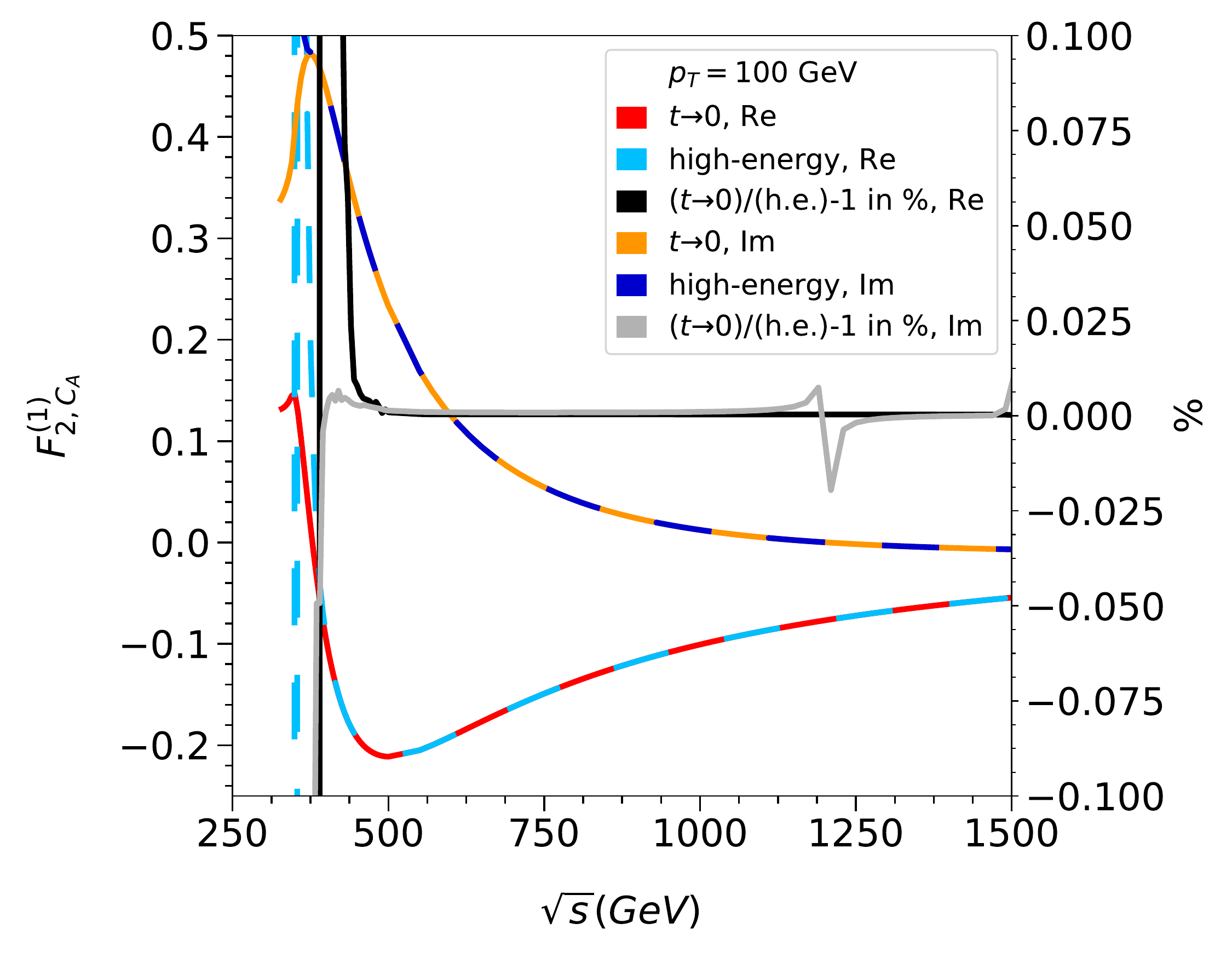} &
    \includegraphics[width=0.31\textwidth]{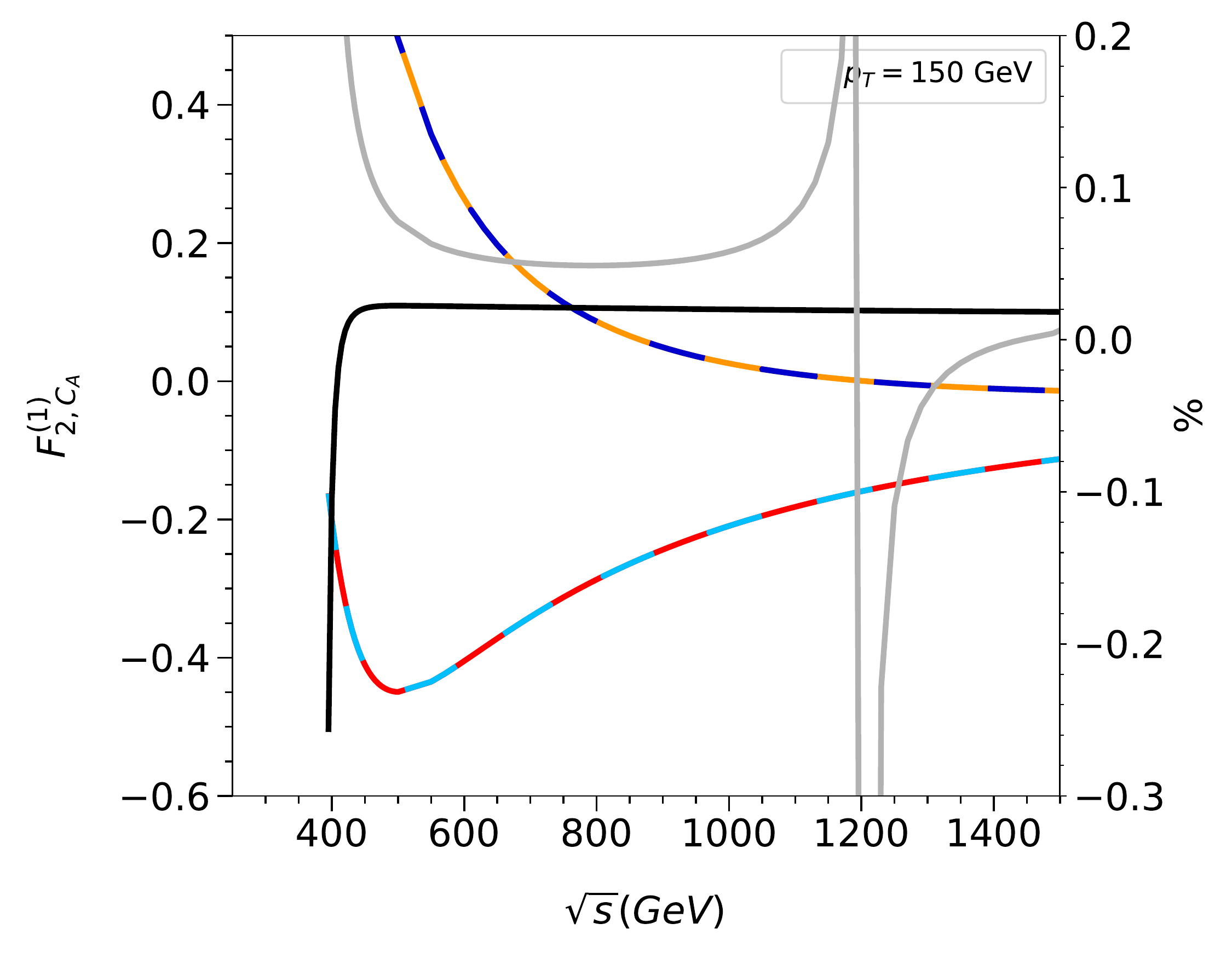} 
    \\
    \includegraphics[width=0.31\textwidth]{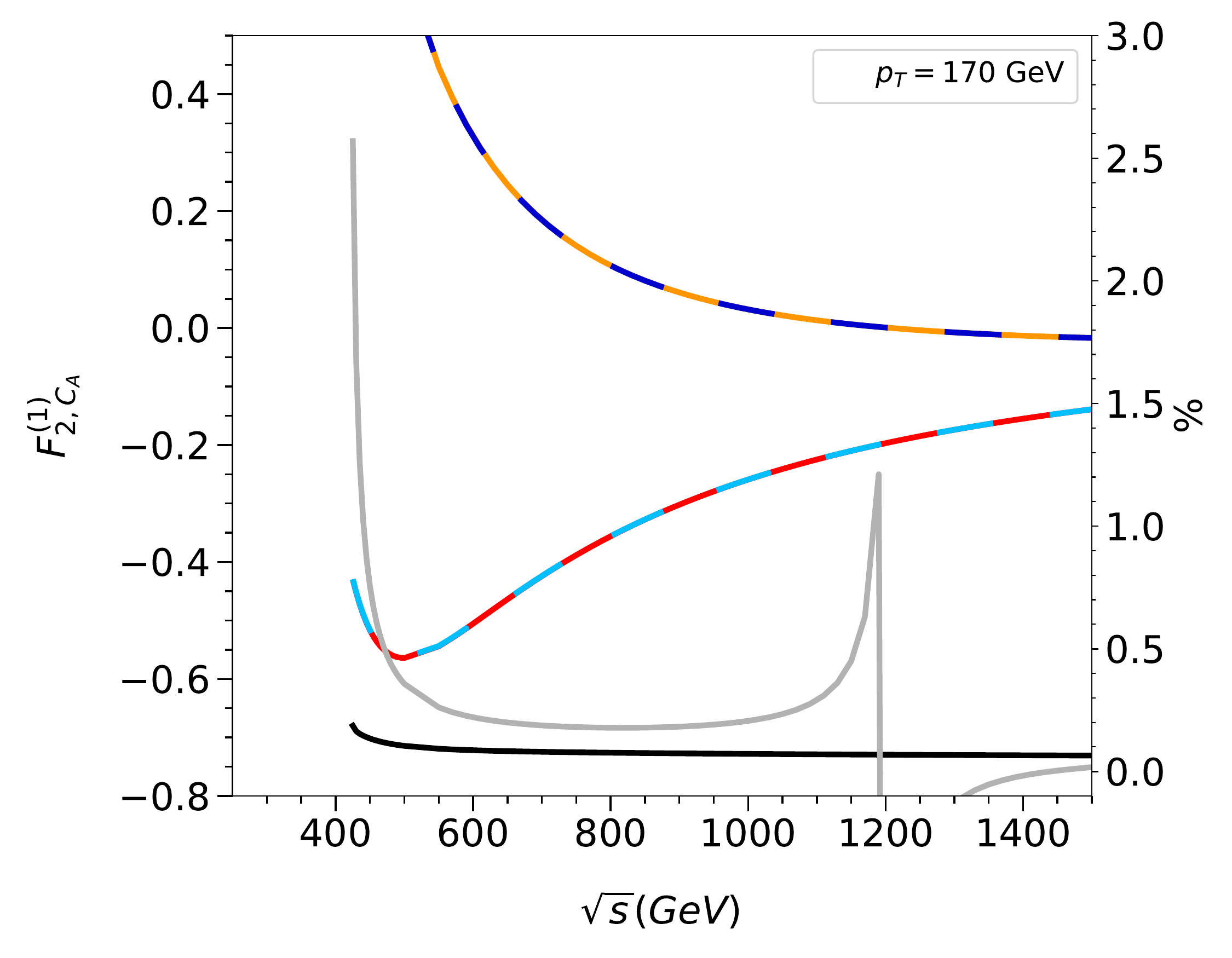} &
    \includegraphics[width=0.31\textwidth]{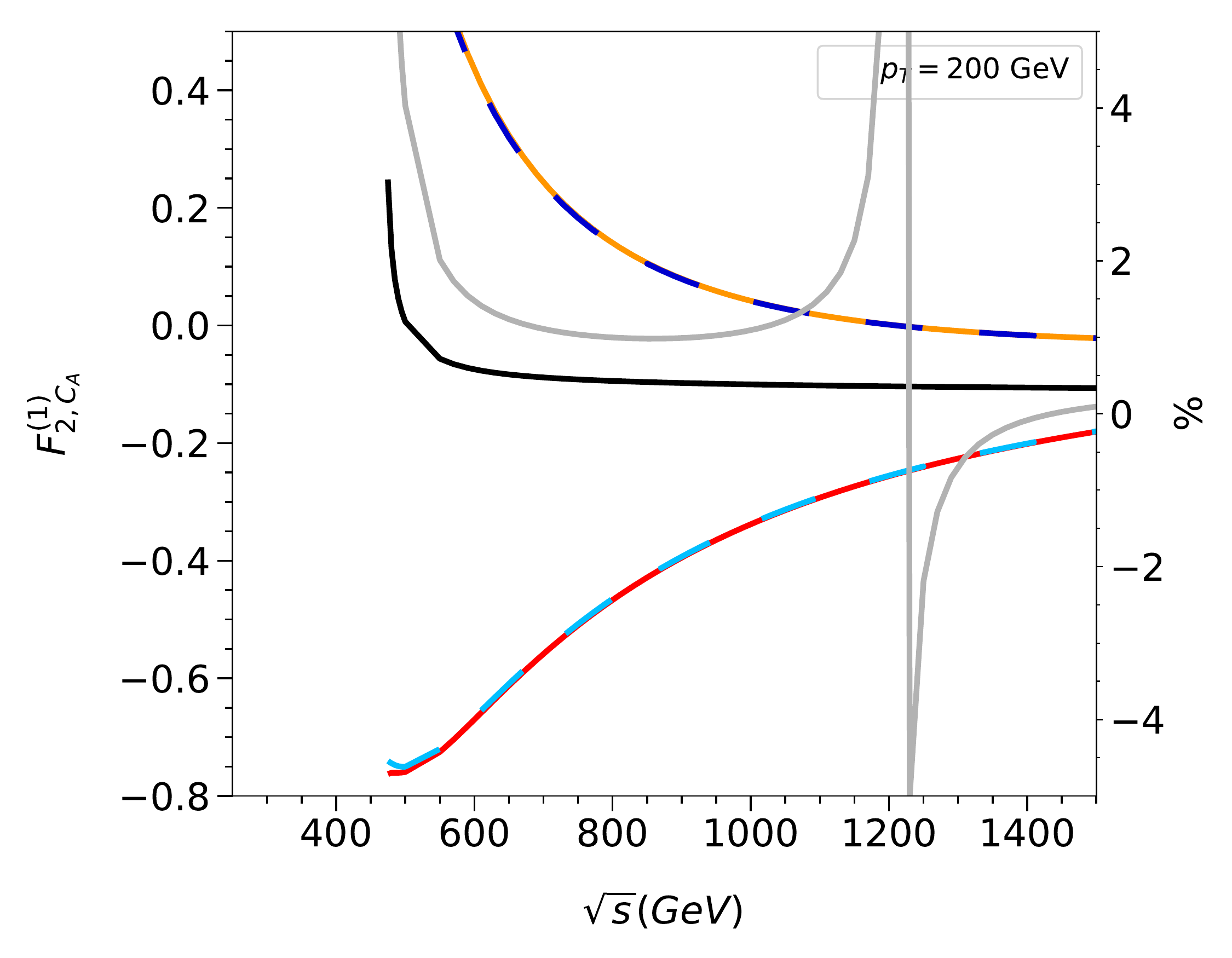} &
    \includegraphics[width=0.31\textwidth]{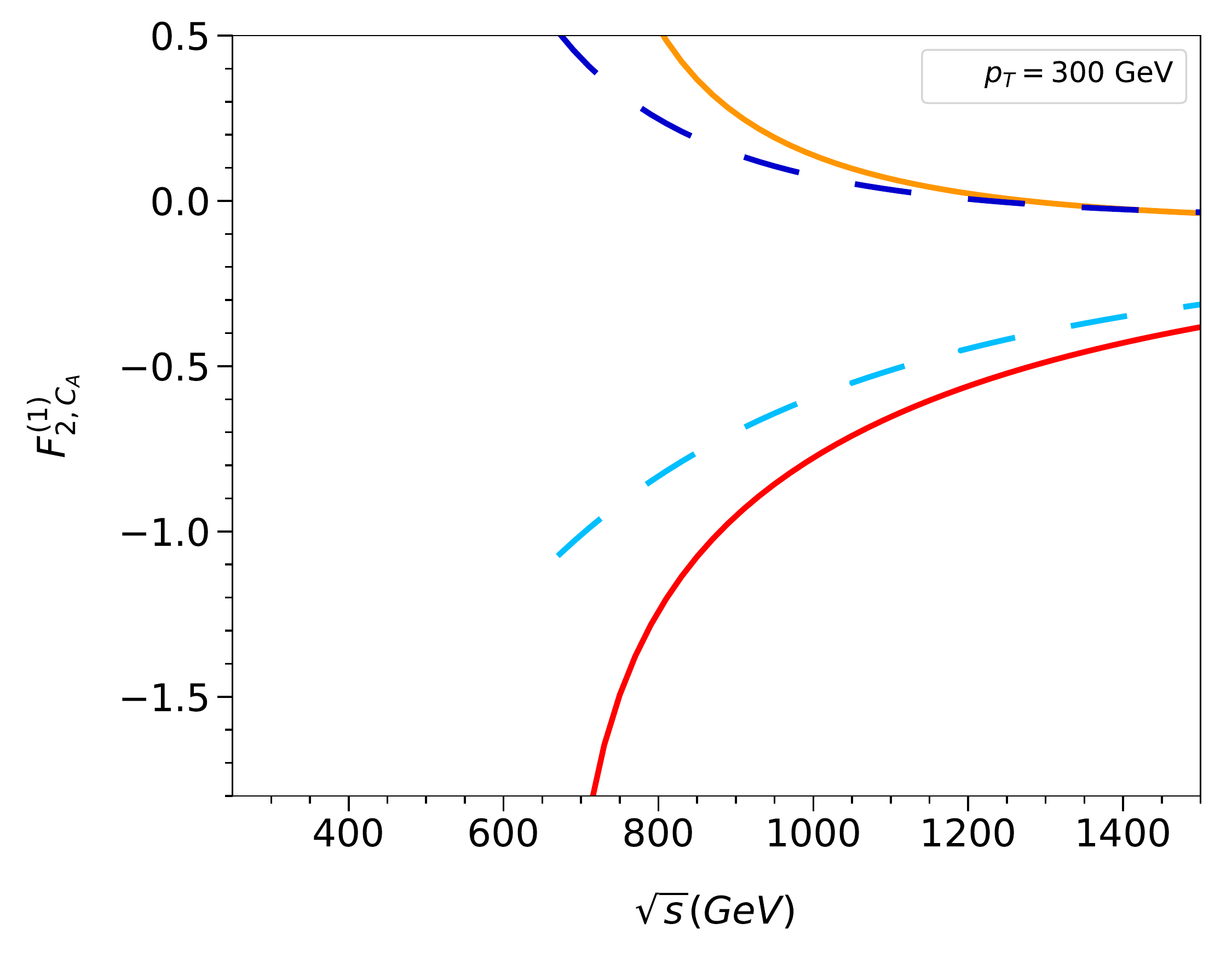} 
  \end{tabular}
  \caption{\label{fig::F22l_cA}$C_A$ contribution to the two-loop form factor
    $F_{\rm box2}^{(1)}$ as a function of $\sqrt{s}$
    for various values of $p_T$.}
\end{figure}

In Figs.~\ref{fig::F12l_cR},~\ref{fig::F12l_cA},~\ref{fig::F22l_cR}
and~\ref{fig::F22l_cA} we show the results for the
two colour factors of the two-loop form factors, for various values of $p_T$, as
a function of $\sqrt{s}$.  For the small-$t$ expansion terms up to $t^5$ are taken
into account and the high-energy expansion includes Pad\'e approximations with
at least $(m_t^2)^{49}$ and at most $(m_t^2)^{56}$ terms. In all cases quartic
terms in $m_H$ are included. Results for the high-energy form factors at the
deeper expansion depths considered here are provided in the ancillary files of this
paper~\cite{progdata}.

An exact result for the form factors is not at our disposal, however, we observe that the
approximations show a very similar behaviour as at one-loop order.  In
particular, we observe that for $100$~GeV$\lesssim p_T \lesssim 200$~GeV there
is a wide range in $\sqrt{s}$ where we find excellent agreement between the two
approximations.
We want to stress that for these $p_T$ values the small-$t$
expansion works well even for larger values of $\sqrt{s}$.  This
is demonstrated by the black and gray curves which show the relative percentage difference
between the small-$t$ and high-energy expansions for the real and
imaginary parts of the form factors, respectively.
For each value of
$100$~GeV$\lesssim p_T \lesssim 170$~GeV there is an overlap region in which the
relative difference is far below 1\%, and mostly even below $0.1\%$.
Note that the spikes in the gray and black curves are related to zeros
of the form factors.

For $p_T>200$~GeV we can rely on the high-energy expansion.  This is supported
by the fact that even for $p_T\approx 100$~GeV the high-energy expansion
agrees with the small-$t$ expansion even for $\sqrt{s}\approx 2 m_t$.  Note
that for $\sqrt{s} < 2 m_t$ the high-energy expansion is not valid for any value
of $p_T$ since no information about the top quark pair threshold is used for
the construction of the approximation. However, for $\sqrt{s} < 2 m_t$ the
small-$t$ approximation is always valid since $p_T$ is kinematically constrained
to be less than about 120~GeV.

For smaller values of $p_T$ the small-$t$ expansion is even more reliable, as can
been seen from the one-loop comparison in Tab.~\ref{tab::F_1loop_10}.

In summary, in Sections \ref{sub::high_energy} and \ref{sub::t0} we demonstrate that
the combination of the small-$t$ and high-energy expansions is sufficient to cover
the whole phase space, and that the final uncertainty is given only by the expansion
in $m_H$ which we estimate to be below 1\%.

In our current implementation in {\tt Mathematica} we have an explicit
dependence on all parameters ($m_t, m_H, s$ and $t$) which allows for
a straightforward change of parameter values or renormalization scheme.
Thus the computing time required to evaluate the form factors is not very
optimized. Nevertheless, it takes just a few seconds to evaluate the
small $t$ expansion. The numerical evaluation of the high-energy expansion
and the subsequent Pad\'e approximation takes between 40 and 50 seconds.
If required a significant speed-up is possible.


\subsection{\label{sub::Vfin}Virtual NLO corrections}

As a final comparison, we construct the infra-red subtracted virtual corrections,
following Ref.~\cite{Heinrich:2017kxx}. They are given by
\begin{eqnarray}
  \widetilde{\mathcal{V}}_{\textnormal{fin}} &=&
  \frac{\alpha_s^2\left(\mu\right)}{16\pi^2}\frac{G_F^2 s^2}{64}
  \left[   C + 2\left( {\tilde{F}_1^{(0)*}}
    \tilde{F}_1^{(1)} + {\tilde{F}_2^{(0)*}} \tilde{F}_2^{(1)} +
    \tilde{F}_1^{(0)} {\tilde{F}_1^{(1)*}}+\tilde{F}_2^{(0)}
          {\tilde{F}_2^{(1)*}} \right) \right] \,,
      \label{eq::Vtil}
\end{eqnarray}
with
\begin{eqnarray}
  C &=& \left[\left|\tilde{F}_1^{(0)}\right|^2+\left|\tilde{F}_2^{(0)}\right|^2\right]
        \left(
        C_A\pi^2 - C_A\log^2\frac{\mu^2}{s}
        \right)
        \,,
\end{eqnarray}
where $\alpha_s$ corresponds to the five-flavour strong coupling constant. 
It is convenient to introduce the $\alpha_s$-independent quantity
\begin{eqnarray}
  \mathcal{V}_{\textnormal{fin}} 
  &=& \frac{\widetilde{\mathcal{V}}_{\textnormal{fin}}}{\alpha_s^2(\mu)}
      \,.
\end{eqnarray}

We use the exact expressions for the one-loop form factors along with the
approximations discussed in the previous section for the two-loop form factors,
to compute $\mathcal{V}_{\textnormal{fin}}$. The triangle and double-triangle
diagrams are included in the form factors, as described in Eq.~(\ref{eq::F_12});
we use exact expressions for the double-triangle diagrams, while for the triangle
diagrams we use the expansions discussed above.

In Ref.~\cite{Davies:2019dfy} the high-energy expansions of
Refs.~\cite{Davies:2018ood,Davies:2018qvx,Mishima:2018olh} have been combined
with the exact, numerical two-loop results of~\cite{Heinrich:2017kxx}, such that
$\mathcal{V}_{\textnormal{fin}}$ can be evaluated at any phase-space point and
costly two-loop numerical integrations are only required in a restricted phase
space, namely for $p_T < 150$~GeV if $\sqrt{s}\geq 700$~GeV and for
$p_T< 200$~GeV if $\sqrt{s}< 700$~GeV. The results of~\cite{Davies:2019dfy}
are collected as data points in {\tt hhgrid}~\cite{hhgrid}. The
high-energy expansion used in~\cite{Davies:2019dfy} only includes terms up to
$m_t^{32}$, in contrast to the much deeper expansions which we consider in this
work.

\begin{figure}[t]
  \begin{tabular}{c}
    \includegraphics[width=0.99\textwidth]{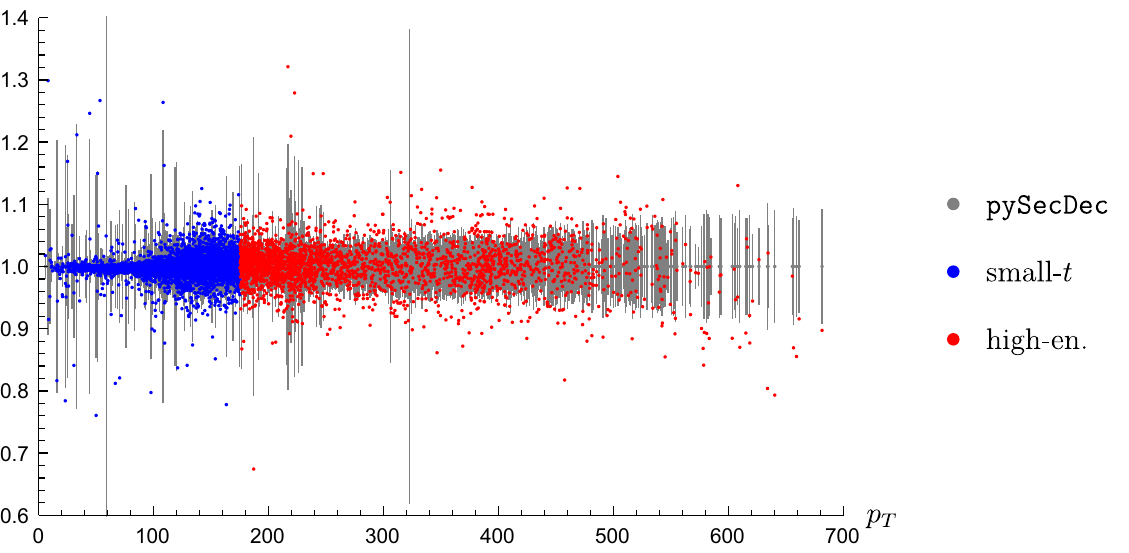}
  \end{tabular}
  \caption{\label{fig::Vfin}${\cal V}_{\rm fin}$ as a function of $p_T$, normalized to the central
  	values of the {\tt pySecDec}-evaluated points of {\tt hhgrid}. We switch from the small-$t$ to
  	the high-energy expansion at $p_T=175$~GeV.}
\end{figure}

In Fig.~\ref{fig::Vfin} we compare our new results for
$\mathcal{V}_{\textnormal{fin}}$ to those obtained using {\tt pySecDec}~\cite{Borowka:2017idc,Borowka:2018goh} in
Ref.~\cite{Davies:2019dfy}.
The grey data points and uncertainties correspond to the {\tt pySecDec} data
points, normalized to their central values. In comparison
the uncertainty of our approximation is negligible.\footnote{The systematic uncertainty
of about $1\%$ due to the expansion in $m_H$ up to quartic order is not shown.} The blue and red data
points are obtained from the small-$t$ and high-energy expansions, where
we normalize to the central values of the {\tt hhgrid} data. This plot may
be compared with Fig.~3 of Ref.~\cite{Davies:2019dfy}.

To quantify the agreement between our approximations and the {\tt pySecDec}
evaluations, the following table describes the proportion of points which are
contained within a number of {\tt pySecDec} error intervals.
\\[.5em]
\mbox{
\begin{tabular}{r|c|c|c}
	{\tt pySecDec} err.~intervals & 1$\sigma$ & 2$\sigma$ & 3$\sigma$ \\
	\hline
	small-$t$ & 0.57 & 0.85 & 0.92 \\
	high-energy & 0.65 & 0.94 & 0.99
\end{tabular}
}
\\[.5em]
We observe that the high-energy expansion demonstrates a
Gaussian behaviour, while the small-$t$ expansion shows a non-Gaussian
disagreement, which we ascribe to the systematic error due to the slower
convergence of the $m_H^2$ expansion in the lower-$p_T$ region, as shown in
Fig.~\ref{fig::mhexp}.

Let us finally compare to the findings of Refs.~\cite{Bellafronte:2022jmo,Degrassi:2022mro}.
In these works the integration over $t$ has been performed and an uncertainty
of $1\%$ is claimed. We present detailed results for the form factors
and find a several-digit agreement in the overlap region for $p_T\approx
100$~GeV to $200$~GeV. On the other hand, the result for the form factors in
Refs.~\cite{Bellafronte:2022jmo,Degrassi:2022mro} suggest a several-percent
difference between the expansions in some cases.

In Refs.~\cite{Bellafronte:2022jmo,Degrassi:2022mro} only 13
high-energy terms have been taken into account to construct a $[6/6]$ Pad\'e
approximant and thus the transition from the small-$p_T$ to the high-energy approximation 
is made at relatively high values of $p_T$ ($p_T\approx 312$~GeV and $340$~GeV
for the choices $\sqrt{s}=900$~GeV and $\sqrt{s}=2000$~GeV in Fig.~3 of
Ref.~\cite{Bellafronte:2022jmo}).
As we show in Figs.~\ref{fig::F11l} and \ref{fig::F21l} the $t\to 0$ expansion does
not perform very well in this region.
In our approach, we use the high-energy expansion at much smaller values of $p_T$ so
this region is well described.
Let us also mention that in Refs.~\cite{Bellafronte:2022jmo,Degrassi:2022mro} only
quadratic $m_H$ terms are taken into account which leads to a few-percent
systematic uncertainty at the level of the form factors.

In the small-$p_T$ expansion in Refs.~\cite{Bellafronte:2022jmo,Degrassi:2022mro} only
a $[1/1]$ Pad\'e approximant is constructed which means that three expansion
terms are available. In our analysis we use terms up to $t^5$, i.e. six
expansion terms; no Pad\'e improvement of the $t\to 0$ expansion is necessary.




\section{Conclusions}
\label{sec:conclusions}

In this paper we consider a $2\to2$ process with massive internal particles,
which is a multi-scale problem and thus notoriously difficult, both in an
analytic and in a numerical approach.  We show that the combination of
analytic expansions in two regions of phase space provides a complete
description of the two-loop virtual amplitude. On the one hand we
consider a deep expansion in the high-energy limit where the internal mass (in
our application, the top quark mass) is small compared to the Mandelstam
variables $s$ and $t$. On the other hand we perform an expansion in $t$ which
again eliminates a scale from the integrand. In both cases we expand in the
mass of the final-state particles.

We discuss in detail the two-loop corrections for $gg\to HH$ and show
that for this process no numerical integration is necessary
to obtain the differential virtual corrections.
Other processes such as $gg\to ZH$ or $gg\to ZZ$ can be treated in
analogy.

Using a similar approach to the one developed in this paper it might be
possible to extend the $t\to0$ expansion to three loops, yielding the
NNLO virtual corrections to this gluon fusion processes.
Possible bottlenecks, which have to be studied in the future, are
huge intermediate expressions and the integration-by-parts reduction
of the expanded amplitudes to master integrals.



\section*{Acknowledgements}  

This research was supported by the Deutsche
Forschungsgemeinschaft (DFG, German Research Foundation) under grant 396021762
--- TRR 257 ``Particle Physics Phenomenology after the Higgs Discovery''
and has received funding from the European Research Council (ERC) under
the European Union's Horizon 2020 research and innovation programme grant
agreement 101019620 (ERC Advanced Grant TOPUP).
The work of GM was supported by JSPS KAKENHI (No. JP20J00328).
The work of JD was supported by the Science and Technology Facilities Council (STFC) under
the Consolidated Grant ST/T00102X/1.






\end{document}